\documentclass[]{aa}  

\usepackage{graphicx}
\usepackage{xcolor}
\usepackage{float}
\usepackage{rotating}
\usepackage{booktabs}
\usepackage{mathtools}
\usepackage{placeins}
\usepackage{bm}
\usepackage{longtable}
\usepackage{adjustbox}
\usepackage{supertabular}
\usepackage{multirow}
\usepackage{tabularx}
\usepackage{url}
\usepackage{esdiff}
\usepackage{amsmath}
\usepackage{siunitx}
\usepackage[switch]{lineno}

\usepackage{hyperref}
\hypersetup{
  colorlinks=true,
    linkcolor=blue,
    citecolor=blue,
    urlcolor=magenta}

\usepackage{natbib}

\usepackage{txfonts}

\begin{document} 


   \title{Tracking active nests in solar-type pulsators:\\ 
   Ensemble starspot modelling of \textit{Kepler} asteroseismic targets}
    
   \titlerunning{Tracking active nests in solar-type pulsators}

   \author{S.N.~Breton\inst{1}
          \and
           A.F~Lanza\inst{1}
           \and 
           S.~Messina\inst{1}
          }
    \institute{INAF – Osservatorio Astrofisico di Catania, Via S. Sofia, 78, 95123 Catania, Italy \\
    \email{sylvain.breton@inaf.it}
    }

   \date{}

 \abstract{
 The satellite Planetary Transits and Oscillations of stars (PLATO), due to be launched late 2026, will provide us with an unprecedented sample of light curves of solar-type stars that will exhibit both solar-type oscillations and signatures of activity-induced brightness modulations. Solar-type pulsators only have moderate levels of activity because high levels of activity inhibit oscillations. This means that these targets represent a specific challenge for starspot modelling. 
 In order to assess the possibilities that PLATO will soon open, we wish to characterise the morphology of active regions at the surface of stars for which we also have a detection of solar-like acoustic oscillations. 
 In this context, we report the results of an ensemble starspot modelling analysis of the Sun and ten solar-type pulsators observed by the \textit{Kepler} satellite. 
 We implement a Bayesian starspot modelling approach based on a continuous-grid model, accounting for the combined starspot and facular contribution to activity-induced brightness modulations.
 From our analysis, we find that several stars of our sample exhibit clear signatures of stable longitudinal active nests while sharing activity levels and convection versus rotation regimes similar to the solar regime. By searching for modulations in the reconstructed starspot coverage, we found significant periodicities that we identify as possible signatures of cyclic modulations similar to the quasi-biennal oscillation or the Rieger cycle. We can infer the corresponding intensity of the magnetic field at the bottom of the convective envelope based on the hypothesis that internal magneto-Rossby waves acting on the tachocline cause these modulations.
 }


   \maketitle

\section{Introduction \label{section:introduction}}

Persistent active nests that are detected at the surface of the Sun \citep[e.g.][]{deToma2000,Berdyugina2003,Usoskin2007} and solar-type stars \citep[e.g.][]{Rodono2000} might be a surface manifestation of giant convection cells \citep{Weber2013} or of low-frequency magneto-inertial waves, in particular, Rossby waves \citep[e.g.][]{Loptien2018,Gizon2021,Zaqarashvili2021} that propagate in the convective envelope and are related to the existence and strength of dynamo processes \citep[e.g.][]{Brun2017b,Zaqarashvili2018}.
The surface manifestation of these waves can in principle be characterised through the detection and follow-up of stellar Rieger cycles \citep{Rieger1984}, allowing us to constrain the amplitude of the dynamo-field strength \citep{Gurgenashvili2016}. 
The frequency of the waves is indeed directly related to the strength of the magnetic field through the Alfvén speed \citep{Zaqarashvili2007}. 
Beyond characterisations in the Sun \citep[e.g.][]{Rieger1984,Lean1989,Carbonell1990,Gurgenashvili2017,Gurgenashvili2021}, evidence of the signature of Rieger-like cycles has been provided in solar analogues through a periodogram analysis of light curves \citep[e.g.][]{Gurgenashvili2022}, the activity index modulation \citep[e.g.][]{Distefano2017}, or starspot modelling \citep[e.g.][]{Lanza2009,Lanza2019}. 
Because of the uncertainties in the parametrisation of analytical models and the limitations of 2D and 3D numerical simulations, it is challenging to model the excitation mechanisms of these inertial waves through turbulent convection \citep{Bekki2022b,Philidet2023}. 

In parallel, it is crucial to focus on stars in which it is possible to detect stochastically excited acoustic modes \citep[p modes, see e.g.][]{Aerts2010,ChristensenDaalsgardLectureNotes} because of the upcoming mission Planetary Transits and Oscillations of stars \citep[PLATO,][]{Rauer2014}, which will provide tens of thousands of light curves of solar-type seismic targets \citep[see e.g.][]{Montalto2021}.
This is a specific challenge because these stars tend to be low-activity targets \citep[e.g.][]{Mathur2019} compared to, for example, the bulk distribution of main-sequence low-mass stars observed by the \textit{Kepler} mission \citep{Borucki2010} , where photometric variability arising from active regions allowed measuring the rotation period \citep[see][]{Santos2019,Santos2021}. 
The explanation for this lies in the fact that high levels of activity inhibit stochastic oscillations \citep[e.g.][]{Chaplin2011b}. 
By enabling the possibility to obtain structural insights \citep[e.g.][]{SilvaAguirre2017,Creevey2017}, as well as a view on activity processes \citep[e.g.][]{Salabert2016,Santos2018,Thomas2019} and rotation measurements \citep[e.g.][]{Benomar2018,Hall2021} independent from photospheric indicators, seismic constraints can nevertheless be successfully combined with other analysis techniques in order to obtain exquisite insights into stellar dynamics.     

Exploring the differences and similarities between the Sun and stars in the surrounding range of stellar parameters represents a way forward to better understand the evolution of our own Solar System and the specificity of extrasolar worlds.   
In this regard, because their convective envelope is very thin and they have a convective core \citep[e.g.][]{Deheuvels2016}, F-type stars occupy a special place among solar-type stars. The lack of reliable calibrators for angular momentum redistribution and loss means that additional observational insights are required to understand their dynamical evolution \citep[e.g.][]{Spada2020,Betrisey2023}. 
Even if the signatures of starspot activity \citep{Mathur2014} and stochastically excited acoustic modes \citep[e.g.][]{Chaplin2011,Appourchaux2012,Lund2017} are still detectable for these stars, their location on the transition path between intermediate- and low-mass stars make them critical targets that can provide us with a better understanding of the dynamical behaviour of both these populations \citep[e.g.][]{Breton2022simuFstars}, in particular because they might open the way to measuring the rotation rate of deep stellar layers of main-sequence solar-type stars \citep{Breton2023}. This information is still missing in the landscape of stellar physics \citep[e.g.][]{Appourchaux2013,Belkacem2022}. In particular, \citet{Mathur2014} suggested that the long-term stability of the photometric rotational modulation observed in F-type stars could result from the existence of active longitudes in the photosphere of these stars, while \citet{Mittag2019} speculated that their activity cycle could be shorter and of different nature than those of cooler solar analogues. 


With the PLATO mission soon to be launched, it is therefore crucial to collect more insight into the activity-induced brightness variations of stars exhibiting solar-type properties. It is pivotal to connect these variations to the internal processes in the stellar convective envelope and below, especially to internal waves. 
In this paper, we use a starspot-modelling approach to investigate the similarities and differences in the photospheric optical signature of the Sun and a selected sample of \textit{Kepler} G- and F-type asteroseismic targets. In particular, we try to detect and study the morphology of active nests in other solar-type pulsators, and we discuss the insights they can provide into the convective envelope dynamics of these stars.
We also search for signature of short-term cycles in the starspot models we compute.
The layout of the paper is as follows. 
In Sect.~\ref{sec:data} we present the stellar sample for our spot-modelling analysis.
In Sect.~\ref{sec:spot_modelling} we describe the details of the spot-modelling procedure we implement here, we explain how longitudinal maps of spots can be constructed, and we search for significant temporal modulations in the starspot coverage. 
Section~\ref{sec:the_sun} presents the results we obtained with the method on a solar time series, and the same analysis is applied on a \textit{Kepler} asteroseismic target in Sect.~\ref{sec:kepler_stars}. In particular, we present evidence for the existence of stable active nests at an activity level and in convection versus rotation regimes similar to the solar activity level and regime. After searching for cyclic modulations, we discuss in Sect.~\ref{sec:discussion} the possible origin of the corresponding periodicities and the physics they might allow us to infer. In particular, we explore whether the connection of these periodicities to prograde or retrograde magneto-Rossby waves acting on the tachocline might allow us to infer the intensity of the magnetic field at the bottom of the convective envelope.
The conclusion and perspectives for this work are given in Sect.~\ref{section:conclusion}.

\section{Considered data \label{sec:data}}

To demonstrate the capabilities of our spot-modelling approach, we considered both solar and stellar light curves. 

\subsection{Solar time series \label{sec:solar_time_series}}

The solar time series we considered was obtained by combining the green and red channels of the instrument called Sun Photometers of the Variability of Solar Irradiance and Gravity Oscillations \citep[VIRGO/SPM,][]{Frohlich1995} on board the Solar Heliospheric Observatory \citep[SoHO,][]{Domingo1995}. 
The red channel observes the Sun at 862~nm, and the green channel observes it at 500~nm.
The composite time series using the two passbands is well suited to performing comparisons with \textit{Kepler} observations \citep[e.g.][]{Basri2010,Salabert2017}.
The long-term stability of VIRGO/SPM is affected by instrumental effects, especially orbital modulations, which are accounted for and corrected in the data we use. 
Nevertheless, the spot-modelling approach that we present in Sect.~\ref{sec:spot_modelling} allows avoiding this issue of long-term stability of the time series by considering independent segments with a length of $\sim$20 days for the modelling \citep[e.g.][]{Lanza2007}.

The time series spans a total duration of 23.7 years, from 23 January 1996 to 30 September 2019, which almost completely cover solar cycles 23 and 24 (only the last months of the solar minimum in cycle 24 in 2019 are missing).
Observations are almost continuous, except for the two large data gaps, one of 106 days in 1998, and the other of 33 days in 1999 \citep[e.g.][]{Garcia2005}. The first gap was due to loss of contact with SoHO, and the second gap was caused an update of the satellite software. 
We note that the gap corresponding to this second interruption is slightly longer in the time series we used. It was 50 days. 

\subsection{Kepler targets \label{sec:description_kepler_targets}}

We considered a set composed of ten G- to F-type stars that were observed by \textit{Kepler}. All of them are confirmed solar pulsators \citep{Chaplin2014} with activity modulations of moderate amplitude that are still sufficient for measuring their surface rotation \citep{Santos2021}. 
The main properties of the targets are summarised in Table~\ref{tab:considered_targets}.
They cover a mass range from 0.97 to 1.61~$\rm M_\odot$ and an effective temperature range, $T_\mathrm{eff}$, from 5270 to 6887~K\footnote{Concerning the case of the effective temperature adopted for KIC~9226926, see the corresponding discussion in \citet{Breton2023}.}. The fastest-rotating stars of the sample are KIC~3733735 and KIC~9226926: Their rotation period is between 2 and 3~days. KIC~8006161 is the only target with a rotation period longer than that of the Sun: it is 31.71~days. 
We computed the Rossby numbers, $Ro$, of the stars after \citet{Corsaro2021}, normalising them by the solar value $\rm Ro_\odot$. Here again, we find that all the stars of our sample have a $Ro/\mathrm{Ro_\odot}$ between 0.29 and 1.08. From the scaling argument presented for instance by \citet{Brun2017a}, we therefore expect the stars in our sample to exhibit a solar-type latitudinal differential rotation in which the equator rotates faster than regions at higher latitudes. 
Although it would be interesting to extend this sample towards slower rotators with higher $Ro$, slow rotators tend to be low-activity stars whose variability is dominated by faculae \citep[e.g.][]{Reinhold2019}. This means that they are not well-suited for an approach such as starspot modelling.
We used the photometric activity indicator \citep[$S_\mathrm{ph}$,][]{Mathur2014b} as a proxy to estimate the amplitude of the brightness modulations that are to be modelled with the starspot modelling. The order of magnitudes covered by the $S_\mathrm{ph}$ is $10^2$-$10^3$~ppm. We recall that the detectability limit for activity-induced brightness modulations is a few tens of a parts per million (ppm) \citep[e.g.][]{Santos2019,Santos2021}.
The stellar inclinations $i$ were taken from \citet{Mathur2014} and \citet{Hall2021}. 
The first come from a cross analysis using the photometric surface rotation period $P_\mathrm{rot}$, spectroscopic $v \sin i$, and the model-derived stellar radius $R_\star$. 
When both measurements were available, we favoured \citet{Hall2021} asteroseismic measurements because they depend less strongly on the model.  
These measurements of $i$ are important for starspot modelling because they allow the model to partially lift the degeneracies in the longitudinal or latitudinal distribution of the spots \citep[see e.g. Fig.~2 from][]{Roettenbacher2013}.

The photometric magnetic activity of KIC~3733735, KIC~6508366, KIC~7103006, KIC~9226926, and KIC~10644253 was studied by \citet{Mathur2014} for a larger sample of F-type solar pulsators. The authors reported evidence for the existence of active longitudes in the case of KIC~3733735 and KIC~9226926, and cyclic modulations in the case of KIC~3733735 and KIC~10644253. 
The frequency shifts induced by magnetic activity in the acoustic oscillations of KIC~10644253 were studied by \citet{Salabert2016}.
Asteroseismic measurements of latitudinal differential rotation were performed by \citet{Benomar2018} for KIC~8006161, KIC~8379927,  KIC~9025370, and KIC~10068307. Finally, it has to be underlined that KIC~8006161 (HD173701) is a metal-rich solar analogue that has drawn a particular level of attention and dedicated analysis efforts in the community. It is one of the rare targets for which we possess both a clear characterisation of a Schwabe-like activity cycle with a duration of $\sim 7.4$ years \citep{Karoff2018} and a four-year-long asteroseismic time series from \textit{Kepler}. In particular, \citet{Bazot2018} reconstructed a butterfly diagram for KIC~8006161  during the span of the \textit{Kepler} mission by analysing the properties of its p-modes.

The light curves we used correspond to the \textit{Kepler} observations with the four-year long cadence, which were calibrated with the KEPSEISMIC method \citep{Garcia2011,Garcia2014,Pires2015} and were filtered with a high-pass filter at 55 days. 
In particular, the KEPSEISMIC method was optimised to maintain the integrity of the activity rotational modulations and solar-like oscillations in the calibrated light curves \citep[e.g.][]{Santos2019,Breton2021}. 

\begin{table*}[ht!]
    \centering
    \caption{Global parameters of the considered \textit{Kepler} targets.}
    \begin{tabular}{ccccccccccc}
    \hline
    \hline
    KIC & Id & $T_\mathrm{eff}$ (K) & $M_\star$~($\rm M_\odot$) & $R_\star$~($\rm R_\odot$) & $P_\mathrm{rot}$ (day) & $Ro / \mathrm{Ro_\odot}$ & $S_\mathrm{ph}$ (ppm) & $i$ ($^o$) & Origin \\
    \hline
    3733735  & 1  & $6676 \pm 80$  & $1.26 \pm 0.06$ & $1.38 \pm 0.03$ & $2.57 \pm 0.19$  & 0.52 & $239 \pm 20$ & $31 \pm 4$ & S \\
    6508366  & 2  & $6331 \pm 77$  & $1.58 \pm 0.03$ & $2.21 \pm 0.02$ & $3.72 \pm 0.33$  & 0.29 & $231 \pm 16$ &  $87 \pm 3$ & A \\
    7103006  & 3  & $6344 \pm 77$  & $1.45 \pm 0.04$ & $1.95 \pm 0.02$ & $4.66 \pm 0.46$  & 0.35 & $377 \pm 23$ &  $57 \pm 15$ & A \\
    8006161  & 4  & $5488 \pm 77$  & $0.97 \pm 0.03$ & $0.93 \pm 0.01$ & $31.71 \pm 3.19$ & 0.97 & $750 \pm 17$ &  $37 \pm 4$ & A \\
    8379927  & 5  & $6067 \pm 120$ & $1.12 \pm 0.03$ & $1.12 \pm 0.01$ & $17.09 \pm 1.31$ & 0.84 & $1188 \pm 39$ & $63 \pm 3$ & A \\
    9025370  & 6  & $5270 \pm 180$ & $0.97 \pm 0.01$ & $1.00 \pm 0.01$ & $23.21 \pm 3.46$ & 0.77 & $251 \pm 7$ &  $68 \pm 19$ & A \\
    9226926  & 7  & $6887 \pm 89$  & $1.34 \pm 0.07$ & $1.60 \pm 0.14$ & $2.20 \pm 0.22$   & 0.70 & $104 \pm 6$ & $50 \pm 6$ & S \\
    10068307 & 8  & $6132 \pm 77$ & $1.61 \pm 0.03$ & $2.16 \pm 0.01$ & $18.73 \pm 1.74$ & 1.08 & $103 \pm 3$ &  $42 \pm 6$ & A \\
    10454113 & 9  & $6177 \pm 77$ & $1.15 \pm 0.03$ & $1.24 \pm 0.01$ & $14.69 \pm 1.01$ & 0.95 & $267 \pm 9$ &  $41 \pm 27$ & A \\
    10644253 & 10 & $6045 \pm 77$ & $1.14 \pm 0.03$ & $1.11 \pm 0.01$ & $10.88 \pm 0.71$ & 0.53 & $348 \pm 14$ & $56 \pm 30$ & A \\
    \hline
    \end{tabular}
    \tablefoot{
    For all stars except for KIC~3733735 and KIC~9226926, $T_\mathrm{eff}$ was taken from \citet{Lund2017}, and $M_\star$ and $R_\star$ from the ASTFIT results from \citet{SilvaAguirre2017}. For KIC~3733735, $T_\mathrm{eff}$ was taken from \citet{Mathur2017}, while $M_\star$ and $R_\star$ were taken from \citet{Breton2023}. For KIC~9226926, all three parameters come from \citet{Mathur2017}.
    $P_\mathrm{rot}$ and $S_\mathrm{ph}$ were taken from \citet{Santos2021}, except for KIC~9226926, for which the reference $P_\mathrm{rot}$ we used is close to the 2.17~days measured by \citet{Mathur2014}. 
    For KIC~3733735, KIC~9226926, we used the stellar inclinations from \citet{Mathur2014}, which were derived by combining spectroscopic measurement of $v \sin i$, and photometric $P_\mathrm{rot}$, and $R_\star$ obtained from stellar modelling. 
    The inclination origin specified in the last column is therefore A for asteroseismic and S for spectroscopic measurements.
    The remaining values were taken from \citet{Hall2021} and were directly obtained from the p-mode parameter extraction performed in the periodogram. $Ro$ was computed using Eqs.~(1) and (5) from \citet{Corsaro2021}. The identifiers provided in the second columns are used to easily identify the targets in the summary figures of this work. 
    }
    \label{tab:considered_targets}
\end{table*}

\section{Starspot modelling \label{sec:spot_modelling}}

We used a starspot-modelling procedure to analyse the light curves of the stars presented in Sect.~\ref{sec:data}. The following subsections present the details of the method we used to model the activity brightness variations at the surface of the stars. 

\subsection{Continuous-grid starspot model \label{sec:model}}

Following \citet{Lanza2007,Lanza2009,Lanza2019}, for example, we modelled the stellar variability through a continuous spherical grid of surface elements with ($\theta_i$, $\phi_j$) coordinates, where $\theta_i$ is the co-latitude of the element, $\phi_j$ is its longitude, and the pair $(i,j)$ is its reference index. Each element has its specific intensity, $I_{i,j}$, parametrised by a starspot filling factor, $f_s$, that corresponds to the surface fraction of the element that is covered by dark spots with a contrast $c_s$. 
We considered that each element is also covered by a facular surface of the surface area fraction $Q f_s$, where $Q$ is the ratio of the faculae-to-spot surface, which is taken as uniform on the grid. Faculae have a contrast $c_f$ that in contrast to $c_s$, is a function of the limb angle, with
\begin{equation}
    c_f = c_{f0} (1 - \mu_{i,j}) \; ,
\end{equation}
where $c_{f0} = 0.115$ is a constant calibrated on the Sun \citep[e.g.][]{Foukal1991}, and the angle $\mu_{i,j}$ is given by 
\begin{equation}
    \mu_{i,j} (t) = \sin i \sin \theta_i \cos [\phi_j + \Omega (t - t_0)] + \cos i \cos \theta_i  \; ,
\end{equation} 
where $\Omega$ is the rotational frequency of the star, and $t_0$ is the time origin of the observations. It should be noted that $\mu_{i,j}$ encompasses the temporal dependence of the model.

In order to reduce the size of the parameter space, we considered common $f_s$ values for neighbouring blocks of surface elements: We typically considered $90\times180$ grids of surface elements, with $18\times18$ $f_s$ grid blocks. 
We considered a quadratic limb-darkening law for the unperturbed intensity, taking the form 
\begin{equation}
    I (\mu_{i,j}) / I (0) = 1 + a (1-\mu_{i,j}) + b (1 - \mu_{i,j})^2 \; . 
\end{equation}
The limb-darkening coefficients we used were interpolated from the model atmosphere parameters derived for \textit{Kepler} by \citet{Sing2010}.
Finally, $I_{i,j}$ is given by
\begin{equation}
    I_{i,j} (f_s, \mu_{i,j}) = \Big[1 + (Q c_f - c_s) f_s\Big] I (\mu_{i,j}) \; ,
\end{equation}
and the total normalised perturbed flux $F$ at a moment $t$ is given by
\begin{equation}
\label{eq:perturbed_flux}
   F (t) = \cfrac{\sum\limits_{i,j} I_{i,j} (f_s, \mu_{i,j}) a_{i,j} v_{i,j} \mu_{i,j}}{\sum\limits_{i,j} I (\mu_{i,j}) a_{i,j} v_{i,j} \mu_{i,j}} \; ,
\end{equation}
where $a_{i,j}$ is the surface element area with indices $(i,j)$, $v_{i,j}$ is its visibility, which is simply one if $\mu_{i,j} > 0$ and zero otherwise. The normalising factor $\sum_{i,j} I (\mu_{i,j}) a_{i,j} v_{i,j} \mu_{i,j}$ is the stellar unperturbed flux.

Following the recommendation by \citet{Basri2020} concerning the reliability of starspot-modelling inferences, \citet{Luger2021} extensively discussed the problem that the majority of the stellar variability lies in a null-space, resulting in a net-zero modulation for unresolved observations, especially for modulations with a high spherical degree $\ell$.  
However, in the case of active longitude, the use of long time-span light curves such as the one acquired by \textit{Kepler} enables us to track the regular emergence of spots with time, as demonstrated for example by \citet{Lanza2007}, who compared spot modelling of a solar photometric time series to the actual observed distribution of sunspots. 

\subsection{Bayesian analysis}

Assuming that the noise in the light curve follows a normal distribution, we can define a likelihood $\mathcal{L}$ of the form
\begin{equation}
    \mathcal{L} (\xi, F_\mathrm{obs,k}) = \prod \limits_k \exp \left( - \frac{\bigr[(F_\mathrm{obs,k} - F_k(\xi)\bigr]^2}{\sigma^2_k} \right) \; , 
\end{equation}
where $F_\mathrm{obs,k}$ is the $k$th observation in the light curve, performed at time $t_k$, with the uncertainty $\sigma_k$, $\xi$ is the set of filling factors over the grid, $f_s$, and $F_j$ is computed according to Eq.(~\ref{eq:perturbed_flux}). 

In order to lift the degeneracy of the continuous-grid spot-modelling problem, \citet{Lanza2007,Lanza2009,Lanza2019} imposed on $\mathcal{L}$ a maximum entropy constraint\footnote{In these papers, the authors choose to define and minimise a $\chi^2$ rather than working with a likelihood to maximise, but this is strictly equivalent.} \citep[see e.g.][for more details]{Lanza2016}. We chose here to adopt a Bayesian approach to obtain constraints on the filling factor distribution. The first step was to define the posterior distribution, $p(\theta, F_\mathrm{obs})$, to sample
\begin{equation}
    p(\theta, F_\mathrm{obs}) = p(F_\mathrm{obs}, \theta) p(\theta) \; ,
\end{equation}
where the probability $p(F_\mathrm{obs}, \theta)$ is the likelihood $\mathcal{L} (\theta, F_\mathrm{obs,k})$, and $p(\theta)$ is the prior distribution of the parameters that are to be sampled. As is traditionally done, we also dismissed the Bayes-formula $p (F_\mathrm{obs})$ denominator term as a normalising factor. 

We considered a truncated normal distribution $\mathcal{TN}$ with parameters for the prior distribution of each $(f_s)$ parameter
\begin{equation}
   f_s (\theta_i, \phi_j) \sim \mathcal{TN} (\mu=0; \sigma; 0; 1) \; ,
\end{equation}
where $\mu$ and $\sigma$ are the mean and standard deviation of the non-truncated distribution corresponding normal, while the truncation bounds were set to 0 and 1. With this choice, similarly to applying a maximum entropy constraint, we favour an unspotted background. 


In order to study the Sun and the selected set of solar-type stars, we used a maximum a posteriori (MAP) algorithm to find the maximum of the $p(\theta, F_\mathrm{obs})$ distribution.
For the purpose of this work, we designed the \texttt{loupiotes} module\footnote{The source code for \texttt{loupiotes} is available at \url{https://gitlab.com/sybreton/loupiotes} while the documentation is hosted at \url{https://loupiotes.readthedocs.io/en/latest}. In French, \textit{loupiotes} is a familiar term referring to a faint lamp that barely lights the surrounding space.}, a Python open-source framework that allows both fast MAP computation and graphic-processing-unit (GPU) accelerated Hamiltonian Monte Carlo \citep[HMC,][]{Duane1987,Neal2011} sampling, and it also allows manipulation and an analysis of \texttt{PyMC}-implemented \citep{pymc} starspot models. 


\subsection{Analysing the complete light curves \label{sec:analysing_complete_lc}}

While we wished to perform a study of four-year \textit{Kepler} light curves and of the complete VIRGO time series presented in Sect.~\ref{sec:data}, the model described in Sect.~\ref{sec:model} does not account for spot evolution. To overcome this issue, we subdivided the light curve into short segments with a length of $3/4 \, P_\mathrm{rot}$ with an overlap of $1/3 \, P_\mathrm{rot}$, and we considered a distinct spot model for each of them. The reference unspotted level of each segment, that is, a normalised flux of 1, was set to be 100~ppm above the maximum value of each segment.

Segments with a length shorter than one $P_\mathrm{rot}$ mean that not all the surface elements spend the same amount of time in the line of sight of the observer. 
Following \citet{Lanza2007}, considering each light curve segment, we therefore corrected for the integrated visibility of each surface element by multiplying each filling factor $f_s$ by the correcting factor 
\begin{equation}
\label{eq:visibility_correction}
    V (\theta_i, \phi_j) = \frac{1}{t_f - t_0} \int_{t_0}^{t_f} \mu_{i,j} (t) v_{i,j} (t)  \mathrm{d}t  \; ,
\end{equation}
where $t_f$ is the reference time of the last observation in the segment. Taking this correction into account, we computed for each segment the longitudinal distribution of the filling factors, $f_s (\phi_j)$, as the mean of the filling factors $f_s (\theta_i, \phi_j)$ at a given longitude $\phi_j$, weighted with respect to the area of each surface element,
\begin{equation}
    f_s (\phi_j) = \frac{\sum\limits_i a_{i,j} V (\theta_i, \phi_j) f_s (\theta_i, \phi_j)}
                        {\sum\limits_i a_{i,j}} \; .
\end{equation}

In order to smooth high-frequency artefacts that might remain, we applied a convolution triangular window in the temporal direction of the $f_s$ longitudinal distributions time series. The chosen width for the convoluting window was $5/3 P_\mathrm{rot}$, that is, given our segment-overlapping choice, we convoluted five segments at a time. 
Because the longitudinal resolution of the starspot modelling is also limited, we also applied a triangular convolution window in the longitudinal direction. The chosen width for the convoluting window was $62^o$ (i.e. a window of 31 bins, so that an odd number of bins was included in the window).
The longitudinal maps we obtained with this method allowed us to follow the appearance and the evolution of starspot nests along the time of observation. Due to the action of latitudinal differential rotation, we expect stable active nests to migrate with time in a reference frame that rigidly rotates with a given reference period. 
We underline here that active nests that migrate eastwards (i.e., with increasing longitude with time) correspond to spots that are located at latitudes where the rotation period is longer than the reference period used in the spot model, while active nests that migrate westwards (i.e. with decreasing longitudes with time) correspond to spots for which the rotation period is shorter than the reference period.

It is finally possible to compute the total spot coverage of the model as the weighted mean of the filling factors $f_s$ with respect to the area of each surface element.
We recall that this total spot coverage has to be considered as relative to an unknown brightness background activity level that does not produce modulations that are visible in the light curve \citep{Luger2021}. Finally, to search for possible cyclic modulation, we followed the approach of \citet{Gurgenashvili2021} and computed the wavelet Morlet transform \citep{Torrence1998,Liu2007} of the spot coverage. We also averaged the wavelet power spectrum along the time axis to obtain the global wavelet power spectrum (GWPS) in order to assess whether the frequencies persisted throughout the whole time series. 
Assuming a white-noise background for the signal, we computed the contours of significant modulations with a confidence level of 90~\%. 
An important point to discuss for the interpretation of this wavelet decomposition is connected to the beating signature in the total spot coverage. The spot coverage is strongly correlated with the $S_\mathrm{ph}$ indicator \citep{Salabert2017}, for which \citet{Mathur2014} showed that beating variations induced by stable active regions located at different latitudes could easily be confused with actual cyclic modulations. We show below (Sect.~\ref{sec:wavelets_kepler}) that we indeed retrieved the same type of beating signature in the wavelet decomposition of the modelled spot coverage.

\section{A difficult case with ground truth: The Sun \label{sec:the_sun}}

In this section, we show the results obtained from the analysis of the VIRGO/SPM composite time series presented in Sect.~\ref{sec:solar_time_series}. In particular, in order to prepare the interpretation of starspot modelling of asteroseismic targets in the following section, we compare the case of a spots-and-faculae model with $Q=9$ and a spots-only model with $Q=0$. The $Q=9$ parametrisation corresponds to the choice made for the reference model used by \citet[][]{Lanza2007}.
Because the maximum possible value for the spot-filling factors depends on the value of $Q$ and decreases as $Q$ increases, we adopted distinct $\sigma$ values for the prior distribution of the spots-and-faculae model and the spots-only model. In the first case, we used $\sigma = 0.01$, and in the latter case, we used $\sigma = 0.1$. 
Studying these two cases allows us to assess the dependence of identified spot features on the spots-to-faculae coverage ratio. We do not know this quantity for the \textit{Kepler} targets, although it should be noted that \citet{Karoff2018} were able to estimate the facular contribution to the variability of  KIC~8006161. 
We considered the Carrington synodic period, $\sim 27.28$~days,  as the choice for $P_\mathrm{rot}$.
This allowed us to straightforwardly identify the longitude of our spot models to Carrington longitudes. 

\begin{figure}[ht!]
    \centering
    \includegraphics[width=0.49\textwidth]{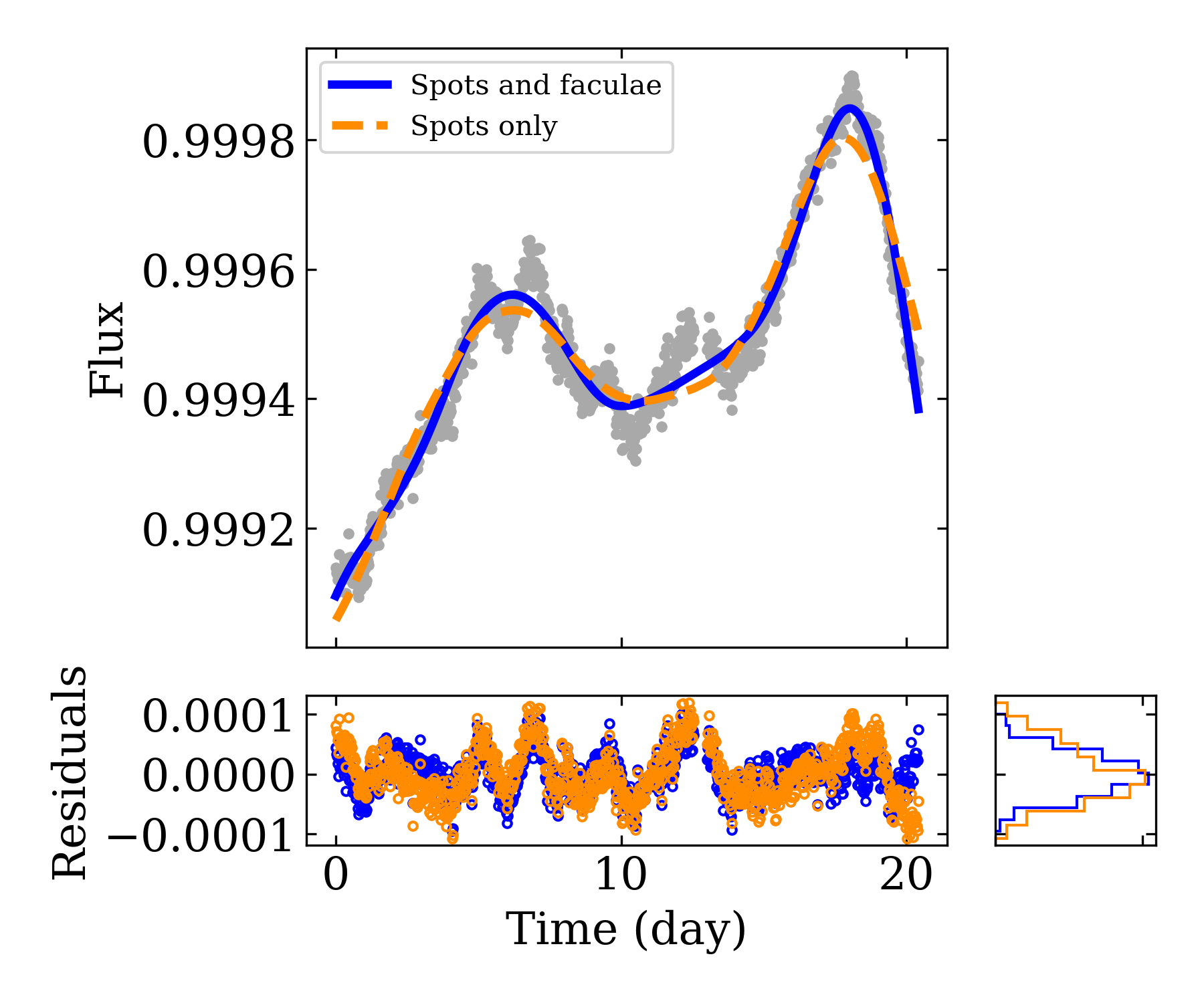}
    \caption{Example of VIRGO/SPM observations compared with the spot models.
    \textit{Top:} Best-fit model obtained with the spots-and-faculae model (blue) and the spots-only model (dashed orange) for the VIRGO/SPM segment spanning from 13 April 2002 to 11 May 2002 (grey dots).
    \textit{Bottom:} Best-fit residuals for the spots-and-faculae model (blue dots) and the spots-only model (orange dots). The histogram distribution of the residuals is shown on the right in the corresponding colours. 
    }
    \label{fig:example_fit_residual}
\end{figure}

We begin by showing in Fig.~\ref{fig:example_fit_residual} an example of best-fit models obtained with the spots-and-faculae model and the spots-only model applied on one segment of the light curve, spanning from 13 April 2002 to 11 May 2002, which is close to the activity maximum of cycle~23. Because the model does not account for spot evolution, some brightness modulations on short timescales are not captured. Nevertheless, the global brightness modulation on the segment extent is fairly well reconstructed. The histograms on the right of the bottom panel show that the residual distribution obtained from the spots-and-faculae models has a lower dispersion than that of the spots-only model. 

\begin{figure*}
    \centering    
    \includegraphics[width=0.33\textwidth]{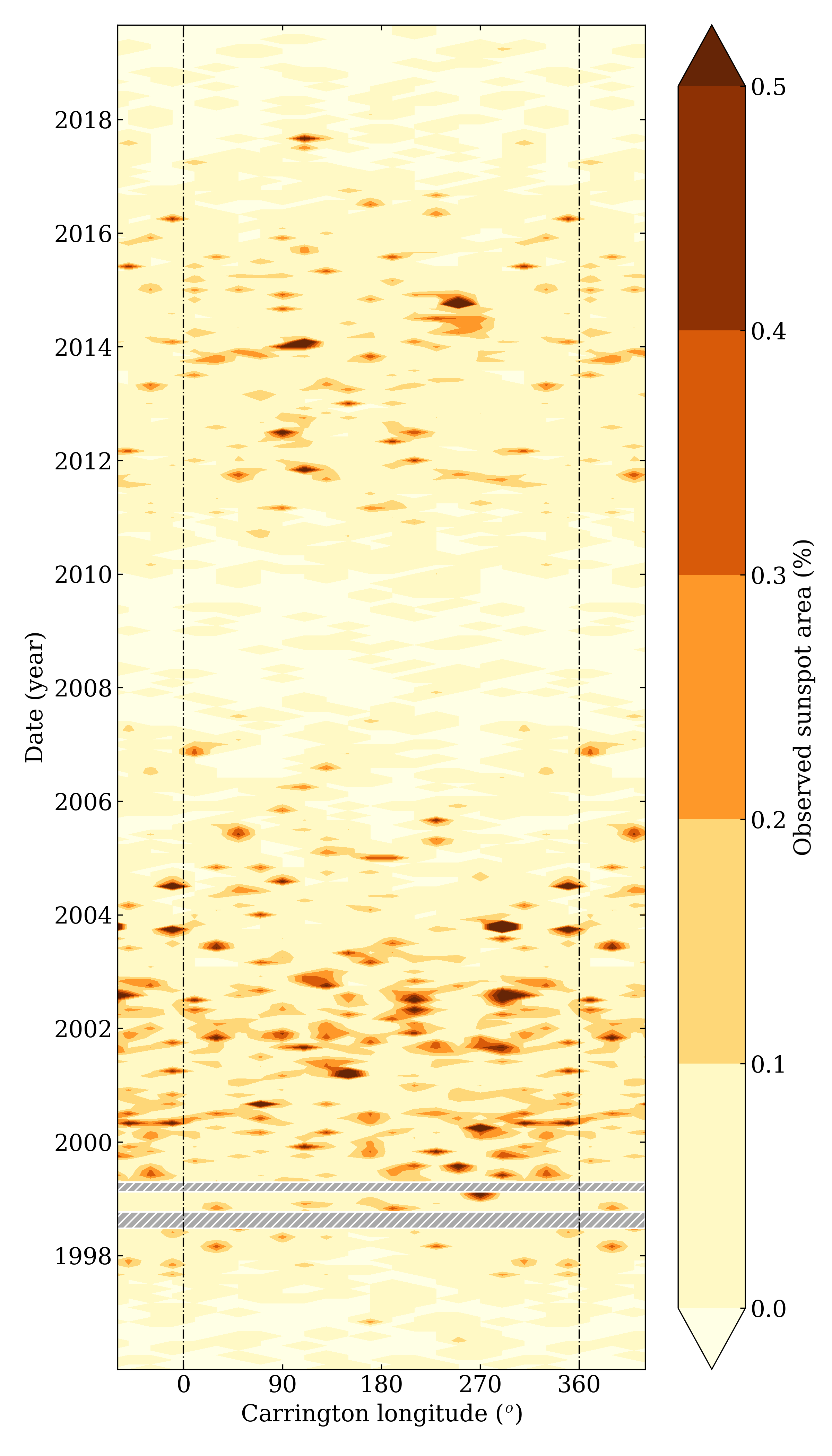}
    \includegraphics[width=0.33\textwidth]{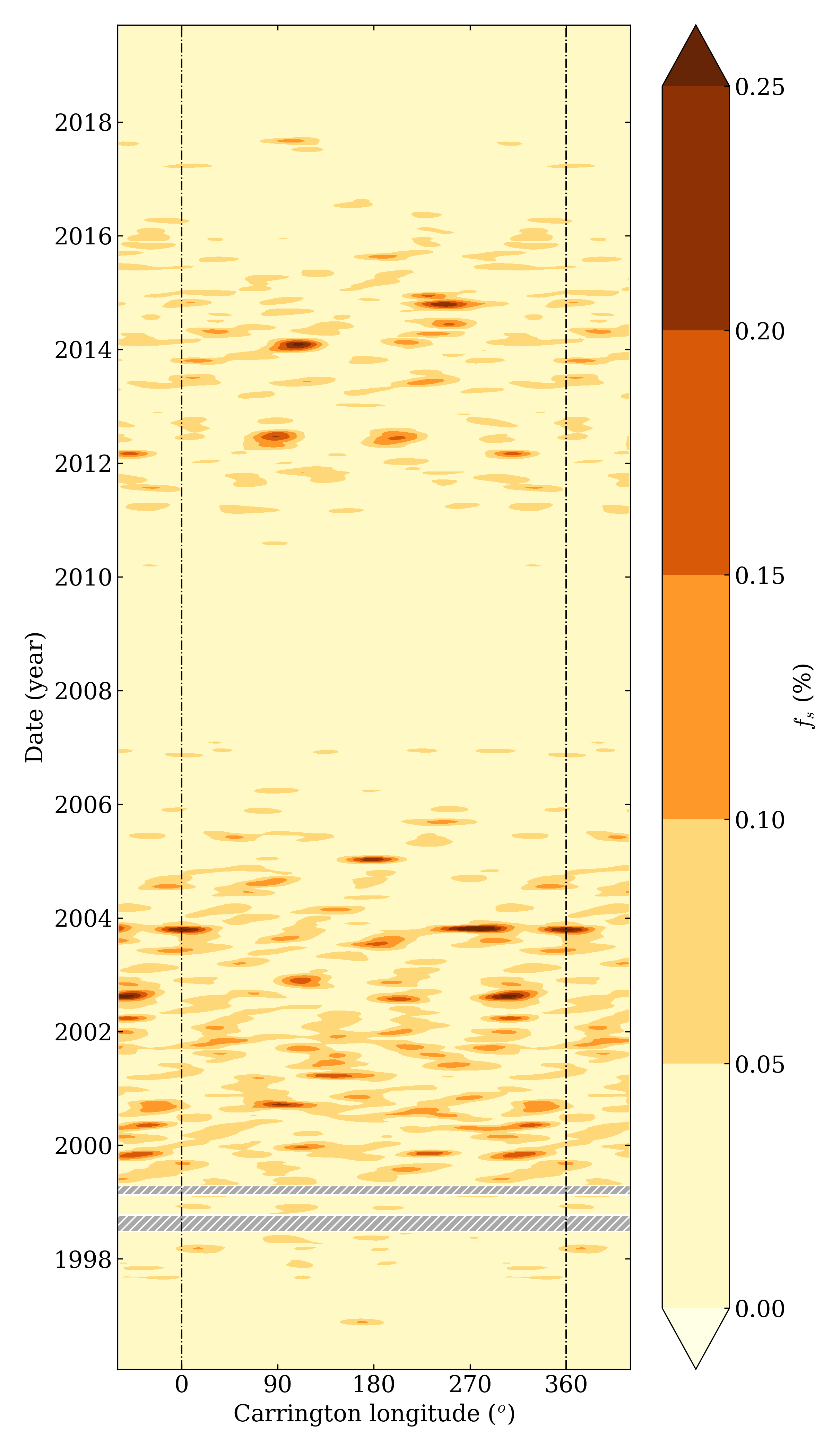}
    \includegraphics[width=0.33\textwidth]{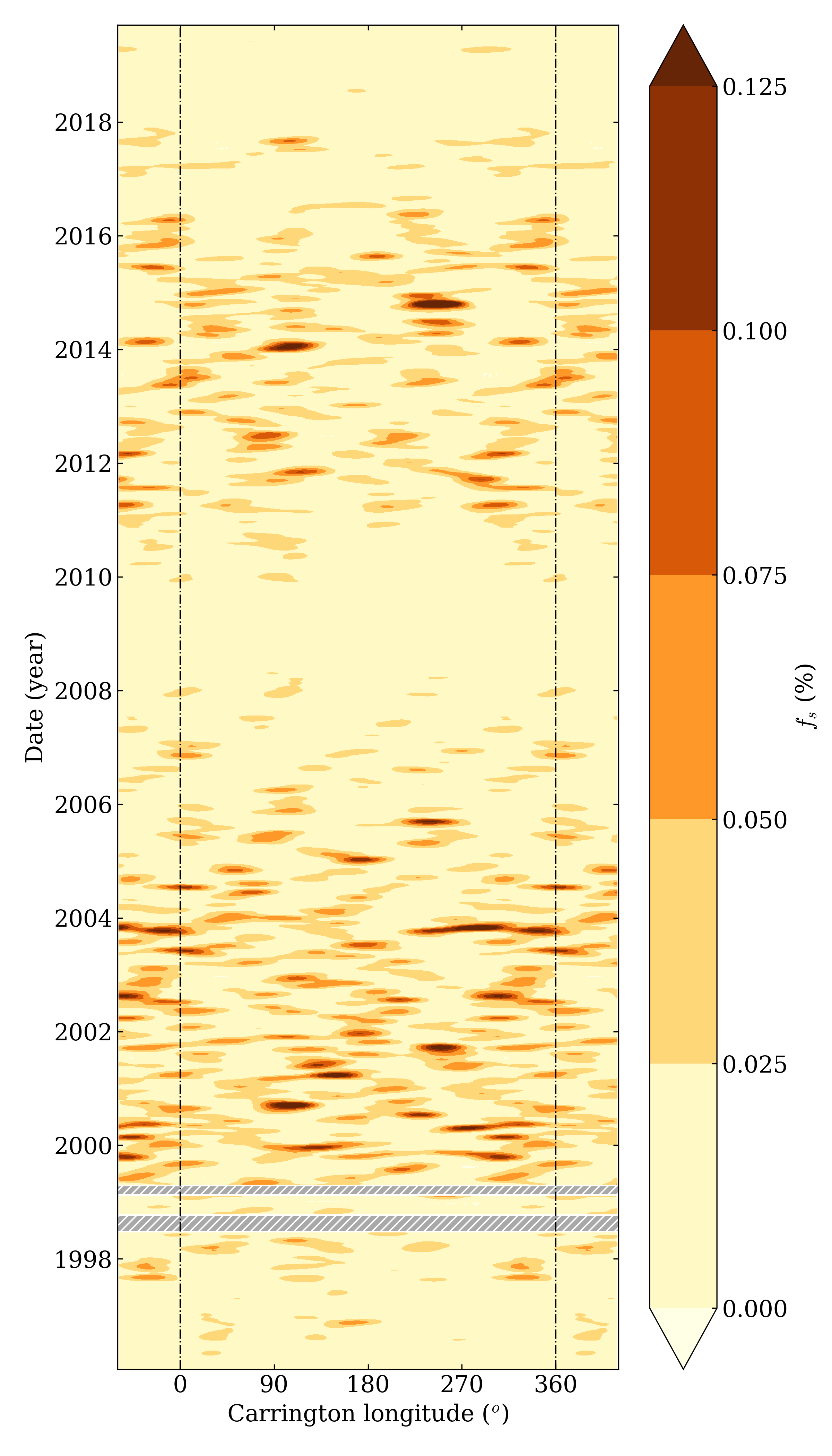}
    \caption{Observed longitudinal sunspot distribution (left) and  longitudinal spot distribution obtained from the analysis of the VIRGO/SPM solar time series for the spots-and-faculae model (middle) and the spots-only model (right) during cycles ~23 and 24. The longitudinal map is extended at each edge to reflect the periodic nature of the distributions. In all three panels, the grey hatched areas highlight the time interval for which VIRGO/SPM observations are lacking.
    }
    \label{fig:solar_longitudinal_distribution}
\end{figure*}

\begin{figure*}[ht!]
    \centering
    \includegraphics[width=0.98\textwidth]{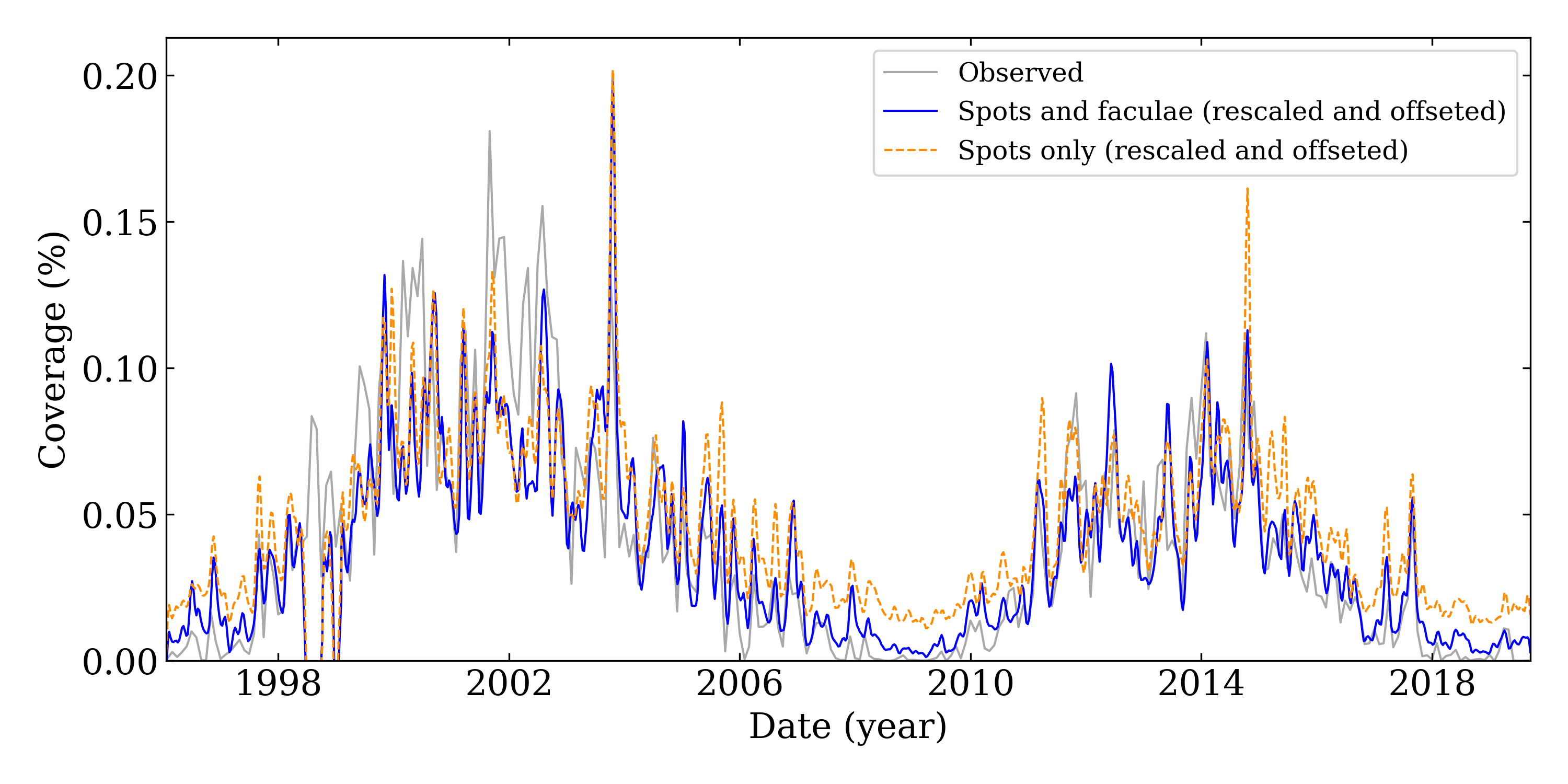}
    \caption{Spot coverage reconstructed from the analysis of the VIRGO/SPM solar time series for the spots-and-faculae model (blue) and the spots-only model (dashed orange) compared to the directly observed coverage (grey). The spots-and-faculae model coverage has been rescaled with a factor 2 and the spot-only model by a factor 4 for a better comparison of the temporal evolution between models and observations.
    An offset of 0.02~\% has been subtracted from the spot-model reconstructed coverage.
    }
    \label{fig:solar_spot_coverage}
\end{figure*}

\begin{figure}[ht!]
    \centering
    \includegraphics[width=0.49\textwidth]{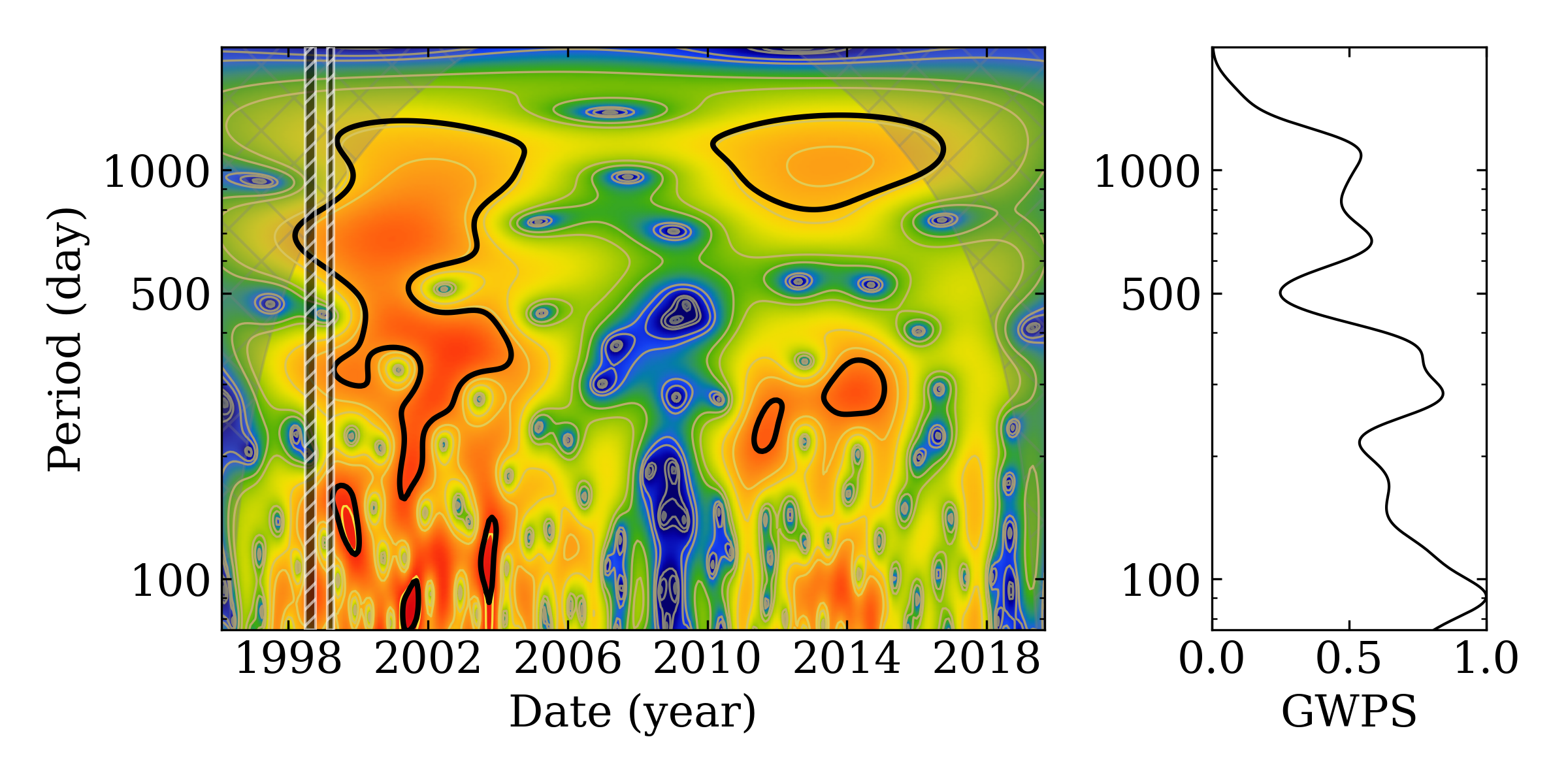}
    \includegraphics[width=0.49\textwidth]{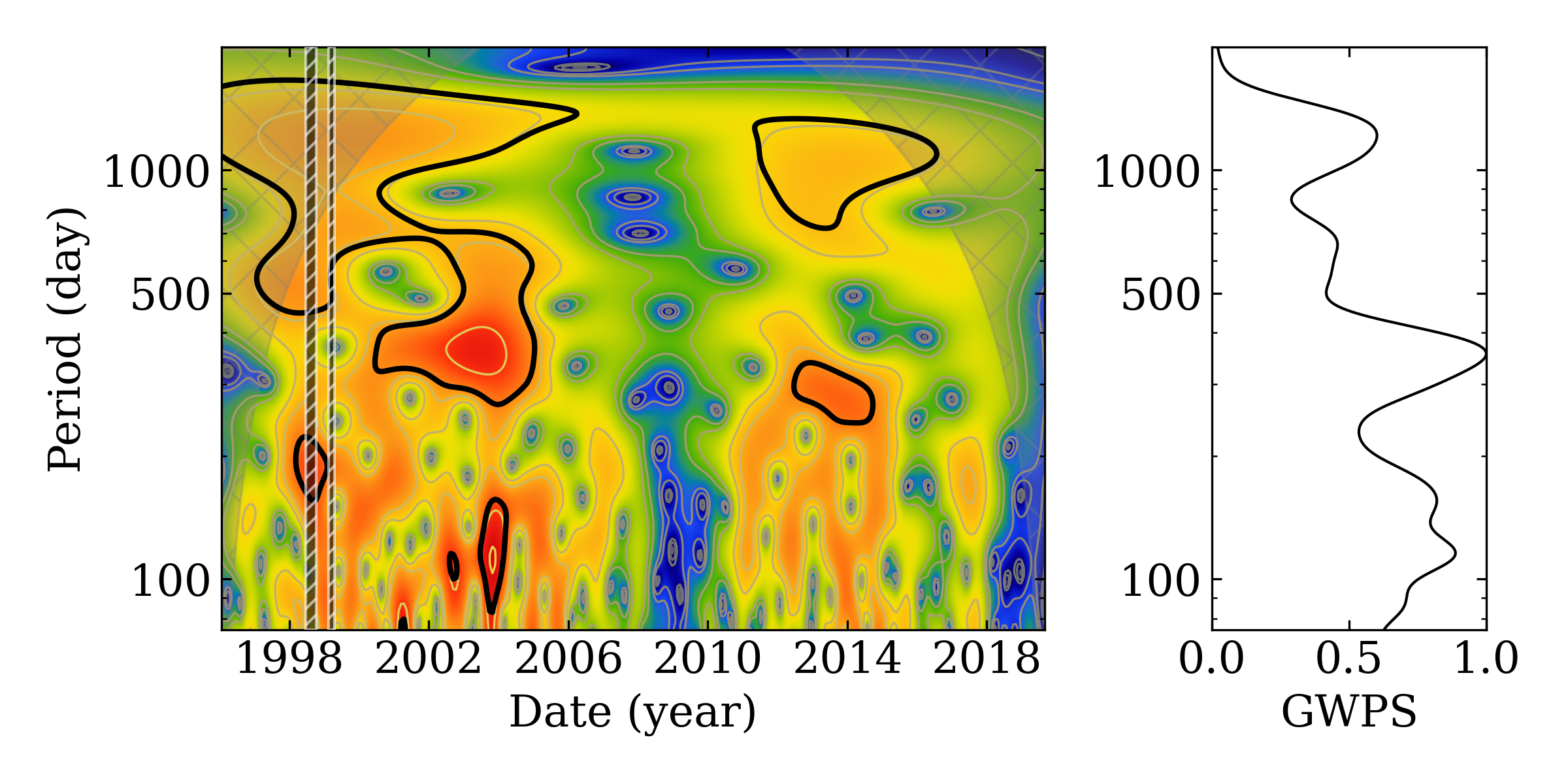}
    \includegraphics[width=0.49\textwidth]{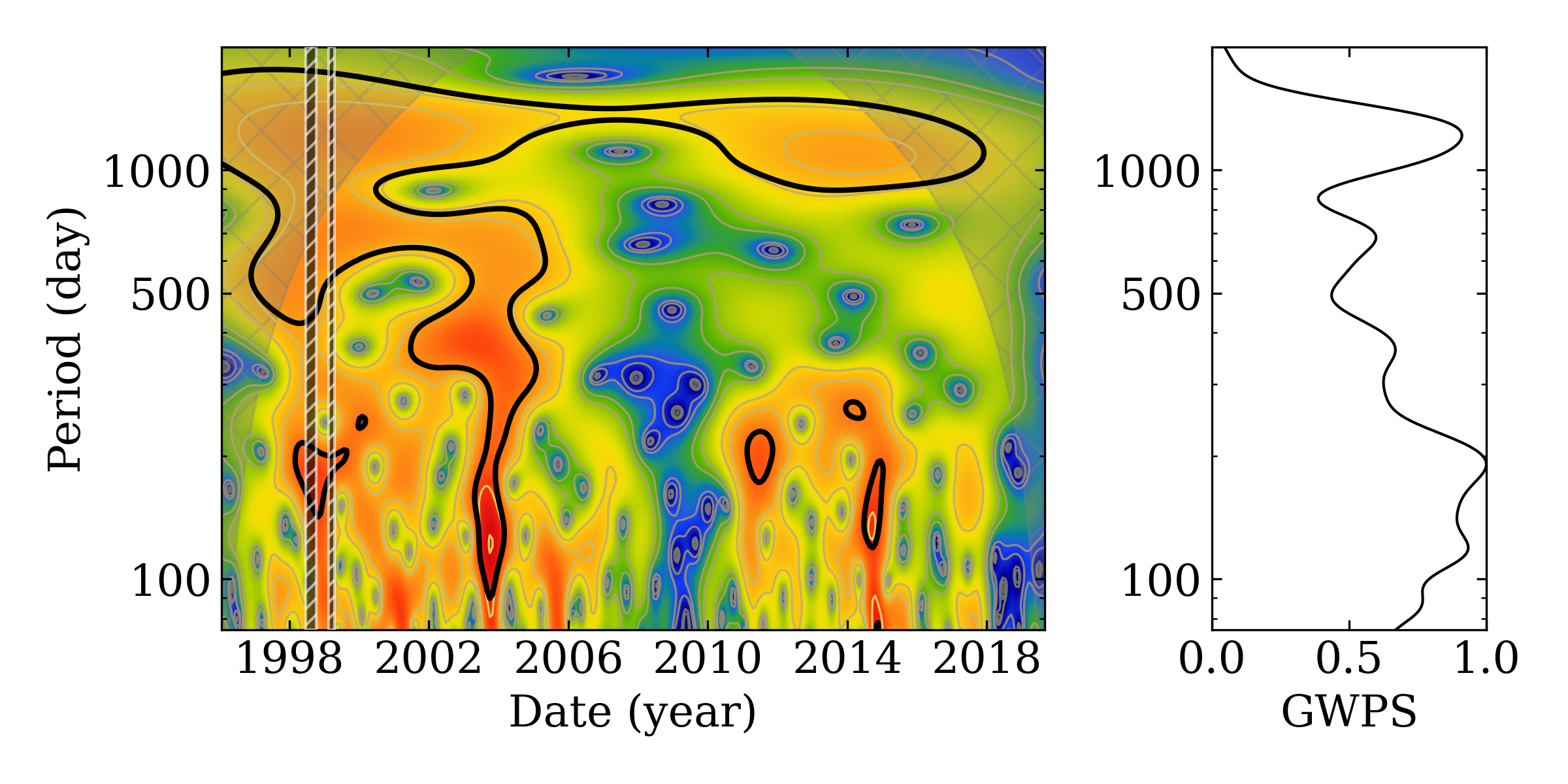}
    \caption{Morlet wavelet transform of the observed solar spot coverage (\textit{top}), the spot coverage reconstructed from the spots-and-faculae model (\textit{middle}), and the spot coverage reconstructed from the spots-only model (\textit{bottom}). In all three rows, the grey hatched area highlights the time interval for which VIRGO/SPM observations are lacking. The cone of influence of the wavelet transform is shown in grey.
    The global wavelet power spectrum is shown in the right panel of each row.}
    \label{fig:solar_wavelets}
\end{figure}

We show in Fig.~\ref{fig:solar_longitudinal_distribution} the longitudinal spot distribution we obtained in each case compared to the observed sunspot distribution. 
The observed distribution was computed from the US Air Force and US National Oceanic and Atmospheric Administration datasets, which are available online\footnote{The ASCII files used to compute the distributions can be downloaded at: \url{http://solarcyclescience.com/activeregions.html}.}.
The variation in the spot coverage with solar magnetic activity is clearly visible in the observed and modelled cases. As expected, the spot coverage increases as cycles 23 and 24 reach their maximum, and then it starts to decrease. The three panels highlight strong similarities between the observed spot distribution and the two distributions reconstructed from spot models. In particular, we recover the patterns of strong active nests that appear during the two cycles.  
We note that the resolution we obtained for the active region is quite high and can resolve spot structures around 20$\rm ^o$, which is significantly smaller than the 54$\rm ^o$ resolution discussed by \citet{Lanza2007}.
This can be explained by the fact that in the Sun, groups of spots are concentrated in narrow longitudinal bands, but we witness a super-resolution effect related to this in our spot model.

In Fig.~\ref{fig:solar_spot_coverage} we show the comparison of the temporal evolution of the total spot coverage for the spots-and-faculae model and for the spots-only model. The coverage of the spots-and-faculae model is multiplied by a factor 2, and the coverage of the spots-only model by a factor 4 to facilitate the comparison to observations. After rescaling, an offset of 0.02\% is also subtracted from the two models. Consistent with the observations, the modelled spot coverage for cycle~24 is lower than that for cycle~23. 

The results we obtain for the wavelet decomposition of the observed sunspot coverage and the two spot models are shown in Fig.~\ref{fig:solar_wavelets}. 
In order to remove the bias that the 11-year modulation would introduce in the decomposition while being entirely located in the cone-of-influence area that is affected by edge effects, we applied a 2-year high-pass filter on the spot coverage time series before computing the wavelet transform.  
The area with a significant excess of power (see Sect.~\ref{sec:analysing_complete_lc}) is encircled by a thick black line. 
In the wavelet decomposition of the observed sunspot coverage, we note two significant excesses of power between 100 and 200~days in 2000 and 2004, which might be a manifestation of the Rieger cycle during cycle~23 \citep[e.g.][]{Gurgenashvili2021}. During cycle~24, the significant modulations with the shortest timescale have periods between 200 and 400 days. Around the maxima of the two cycles, modulations at lower frequencies are visible between 500 and 1500 days and can be interpreted as the manifestation of the quasi-biennal oscillation in sunspot coverage \citep[e.g][]{Kostyuchenko2021}. 
The reconstructed spot coverage exhibits similar features, although some discrepancies are also visible. Nevertheless, we recover the quasi-biennal oscillation modulation in the spots-and-faculae and in the spots-only model at low frequency. 

In summary, with the possibility of comparing the modelling to a ground-truth reference, we underline that the variability properties of the Sun make the Sun an interesting but difficult case study for starspot modelling techniques. An important point to underline, as already noted by \citet{Lanza2007}, is that the results obtained for the spots-and-faculae model and the spots-only model are similar. This allows us to draw conclusions from the starspot model that do not rely on the unknown faculae-to-spots area ratio for stars other than the Sun. 

\section{\textit{Kepler} asteroseismic targets \label{sec:kepler_stars}}

After demonstrating the capabilities of our approach in the case of a solar time series, we now study the sample of \textit{Kepler} asteroseismic targets described in Sec.~\ref{sec:description_kepler_targets}. We start by considering the morphology of the active nests detected with the spot modelling, and then we consider the cyclicity of the spot coverage variations. 

\subsection{Signature of active nests}

\begin{figure*}[ht!]
    \centering
    \includegraphics[width=0.49\textwidth]{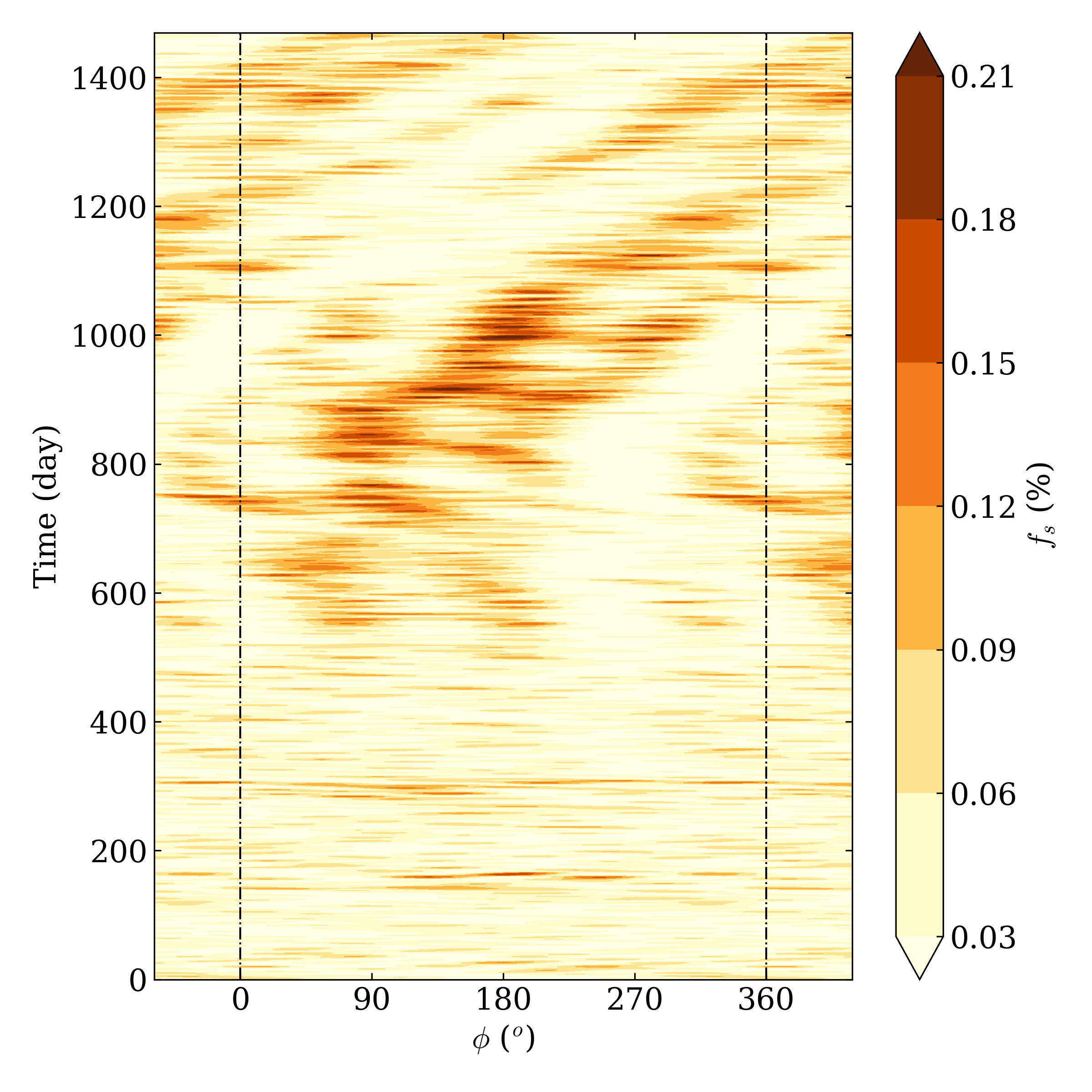}
    \includegraphics[width=0.49\textwidth]{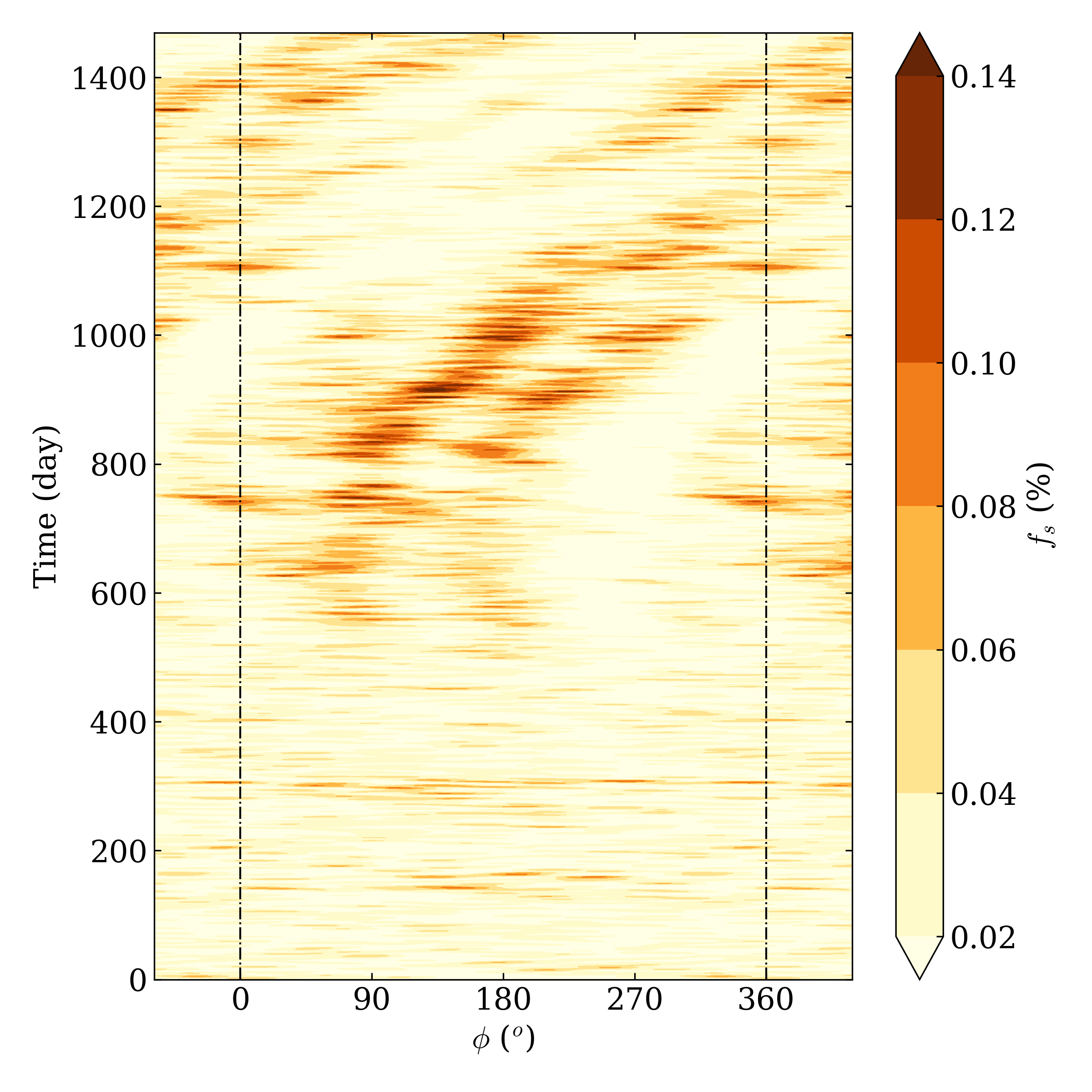}
    \caption{Longitudinal distribution for KIC~3733735 for the spots-and-faculae model (left) and the spots-only model (right).}
    \label{fig:kic3733735_longitude_map}
\end{figure*}

\begin{figure*}[ht!]
    \centering
    \includegraphics[width=0.49\textwidth]{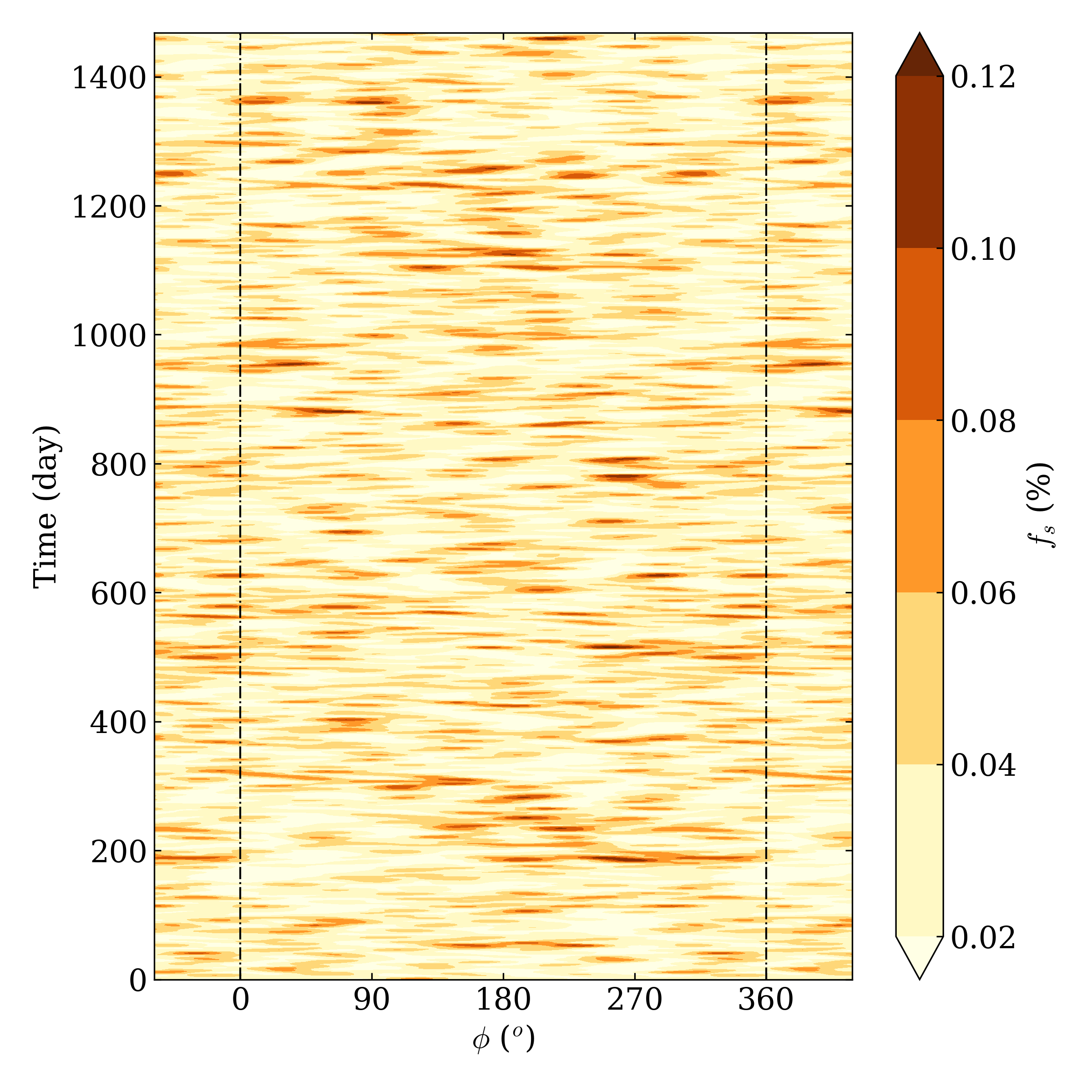}
    \includegraphics[width=0.49\textwidth]{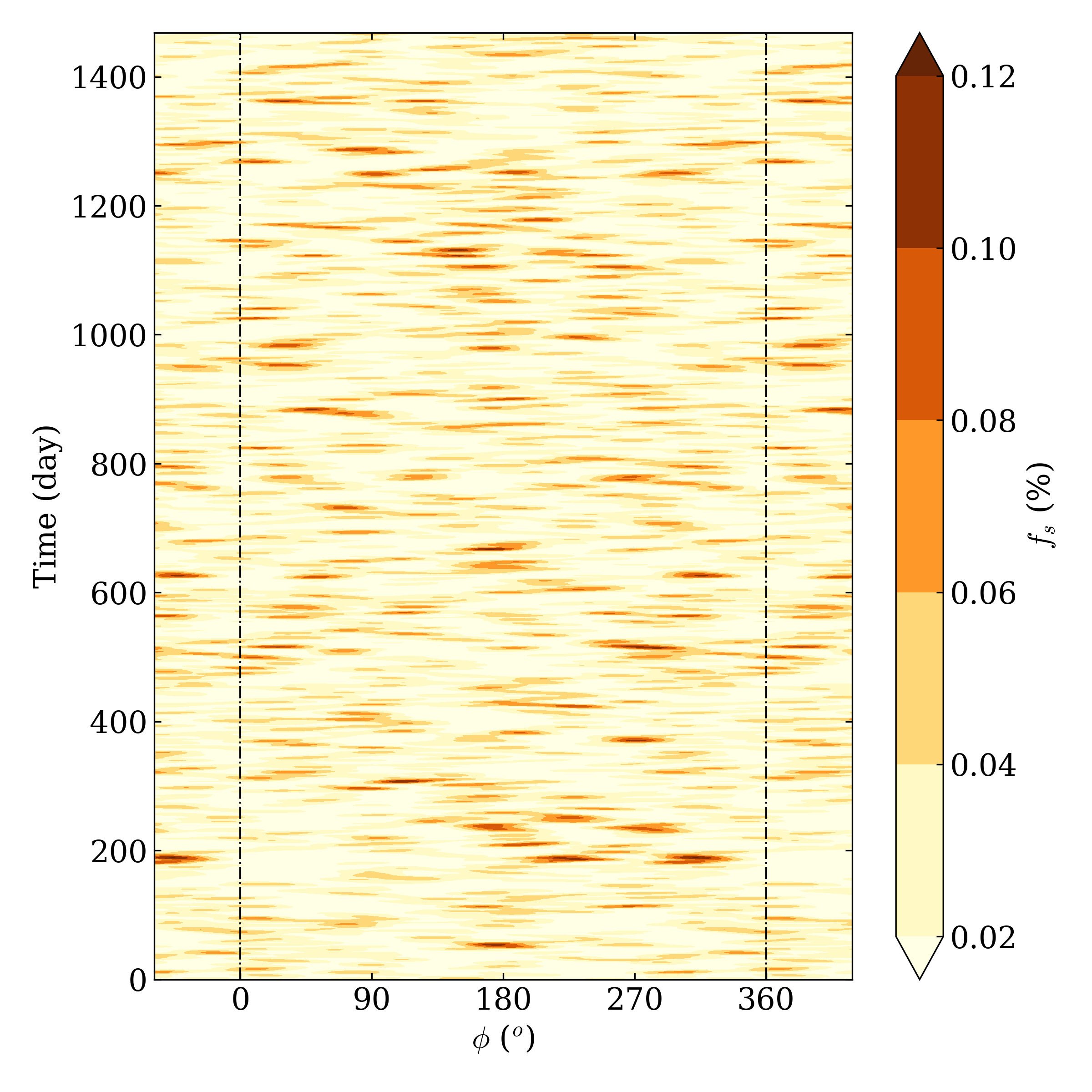}
    \caption{Longitudinal distribution for KIC~6508366 for the spots-and-faculae model (left) and the spots-only model (right).}
    \label{fig:kic6508366_longitude_map}
\end{figure*}

\begin{figure*}[ht!]
    \centering
    \includegraphics[width=0.49\textwidth]{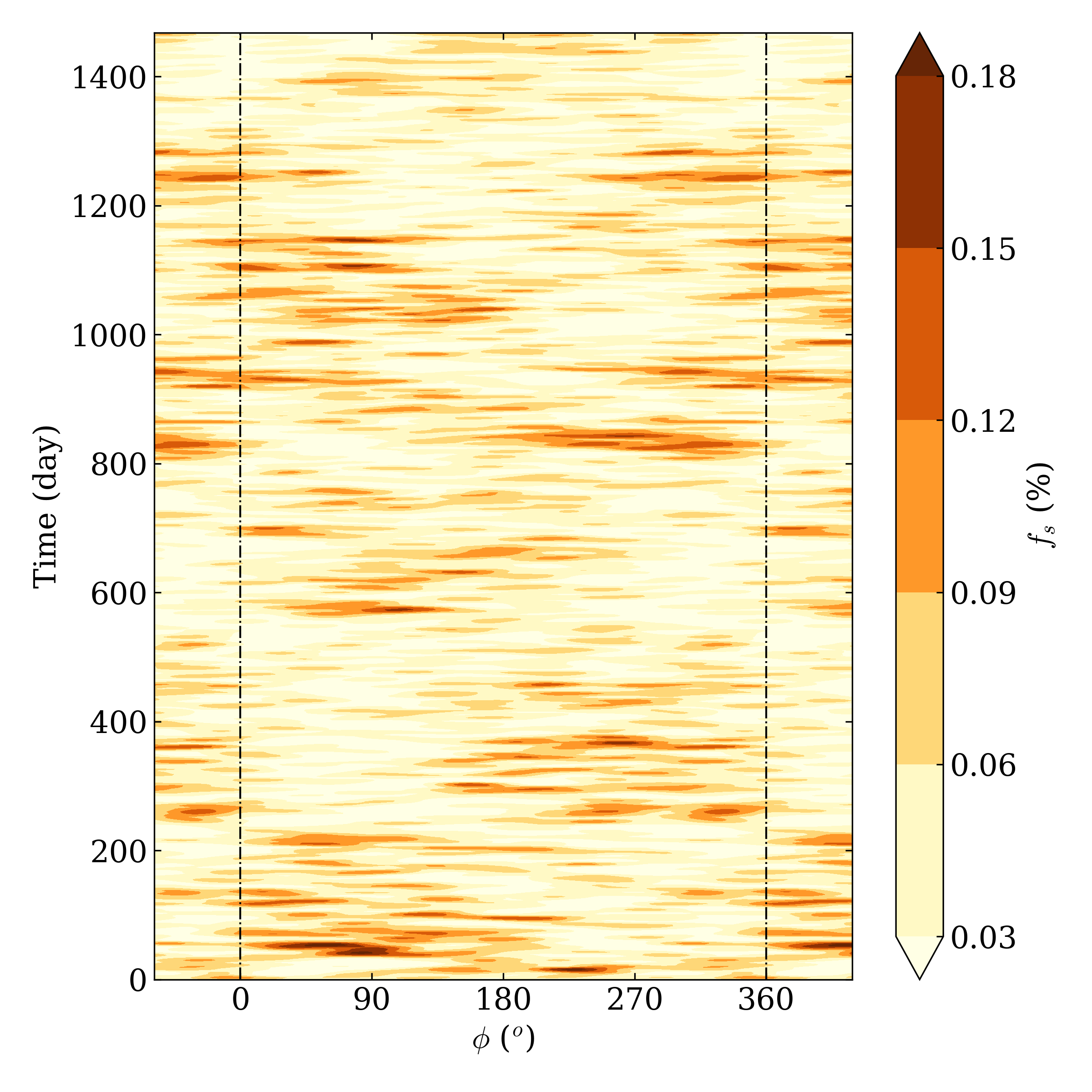}
    \includegraphics[width=0.49\textwidth]{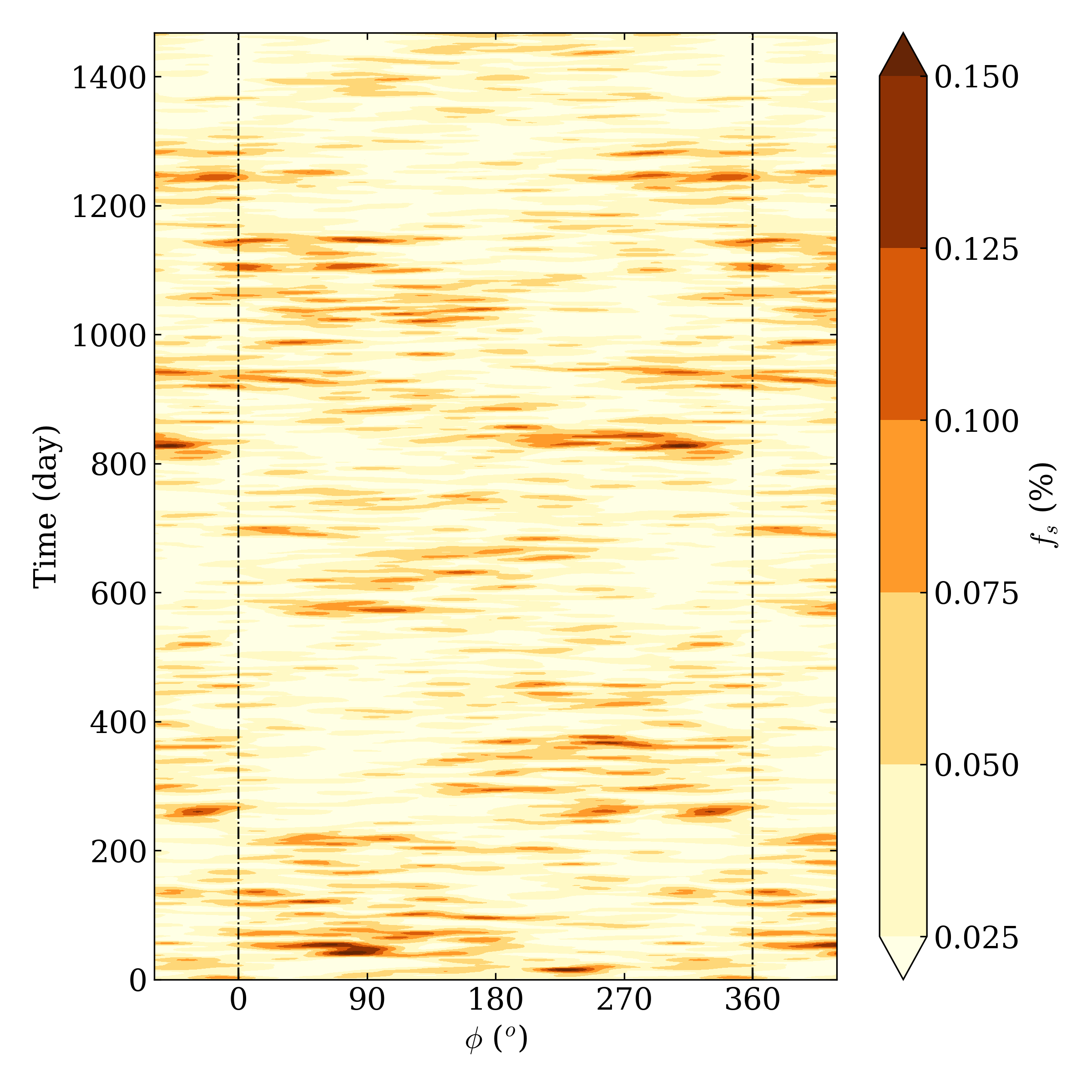}
    \caption{Longitudinal distribution for KIC~7103006 for the spots-and-faculae model (left) and the spots-only model (right).}
    \label{fig:kic7103006_longitude_map}
\end{figure*}

\begin{figure*}[ht!]
    \centering
    \includegraphics[width=0.49\textwidth]{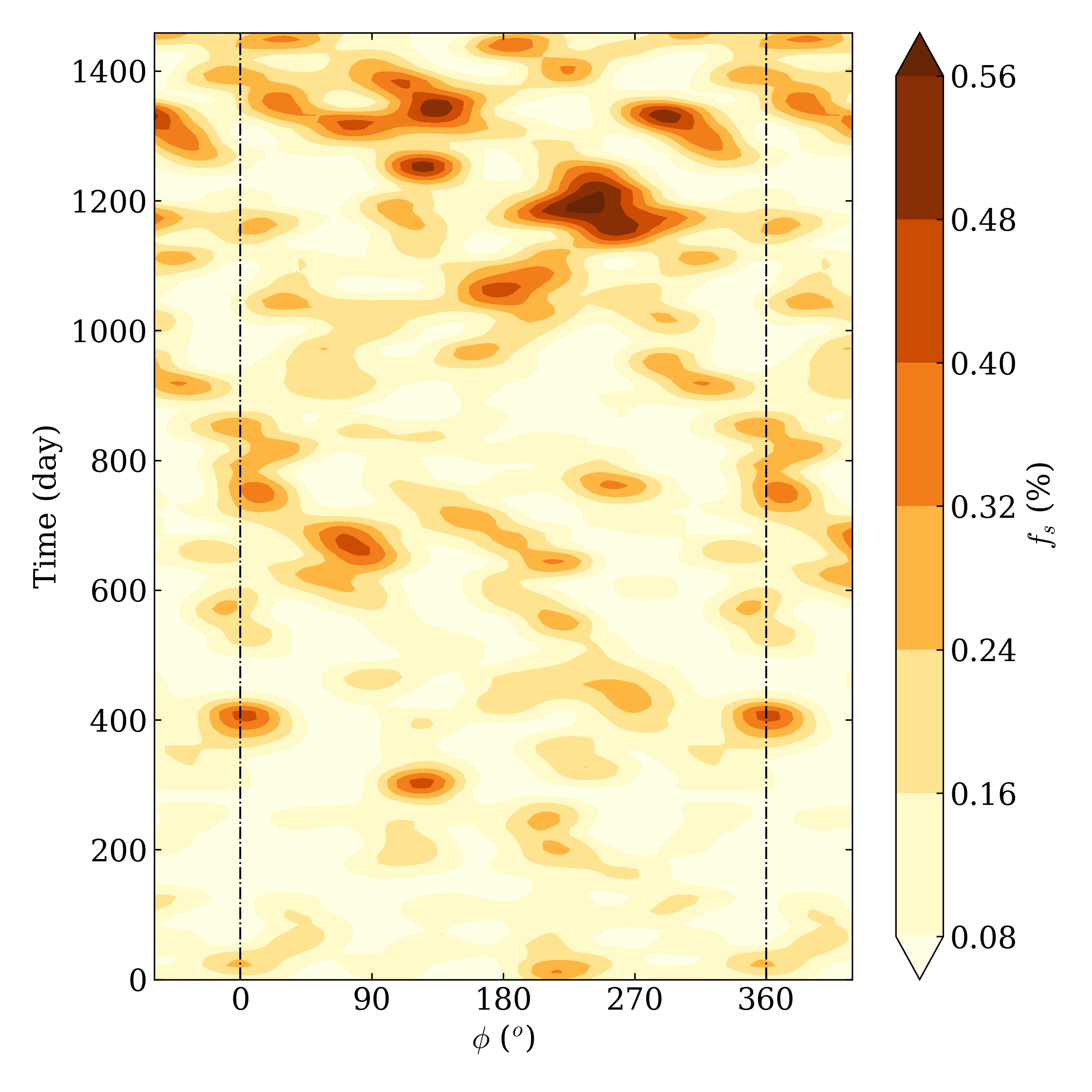}
    \includegraphics[width=0.49\textwidth]{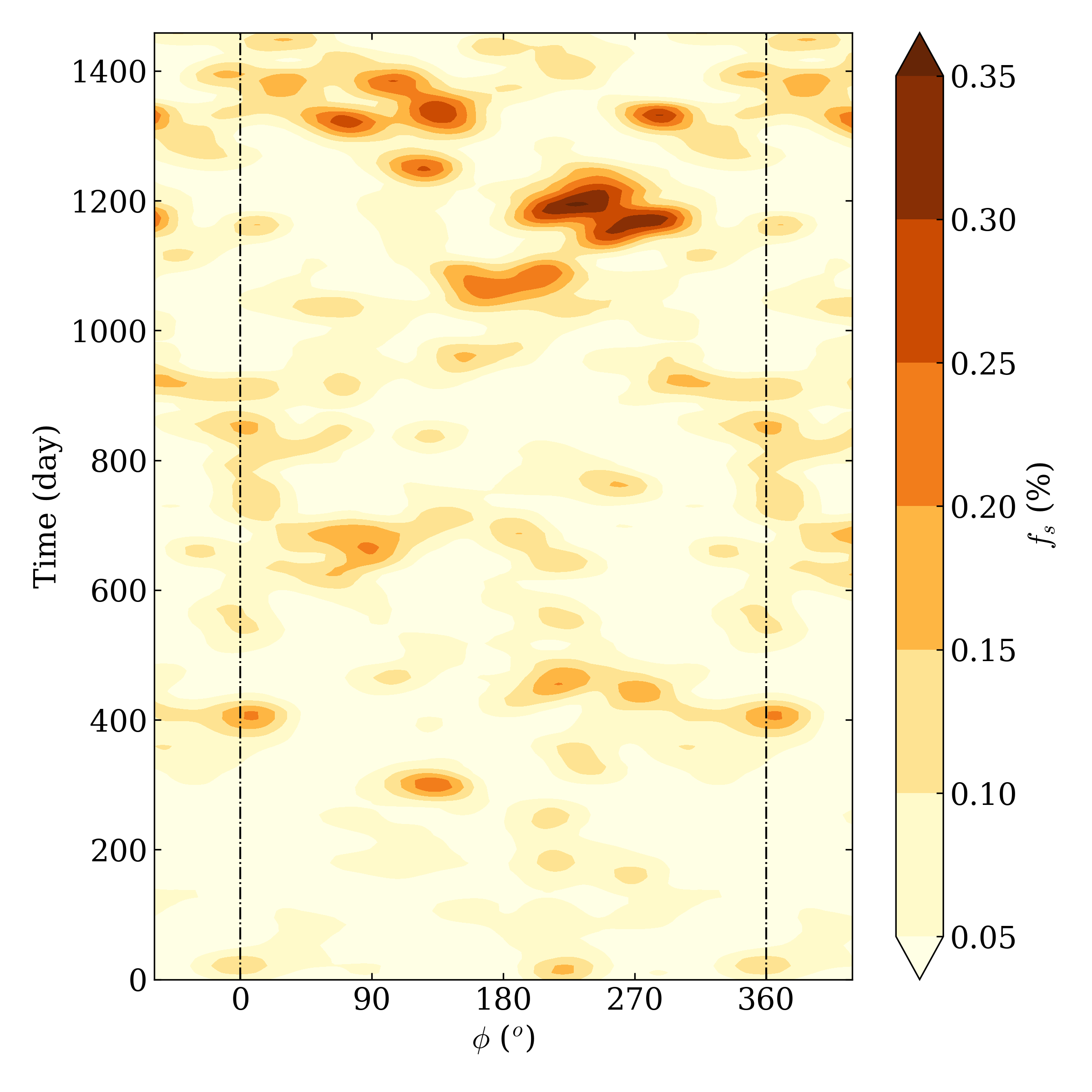}
    \caption{Longitudinal distribution for KIC~8006161 for the spots-and-faculae model (left) and the spots-only model (right).}
    \label{fig:kic8006161_longitude_map}
\end{figure*}

\begin{figure*}[ht!]
    \centering
    \includegraphics[width=0.49\textwidth]{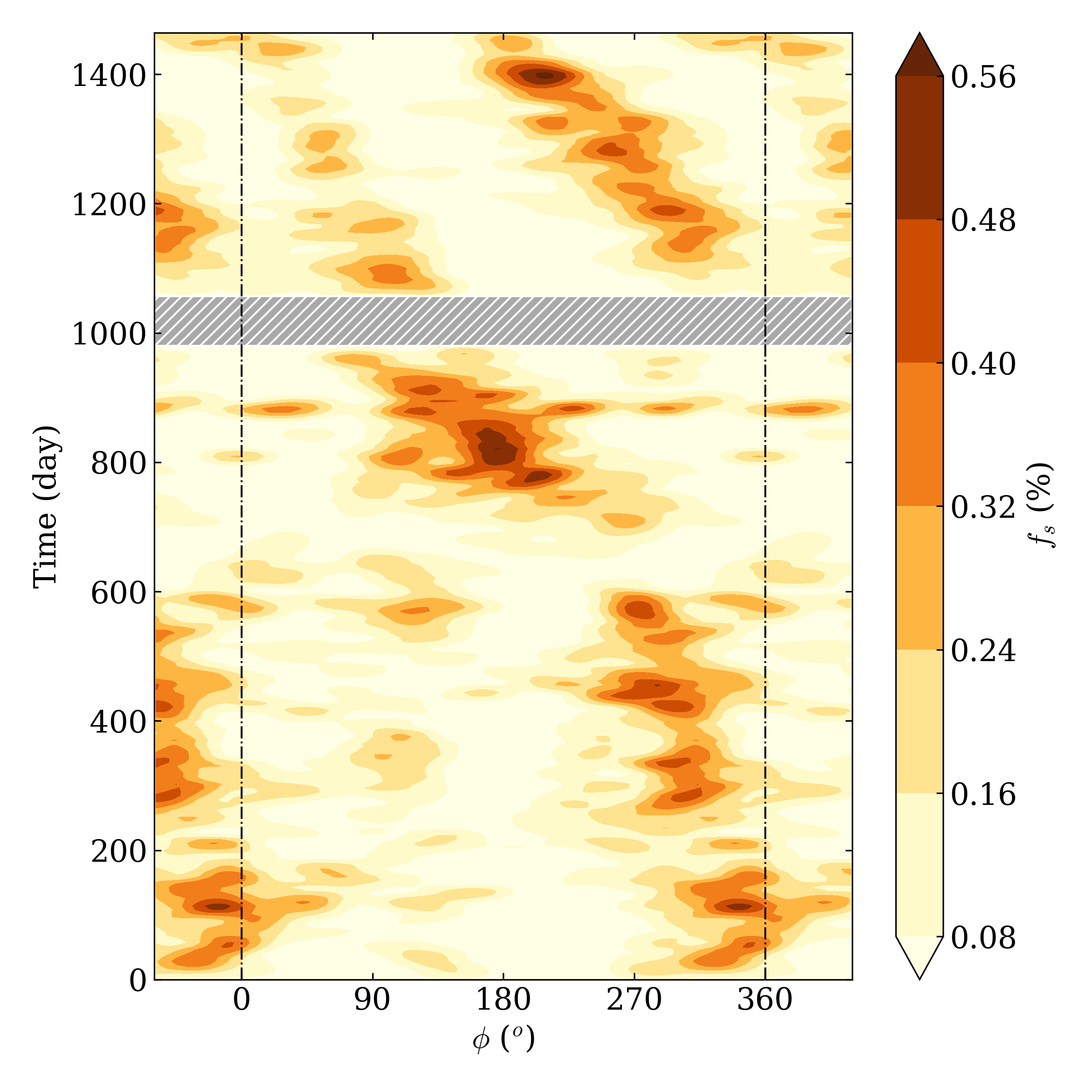}
    \includegraphics[width=0.49\textwidth]{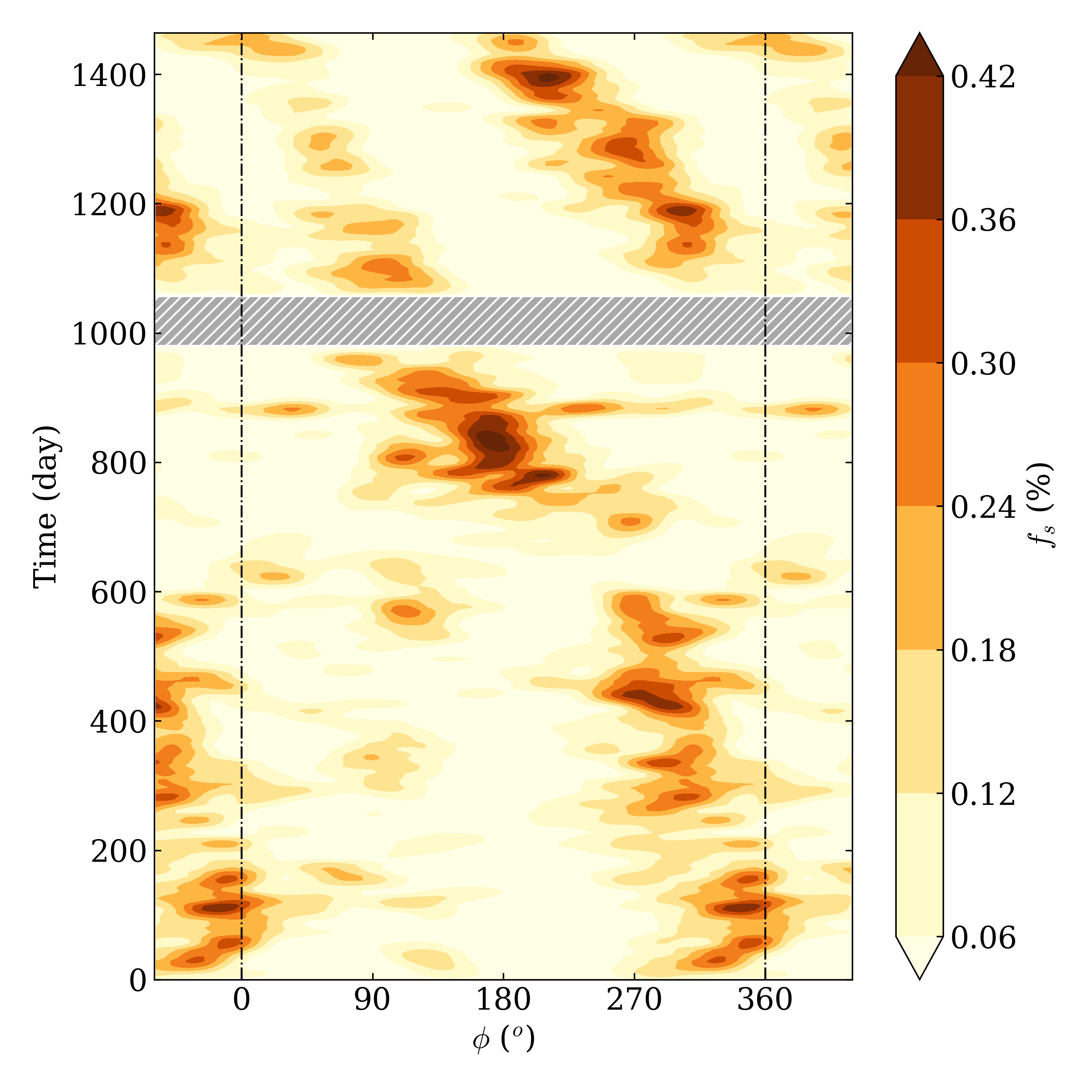}
    \caption{Longitudinal distribution for KIC~8379927 for the spots-and-faculae model (left) and the spots-only model (right). The hatched grey area signals the interval for which \textit{Kepler} data are missing.}
    \label{fig:kic8379927_longitude_map}
\end{figure*}

\begin{figure*}[ht!]
    \centering
    \includegraphics[width=0.49\textwidth]{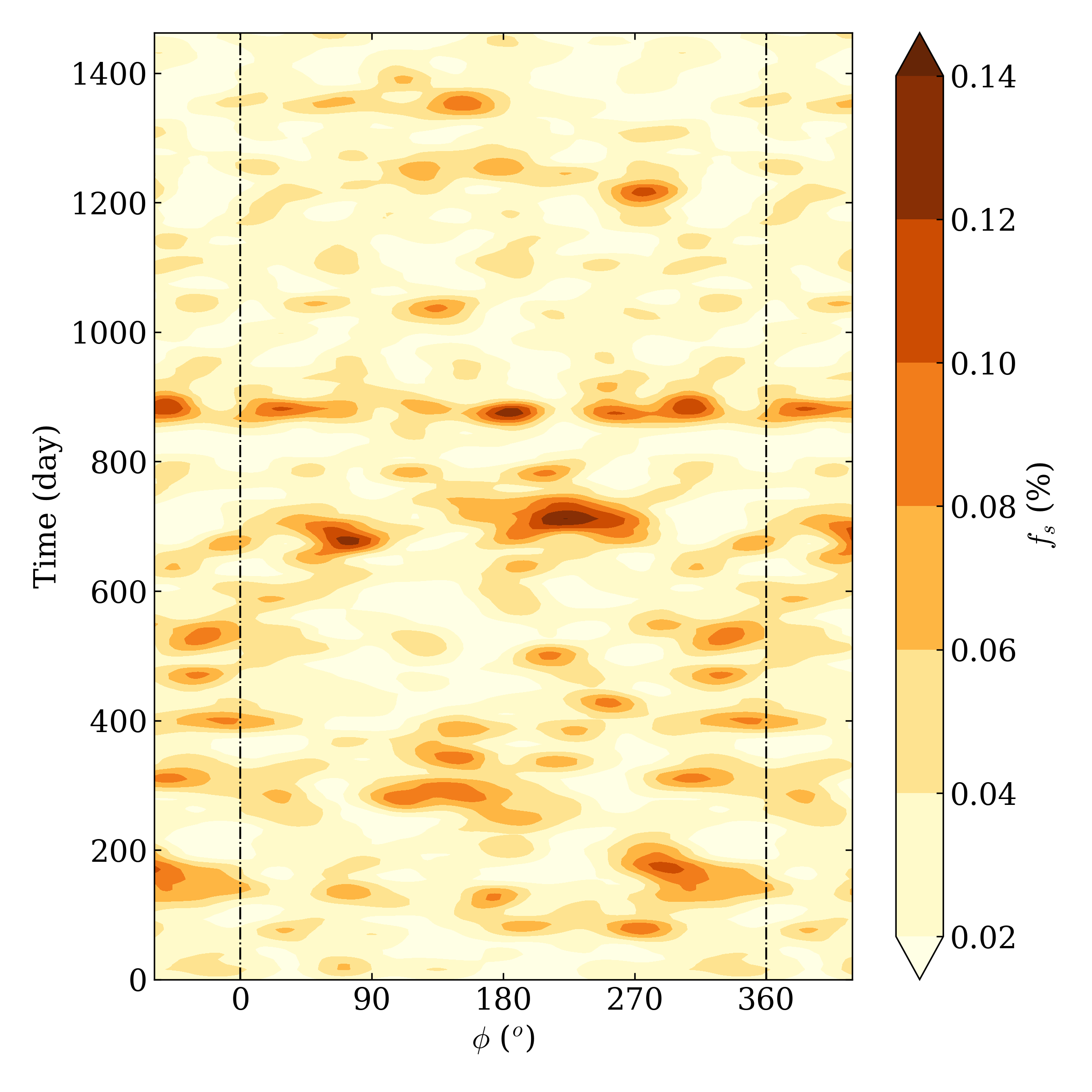}
    \includegraphics[width=0.49\textwidth]{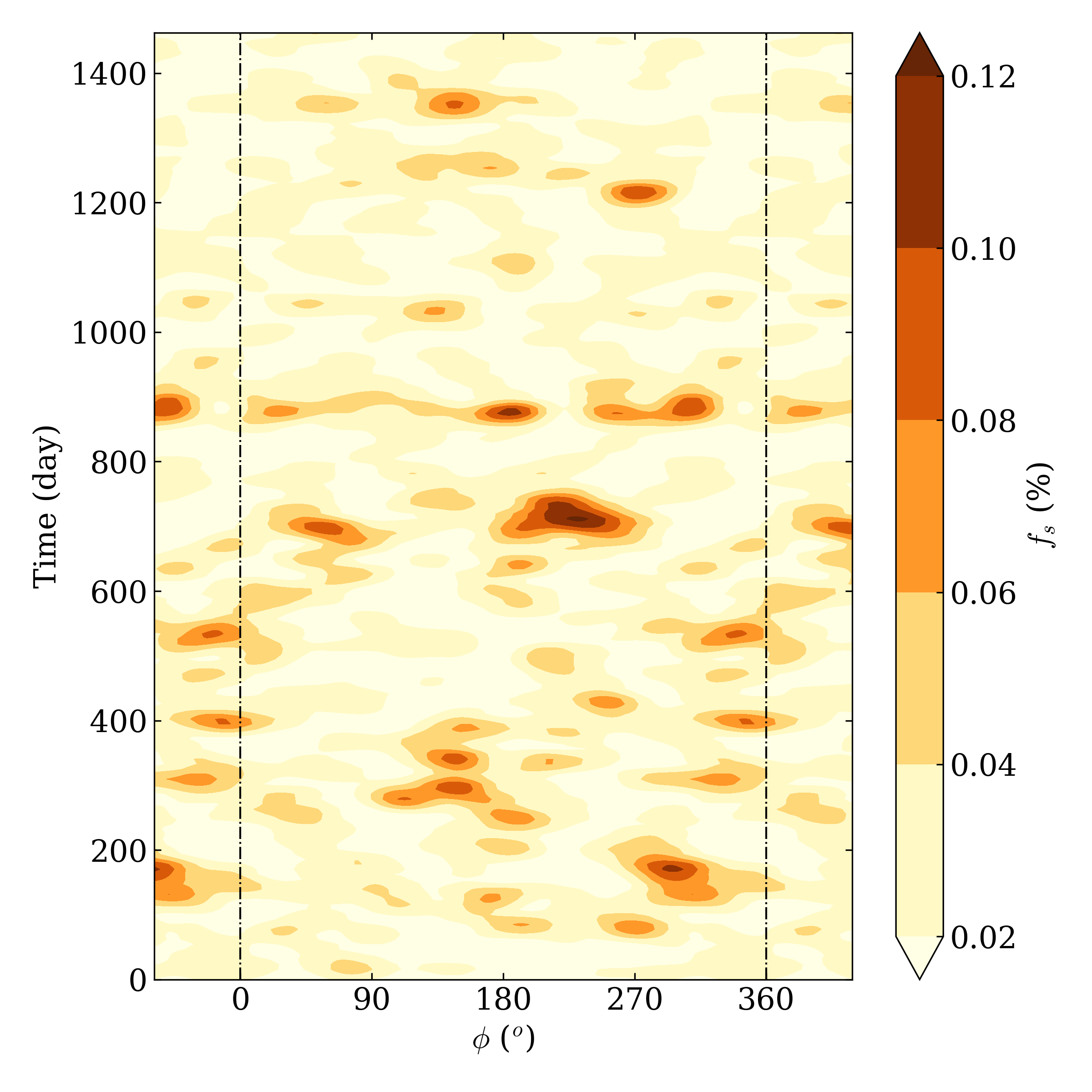}
    \caption{Longitudinal distribution for KIC~9025370 for the spots-and-faculae model (left) and the spots-only model (right).}
    \label{fig:kic9025370_longitude_map}
\end{figure*}

\begin{figure*}[ht!]
    \centering
    \includegraphics[width=0.49\textwidth]{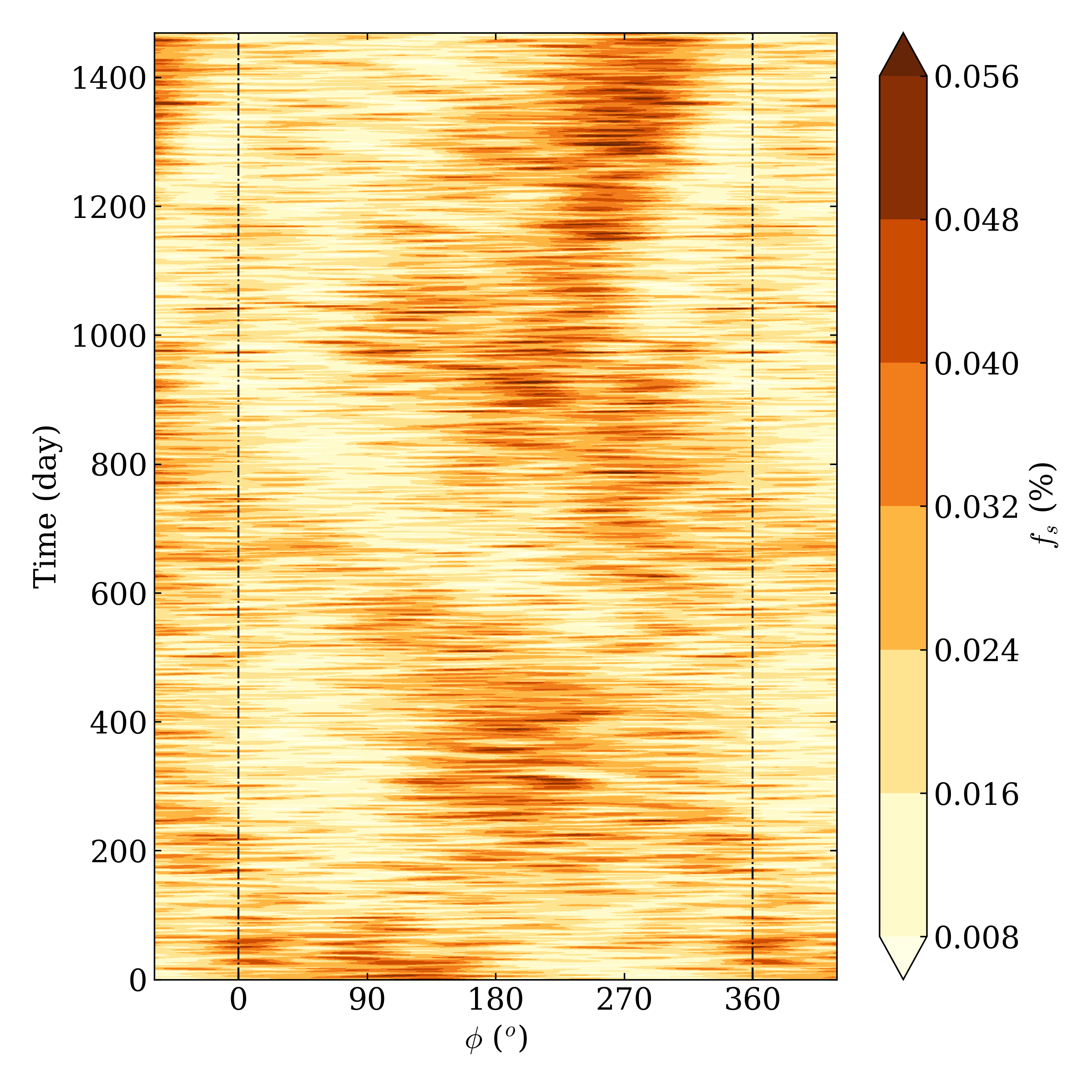}
    \includegraphics[width=0.49\textwidth]{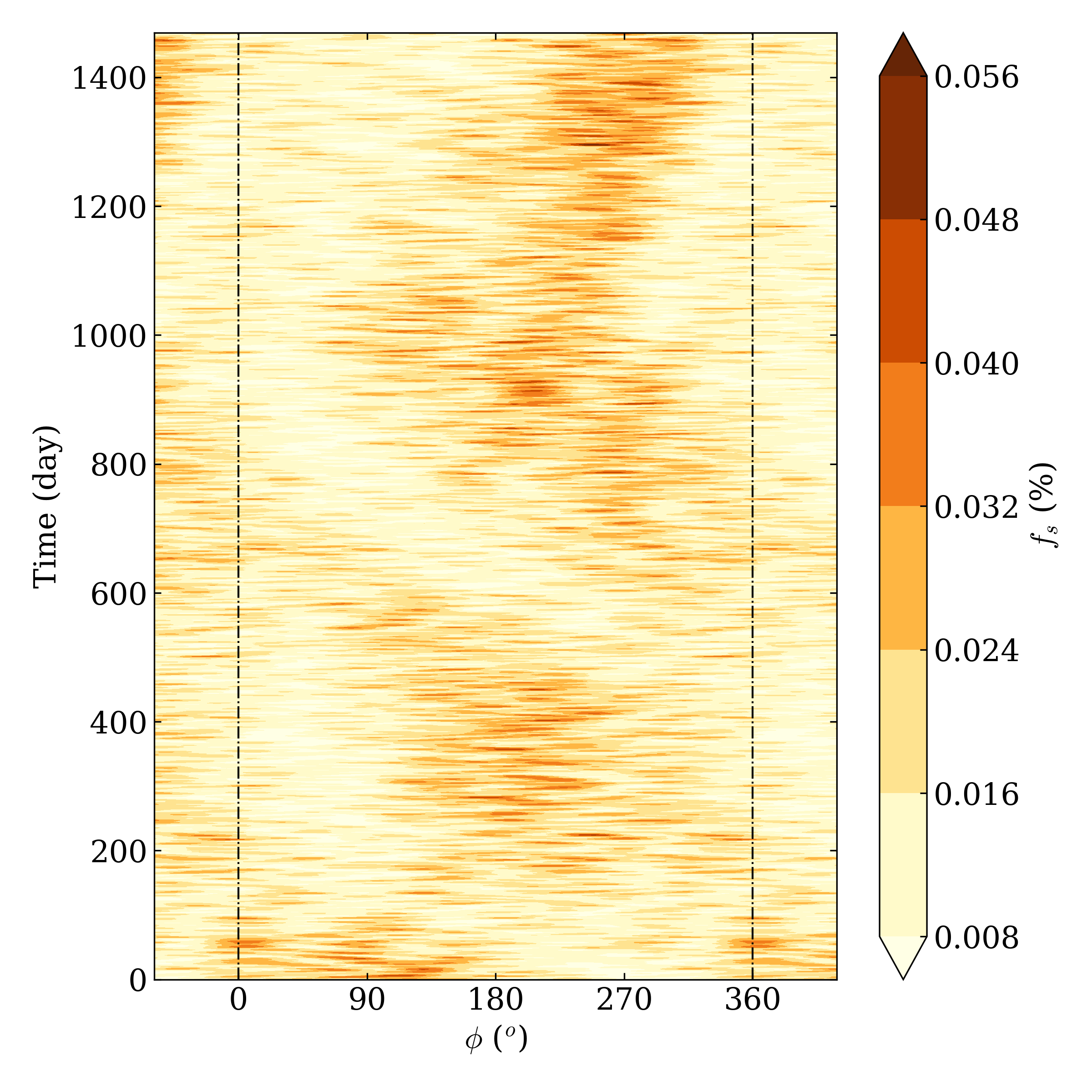}
    \caption{Longitudinal distribution for KIC~9226926 for the spots-and-faculae model (left) and the spots-only model (right).}
    \label{fig:kic9226926_longitude_map}
\end{figure*}

\begin{figure*}[ht!]
    \centering
    \includegraphics[width=0.49\textwidth]{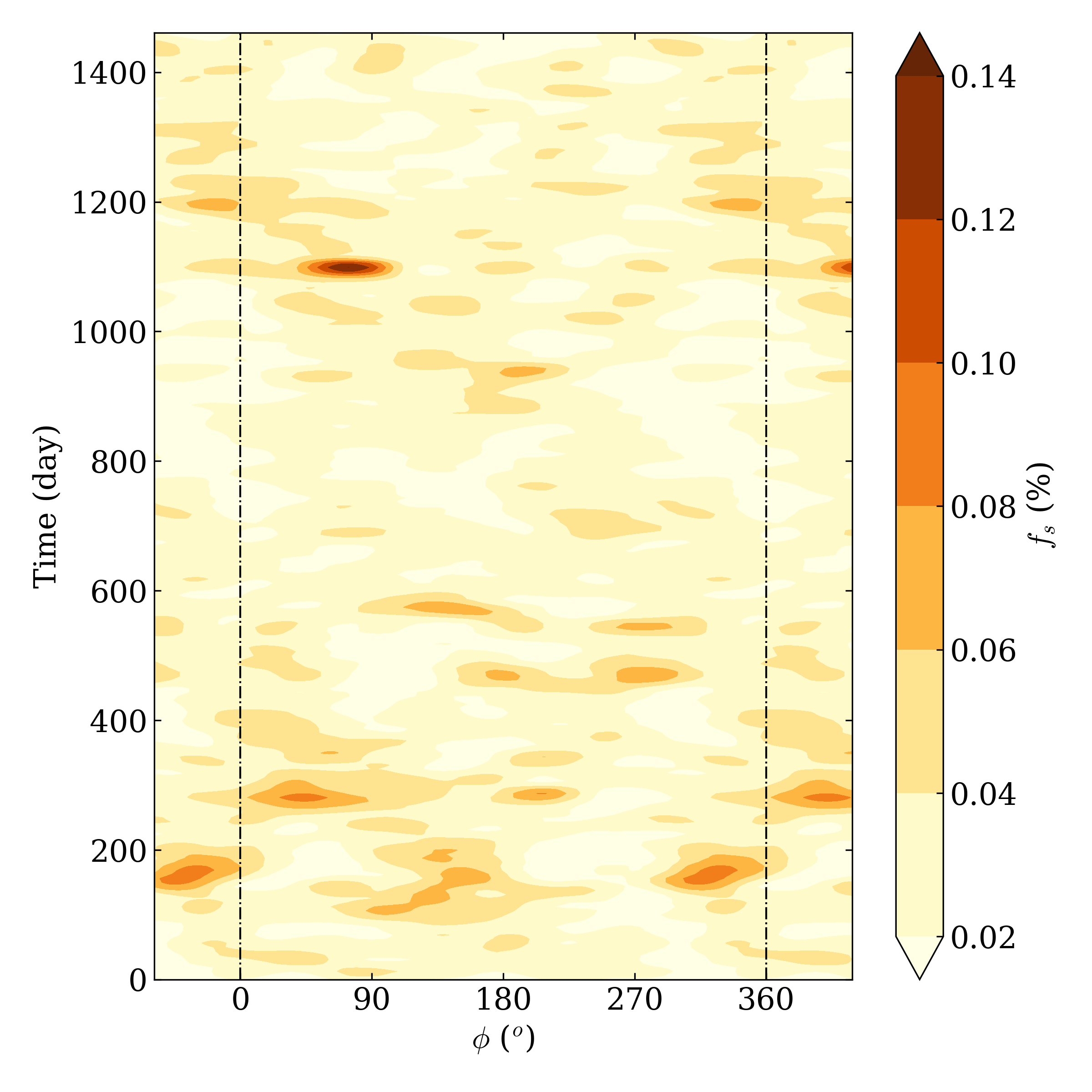}
    \includegraphics[width=0.49\textwidth]{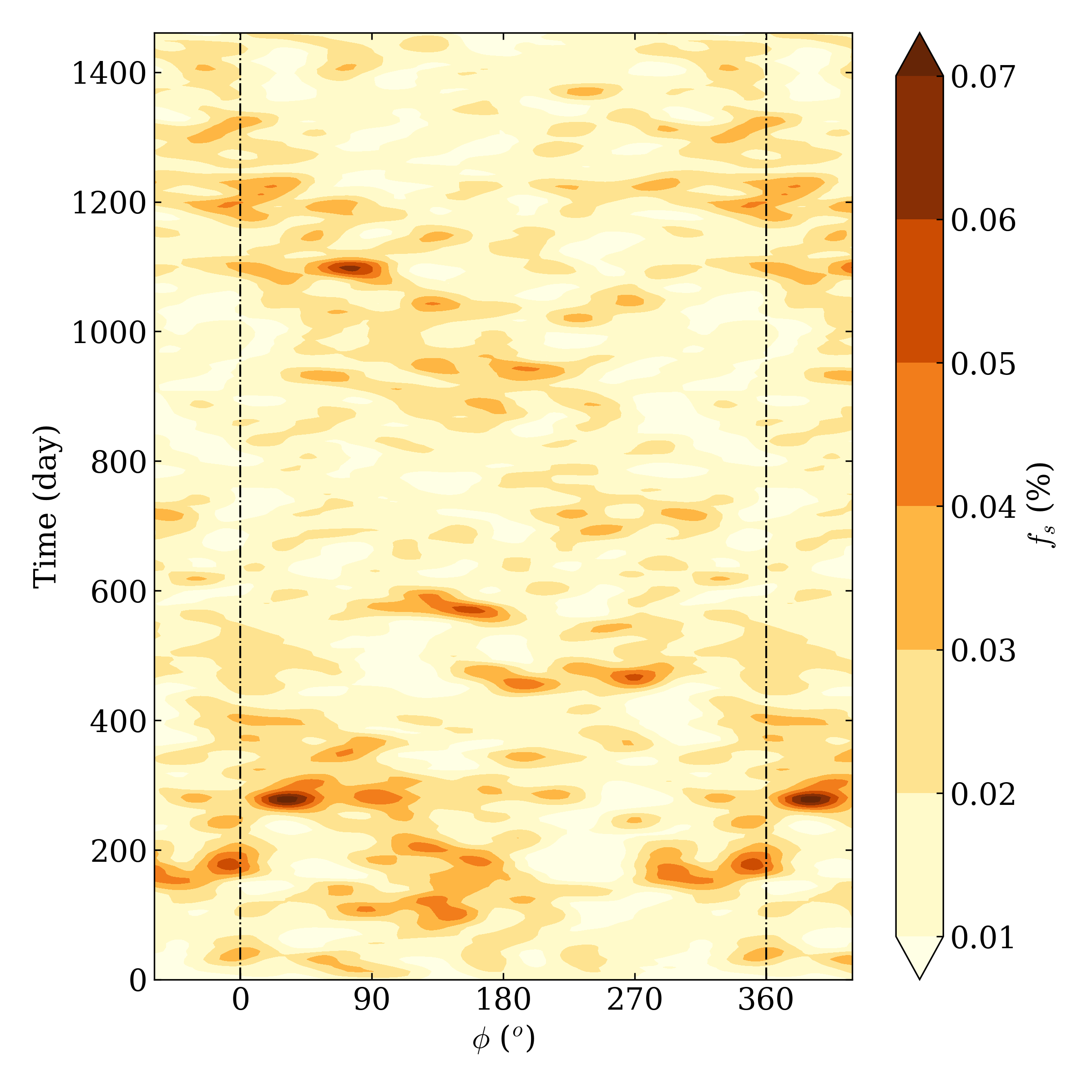}
    \caption{Longitudinal distribution for KIC~10068307 for the spots-and-faculae model (left) and the spots-only model (right).}
    \label{fig:kic10068307_longitude_map}
\end{figure*}

\begin{figure*}[ht!]
    \centering
    \includegraphics[width=0.49\textwidth]{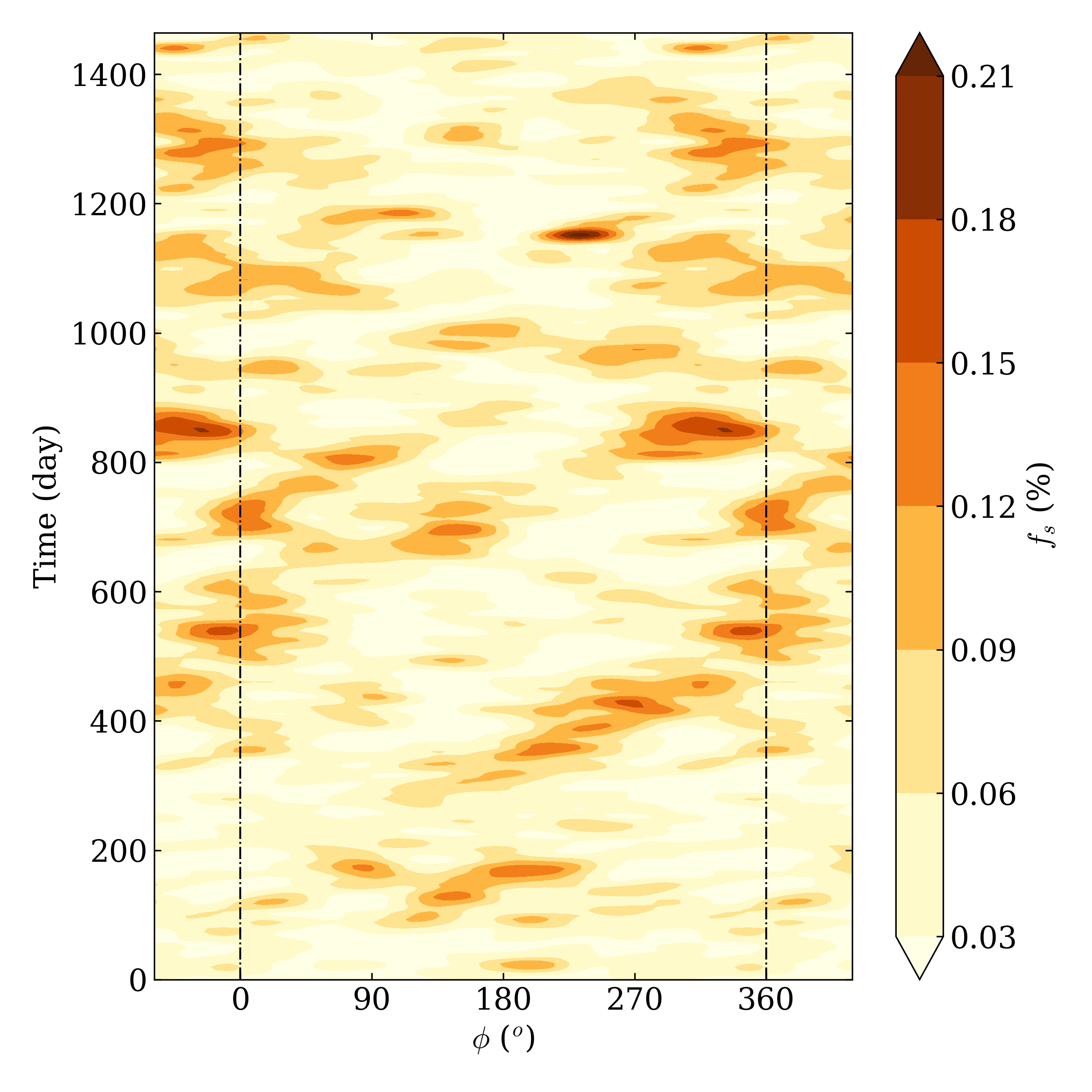}
    \includegraphics[width=0.49\textwidth]{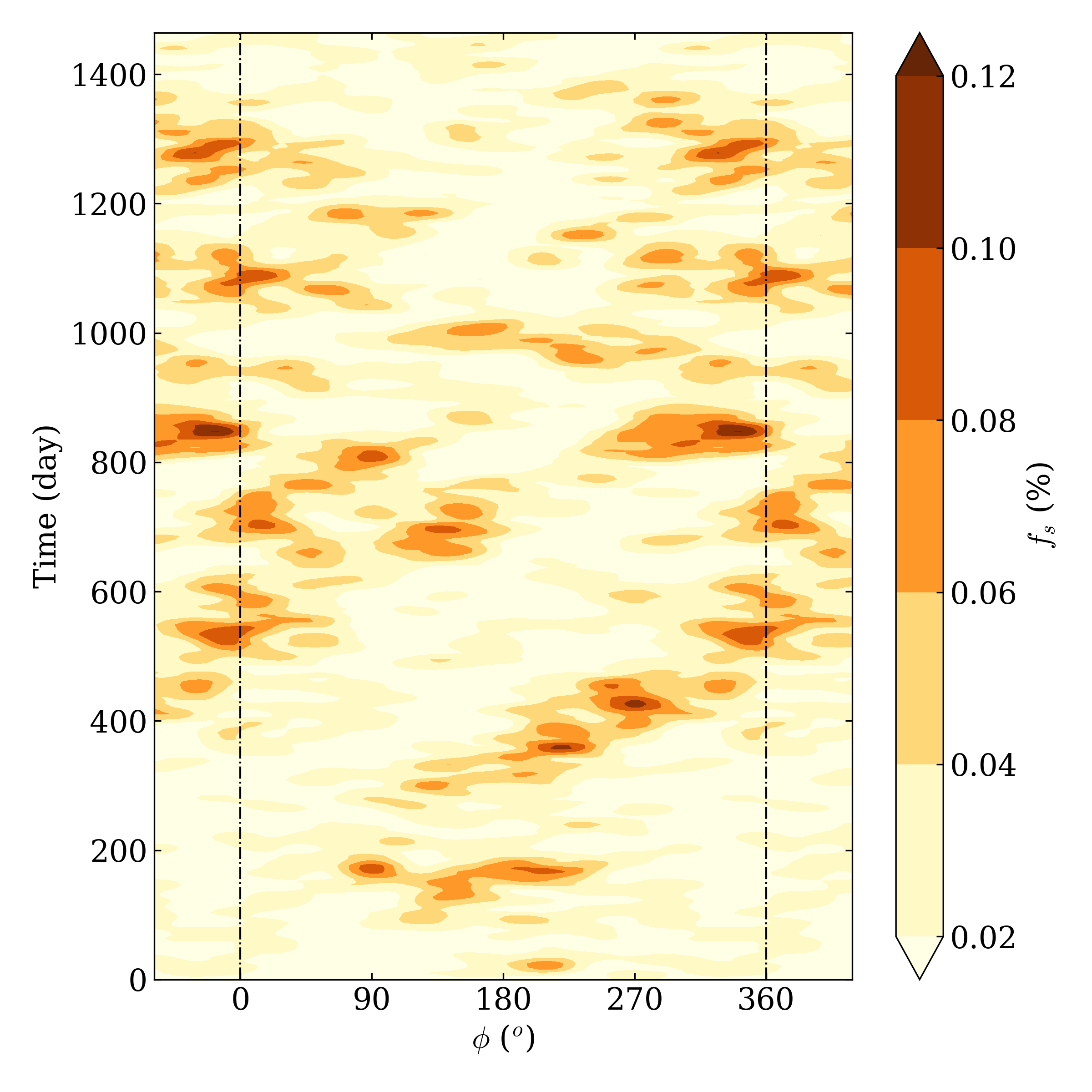}
    \caption{Longitudinal distribution for KIC~10454113 for the spots-and-faculae model (left) and the spots-only model (right).}
    \label{fig:kic10454113_longitude_map}
\end{figure*}

\begin{figure*}[ht!]
    \centering
    \includegraphics[width=0.49\textwidth]{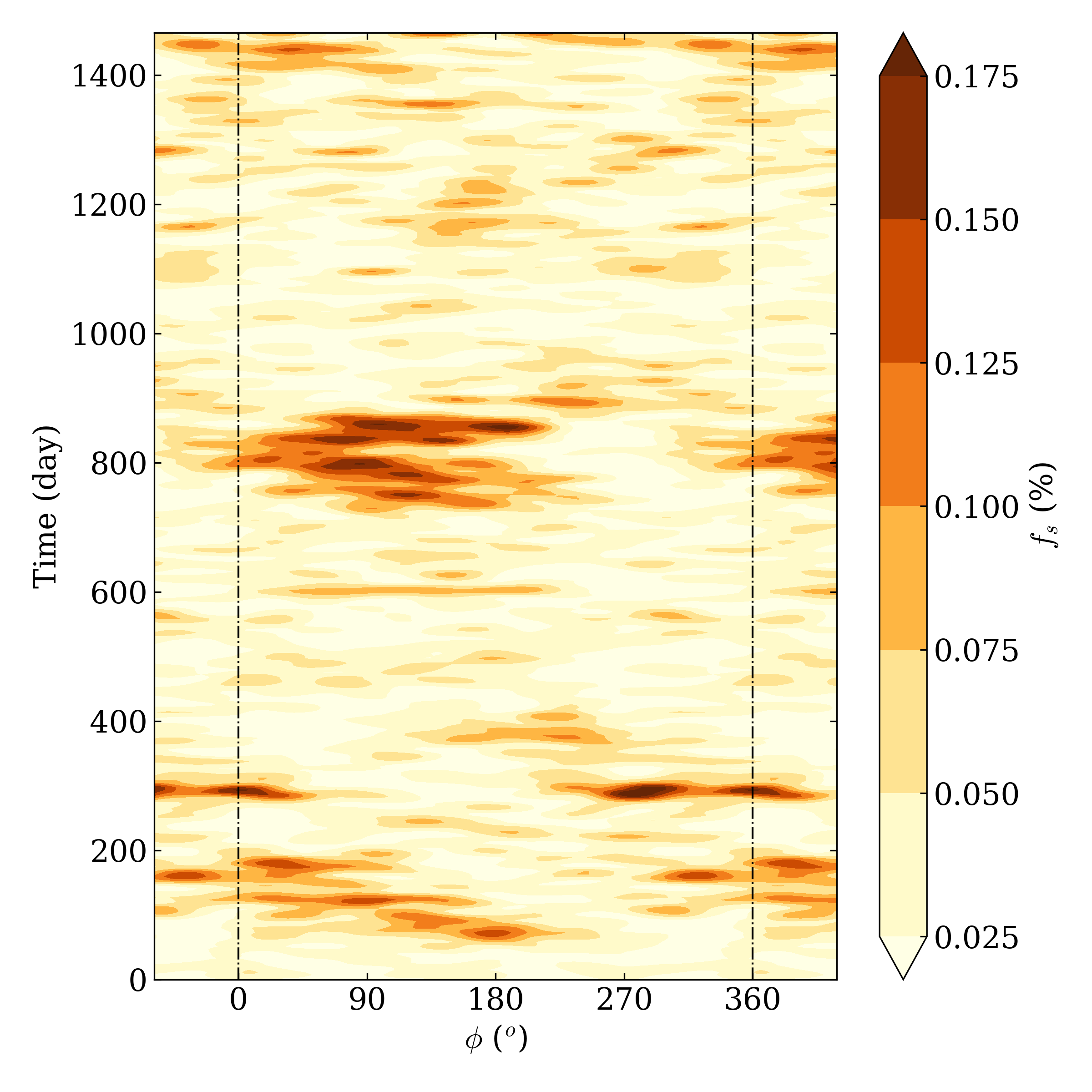}
    \includegraphics[width=0.49\textwidth]{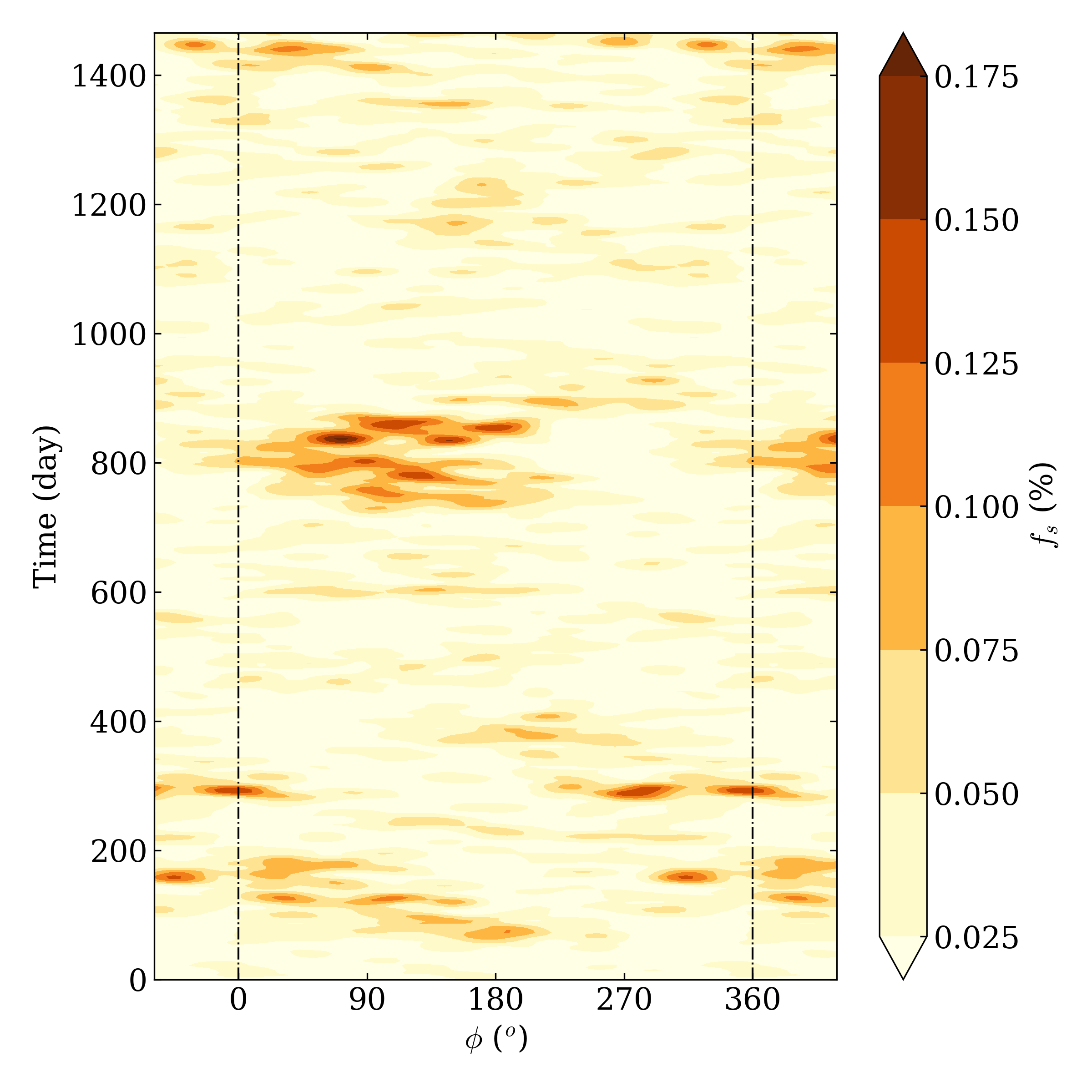}
    \caption{Longitudinal distribution for KIC~10644253 for the spots-and-faculae model (left) and the spots-only model (right).}
    \label{fig:kic10644253_longitude_map}
\end{figure*}

We start by presenting the longitudinal distribution maps we obtained for the different targets, considering the spots-and-faculae model and the spots-only model. These maps are shown in Fig.~\ref{fig:kic3733735_longitude_map} to \ref{fig:kic10644253_longitude_map} for KIC~3733735, KIC~6508366, KIC~7103006, KIC~8006161, KIC~8379927, KIC~9025370, KIC~9226926, KIC~10068307, KIC~10454113, and KIC~10644253.


\begin{figure}
    \centering
    \includegraphics[width=0.49\textwidth]{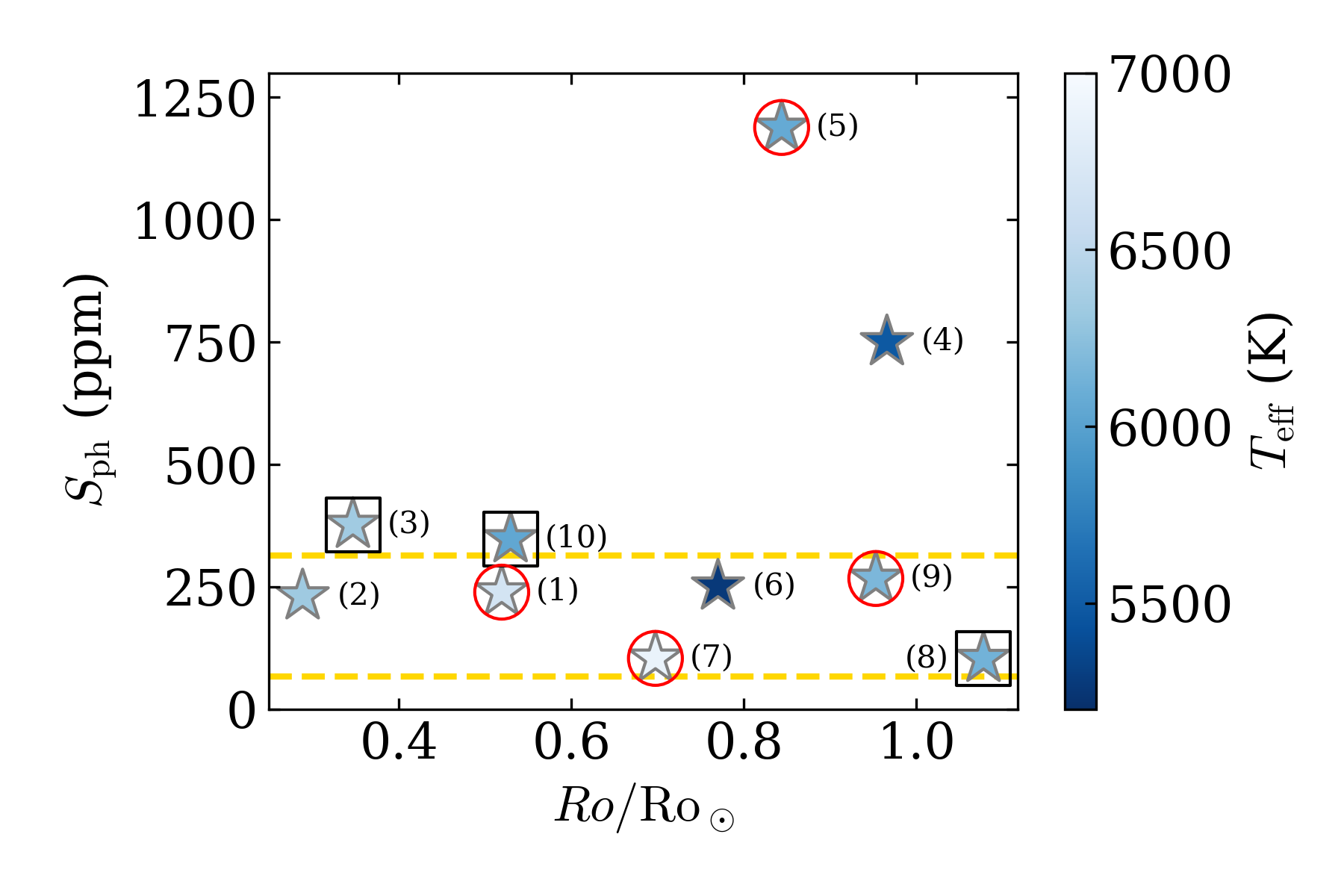}
    \caption{$Ro/ \rm Ro_\odot$ vs. $S_\mathrm{ph}$ diagram for the \textit{Kepler} stars in our sample. The effective temperature, $T_\mathrm{eff}$, is color-coded. The stars for which we have a clear detection of longitudinal active nests are highlighted with red circles, and the stars with a possible signature are highlighted with black squares. The minimum and maximum $S_\mathrm{ph}$ values computed by \citet{Salabert2016b} for the Sun are shown by horizontal dashed yellow lines. The star identifiers of the figure correspond to those given in Table~\ref{tab:considered_targets}.}
    \label{fig:summary_active}
\end{figure}

After about 550 days of \textit{Kepler} observations, KIC~3733735 (see Fig.~\ref{fig:kic3733735_longitude_map}) experienced the activation of two active nests separated by approximately 90 degrees and migrating westwards for about 100~days and then eastwards. These two nests persisted until at least the end of the nominal \textit{Kepler} mission.
It is interesting to note that their longitudinal migration only began after about 200 days, suggesting a latitudinal migration of the nest in the meantime. Over the last 200 days of observation, a third nest seems to appear westwards of them. 

Although some short-lived active regions can be identified, KIC~6508366 (see Fig.~\ref{fig:kic6508366_longitude_map}) does not show evidence of stable regions of active nests during the four-year extent of its \textit{Kepler} light curve.

The map we obtain for KIC~7103006 (see Fig.~\ref{fig:kic7103006_longitude_map}) suggests alternating sporadic nest activation between two areas that are separated by approximately 180 degrees. The position of these area seems to remain relatively stable with time with respect to the reference frame of the model. 

The starspot map for KIC~8006161 (see Fig.~\ref{fig:kic8006161_longitude_map}) is quite similar to the map obtained in the solar case. During the first 100 days of observation, short-lived active nests are mainly observed. It should nevertheless be noted that the longitudes between 180 and 270 degrees appears to be more consistently active, although some features are clearly visible between 0 and 150 degrees as well.
The star exhibits a clear increase in the coverage in the final 100 days of observations. This is consistent with the evidence presented by \citet{Karoff2018}, who reported that its activity level increased during the \textit{Kepler} mission as the star was on the rising sequence of its activity cycle. A first large active nest appears in our spot maps after 1100 days of observations, between $\phi = 180$ and $\rm 270^o$, a second nest is visible between $\phi = 60$ and $\rm 180^o$ after about 1250 days of observation. 

KIC~8379927 (see Fig.~\ref{fig:kic8379927_longitude_map}) has a stable active nest that can be followed in its westwards migration during the whole extent of the \textit{Kepler} mission. 
Around $\rm \phi = 90^o$, we note the occasional appearance of group of spots with a coverage that is significantly lower than for the main nest.   
Interestingly, the coverage of the first main active nest seems to decrease in the last 100 days, while a second active nest, separated from the first nest by $\sim 180$ degrees, appears and also migrates westwards.
It should be noted that for this star, the light curve has an observation gap of about 80 days after about 1000~days of \textit{Kepler} observations. 

In KIC~9025370 (see Fig.~\ref{fig:kic9025370_longitude_map}), which is one of the slower rotators of our sample, active features do not persist during more than a few dozen days, that is, no more than several stellar rotations. 

KIC~9226926 (see Fig.~\ref{fig:kic9226926_longitude_map}) is also a target with low-amplitude brightness variations. Nevertheless, it is still possible to distinguish a pattern of stable active nests on the longitudinal map that drift eastwards with time. 
The activity intensity of the first nest seems to weaken at around 600 days, but increases again shortly after this event. Its signature disappears again after 1000 days of observations, however.
Nevertheless, after 800 days of observations, a second nest appears that is shifted westwards from the initial nest and persists until the end of the light curve. 

KIC~10068307 (see Fig.~\ref{fig:kic10068307_longitude_map}) is one of the stars with the lowest brightness variability in our sample. The spots-only model seems to provide evidence for the presence of an active nest appearing close to $\rm \phi = 180^o$ at the beginning of the observations and migrating westwards, but the corresponding pattern is less clearly visible in the spots-and-faculae model. 

KIC~10454113 (see Fig.~\ref{fig:kic10454113_longitude_map}) shows evidence of trails of active nests directed eastwards. These trails become more sporadic in the last 100 days of observations. 

Finally, the main feature of KIC~10644253 (see Fig.~\ref{fig:kic10644253_longitude_map}) is the activation of a strong active nest after 750 days of observations. The region persists for about 100-150 days and then disappears. We note that shorter-lived features are also visible during the first 100 days of observations.  

As in Sect.~\ref{sec:the_sun}, it is important to note again the strong similarities of the longitudinal maps constructed from the spots-and-faculae model and from the spots-only model. 
To summarise the analysis presented above, we represent the $Ro$ versus $S_\mathrm{ph}$ diagram for our sample in Fig.~\ref{fig:summary_active}. We highlight the location of the stars for which we have a clear detection of one nest or more than one stable longitudinal active nests (KIC~3733735, KIC~8379927, KIC~9226926, and KIC~10454113) and the location of the stars for which we have a possible detection (KIC~7103006, KIC~10068307, and KIC~10644253).
This diagram shows longitudinal active nests for different levels of photospheric activity and for different convection versus rotation regimes (characterised here by the $Ro$ value). 
In particular, when we compare the photospheric $S_\mathrm{ph}$ of the stars in our sample with the minimum and maximum solar values during cycles~23 and 24 \citep[using the values computed by][]{Salabert2016b}, we note that three of the stars with a clear stable longitudinal active nest (KIC~3733735, KIC~9226926, and KIC~10454113) and one star with a possible detection (KIC~7103006) are included in the corresponding interval.
The distribution of stars with detected longitudinal stable active nests in the diagram therefore supports the hypothesis that this phenomenon takes place ubiquituously even in moderately active rotators.   
We therefore underline that the different mechanisms that are able to form stable longitudinal active nests in solar-type stars need to be clarified, in particular, the possibility that some low-frequency magneto-inertial waves propagate in the envelope and shape convection in a way that favours magnetic flux emergence at certain longitudes.

Finally, we emphasise that the migration with time of the observed active longitudes is evidence for latitudinal differential rotation in the corresponding stars. 
We recall that the $Ro/\mathrm{Ro}_\odot$ estimates we computed in Sect.~\ref{sec:data} suggest that all the stars we considered probably exhibit solar-like differential rotation.
However, it is not straightforward to confirm that the regime is indeed solar-like or anti-solar from the active longitude diagram alone. 
The reference periods we used for our spot models have to be interpreted as an average obtained considering photometric modulation during the whole extent of the \textit{Kepler} mission \citep{Santos2021}, and they do not necessarily correspond to the equatorial rotation period. 
For example, in the case of KIC~3733735, where the active longitudes shift slightly westwards at first and then shift eastwards, we can interpret this by considering that the change in direction intervenes when the active regions cross the latitude that corresponds to the reference period. Under the hypothesis that the differential rotation is solar-like, this means that until up to $\sim$650 days of observations, the active regions are located above the reference period latitude, and then they lie below it.

\subsection{Wavelet analysis \label{sec:wavelets_kepler}}

\begin{figure}[ht!]
    \centering
    \includegraphics[width=0.49\textwidth]{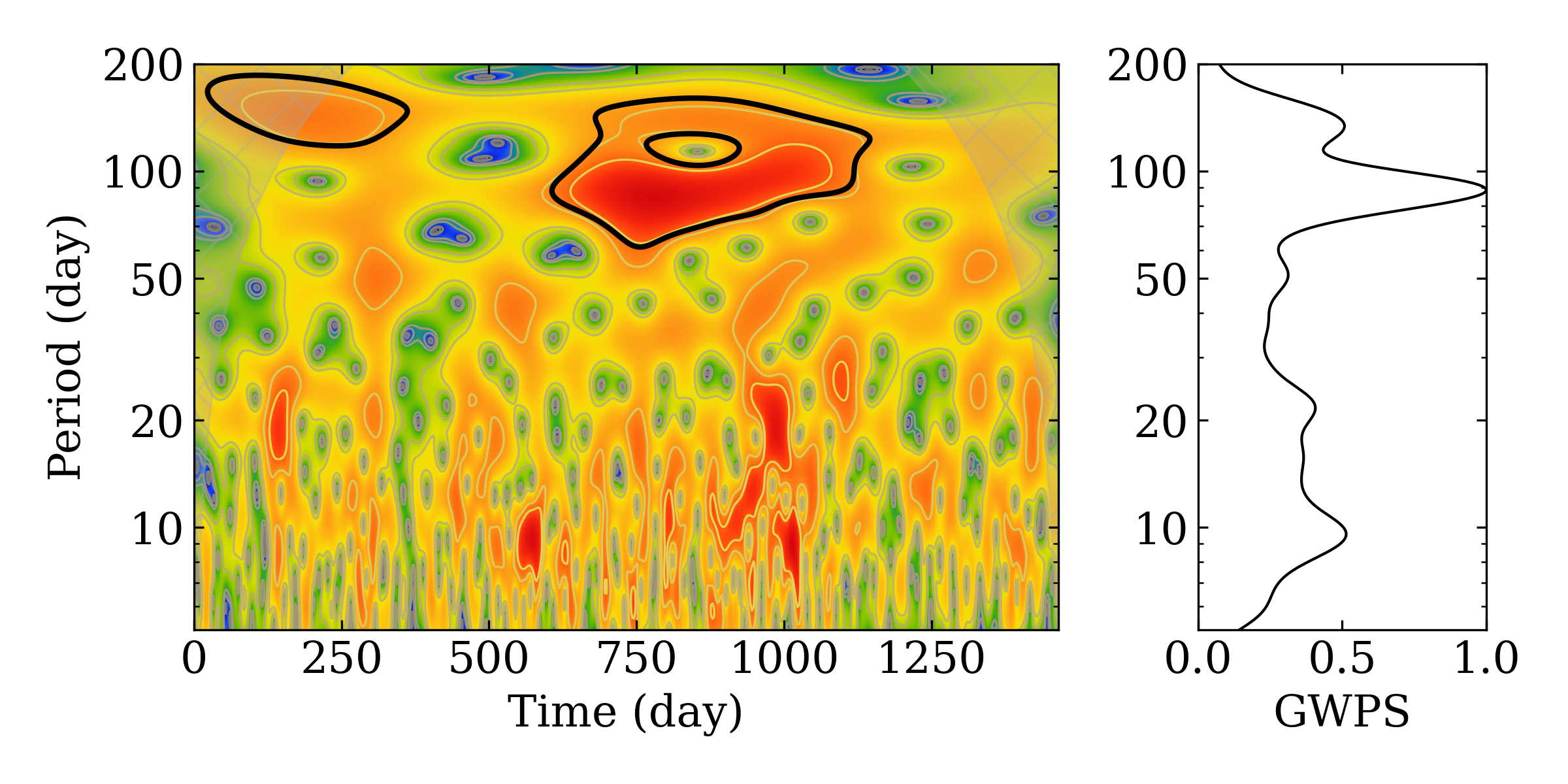}
    \includegraphics[width=0.49\textwidth]{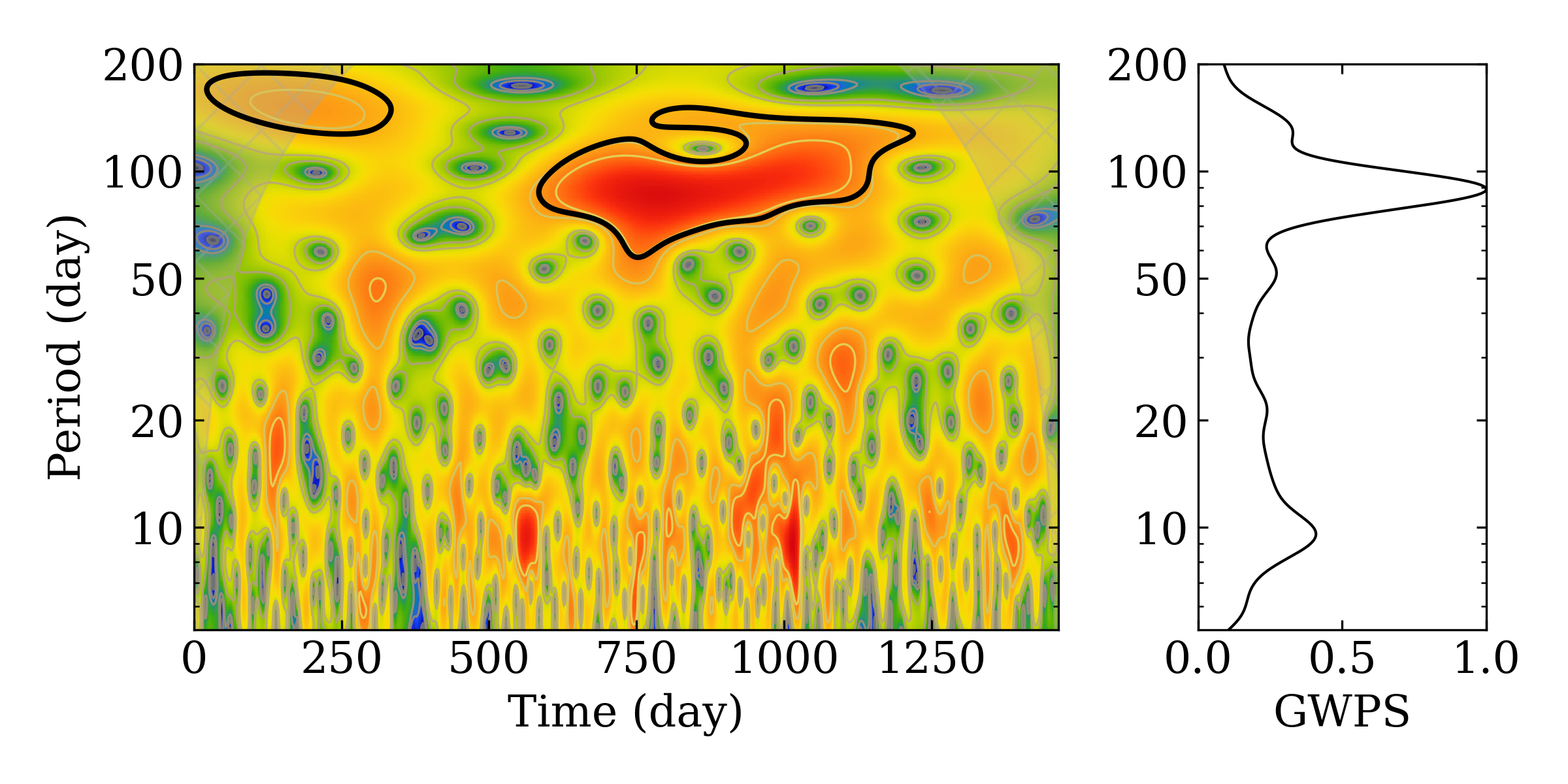}
    \caption{Morlet wavelet transform of the spot coverage reconstructed from the spots-and-faculae model (\textit{top}), and the spot coverage reconstructed from the spots-only model (\textit{bottom}) for KIC~3733735. The cone of influence of the wavelet transform is shown in grey. The global wavelet power spectrum is shown in the right panel of both rows.}
    \label{fig:kic3733735_wavelet}
\end{figure}

\begin{figure}[ht!]
    \centering
    \includegraphics[width=0.49\textwidth]{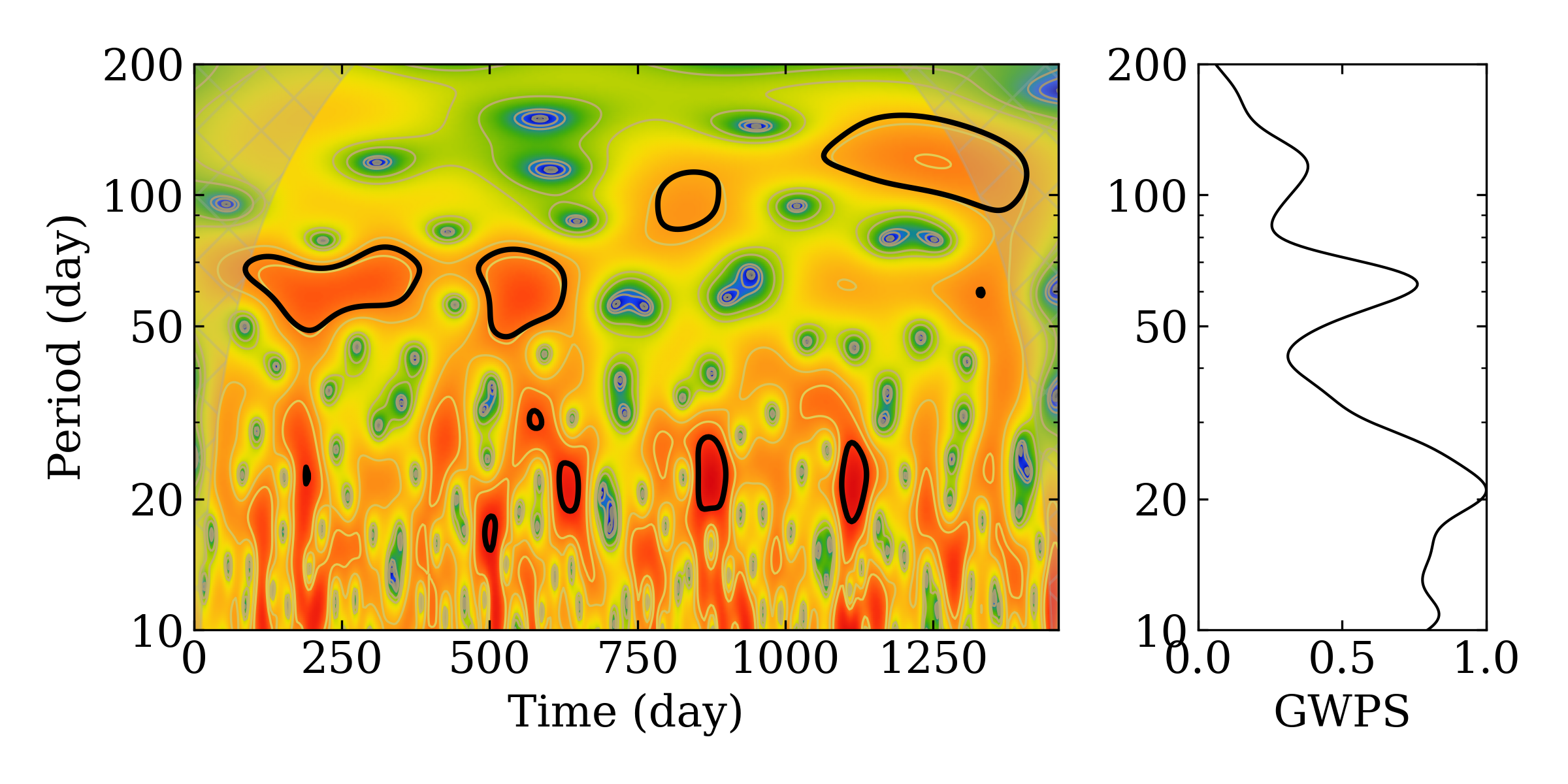}
    \includegraphics[width=0.49\textwidth]{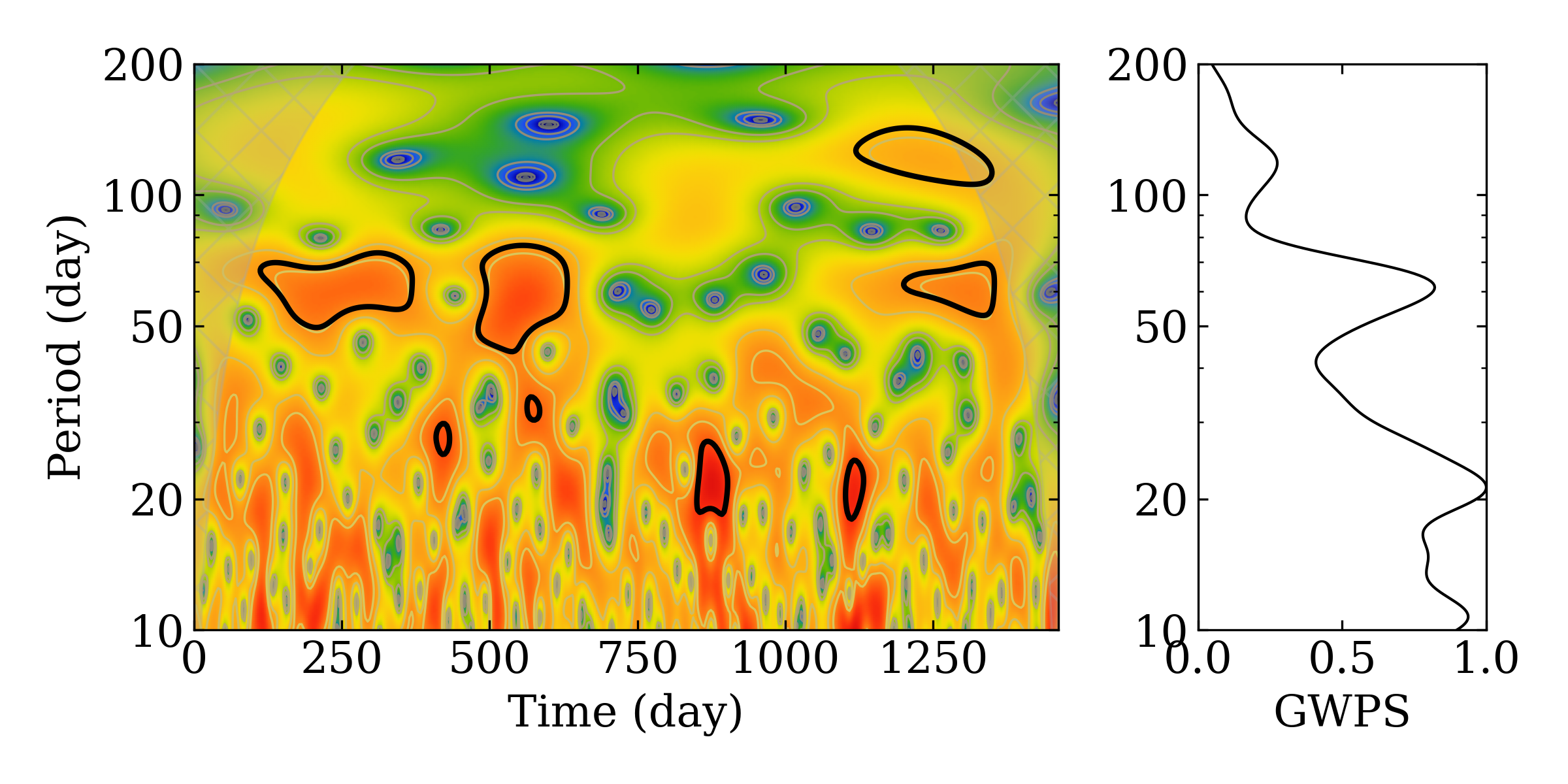}
    \caption{Same as Fig.~\ref{fig:kic3733735_wavelet} for KIC~6508366.}
    \label{fig:kic6508366_wavelet}
\end{figure}

\begin{figure}[ht!]
    \centering
    \includegraphics[width=0.49\textwidth]{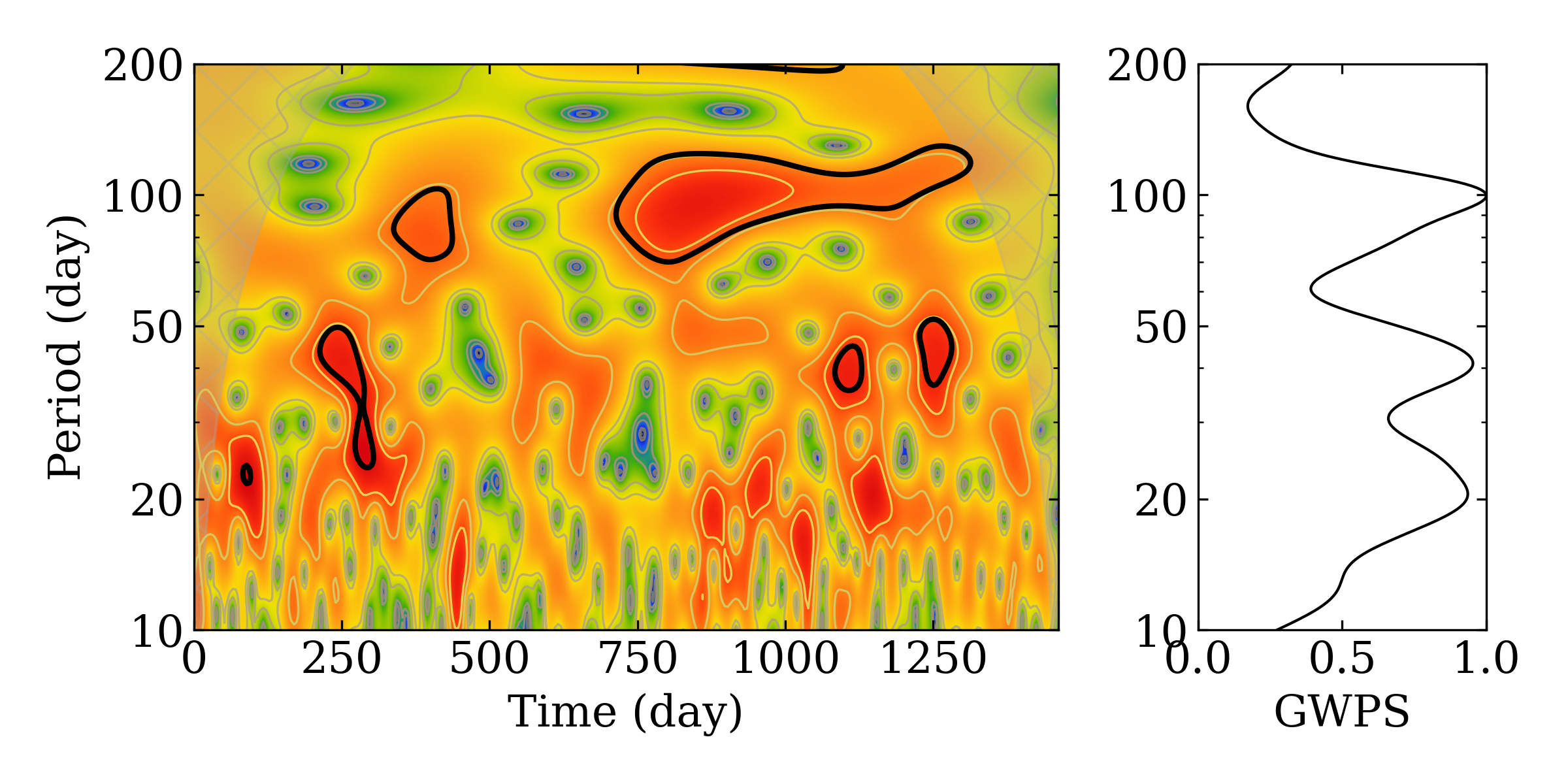}
    \includegraphics[width=0.49\textwidth]{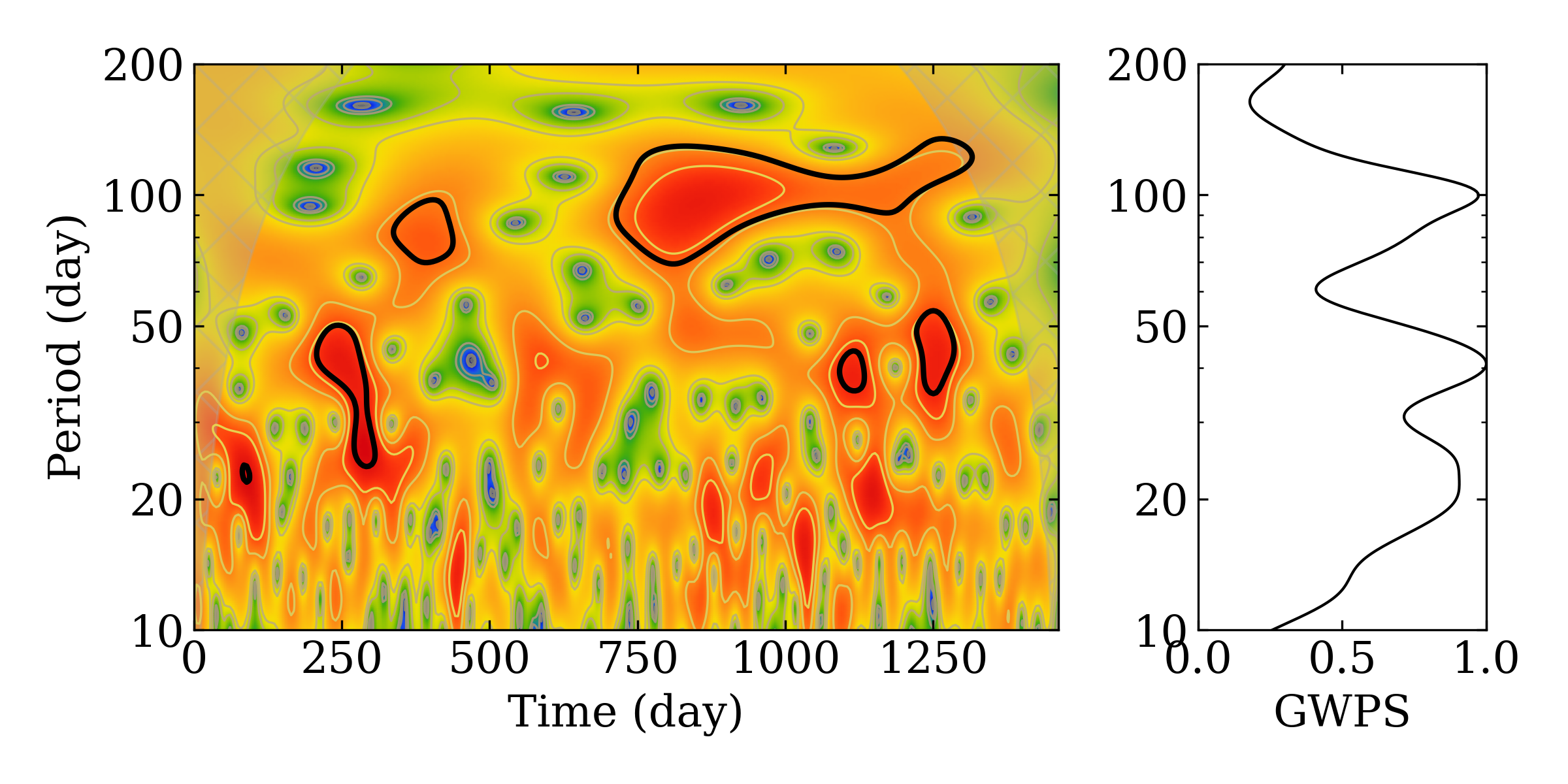}
    \caption{Same as Fig.~\ref{fig:kic3733735_wavelet} for KIC~7103006.}
    \label{fig:kic7103006_wavelet}
\end{figure}

\begin{figure}[ht!]
    \centering
    \includegraphics[width=0.49\textwidth]{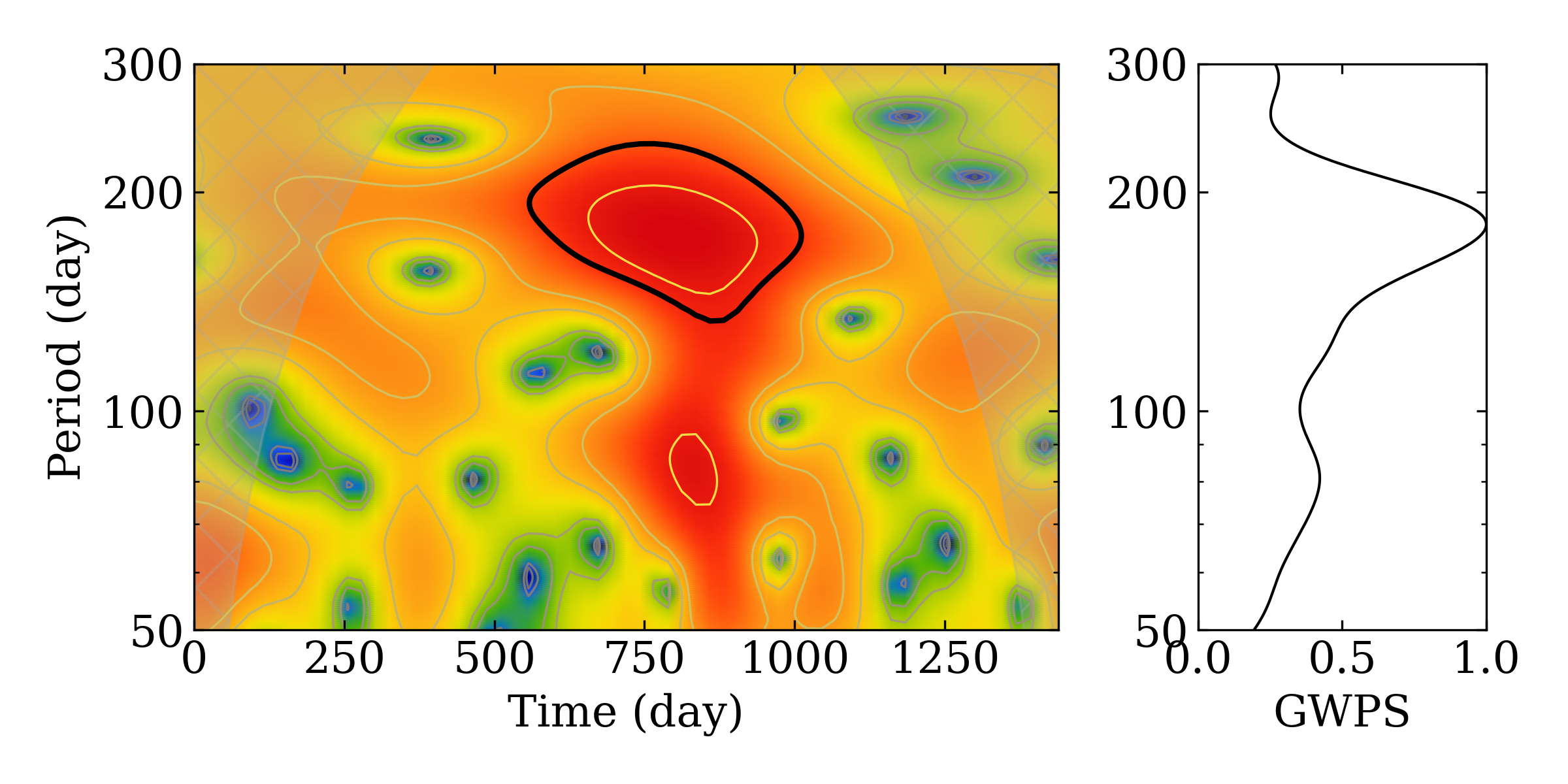}
    \includegraphics[width=0.49\textwidth]{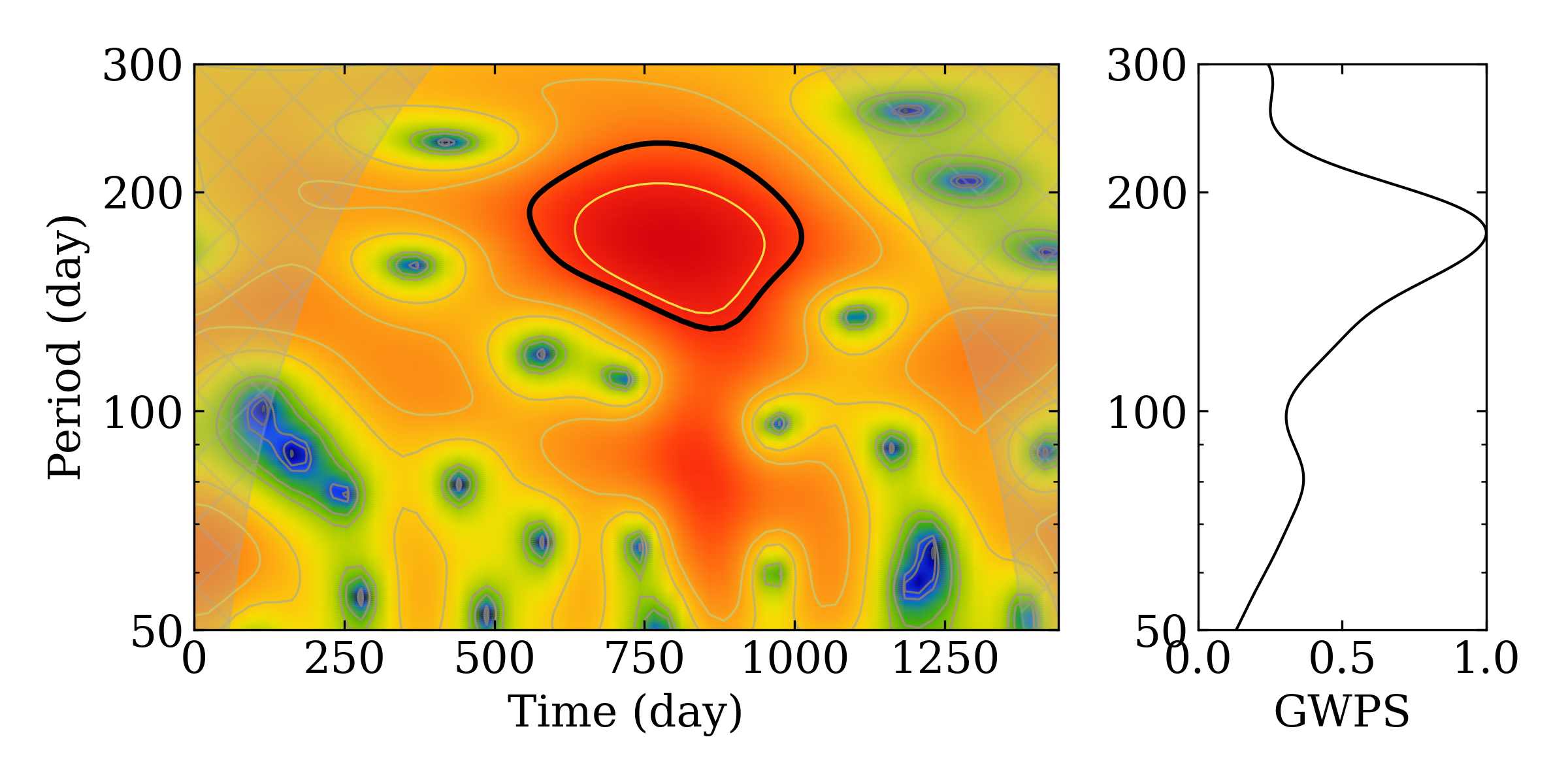}
    \caption{Same as Fig.~\ref{fig:kic3733735_wavelet} for KIC~9025370.}
    \label{fig:kic9025370_wavelet}
\end{figure}

\begin{figure}[ht!]
    \centering
    \includegraphics[width=0.49\textwidth]{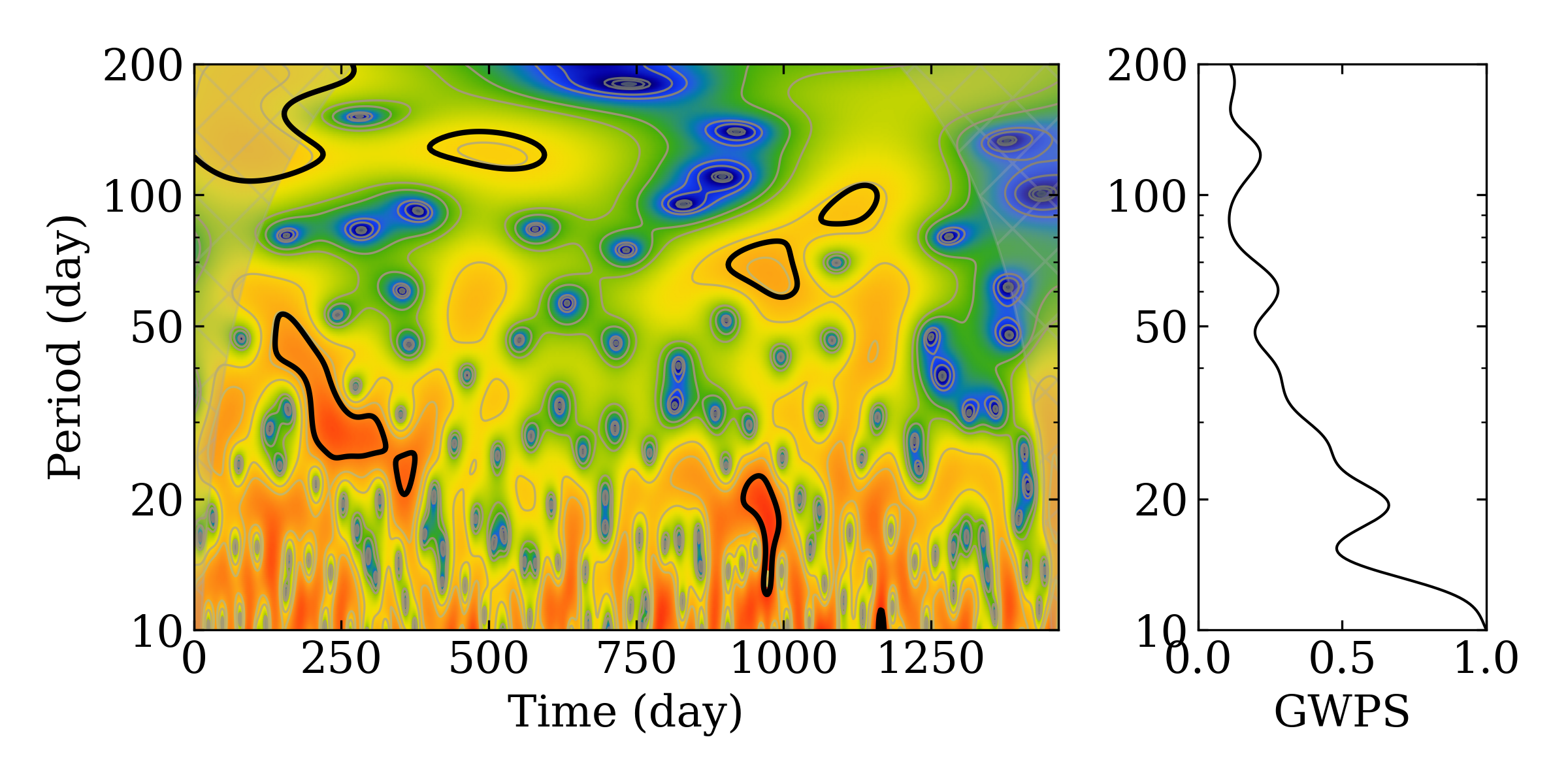}
    \includegraphics[width=0.49\textwidth]{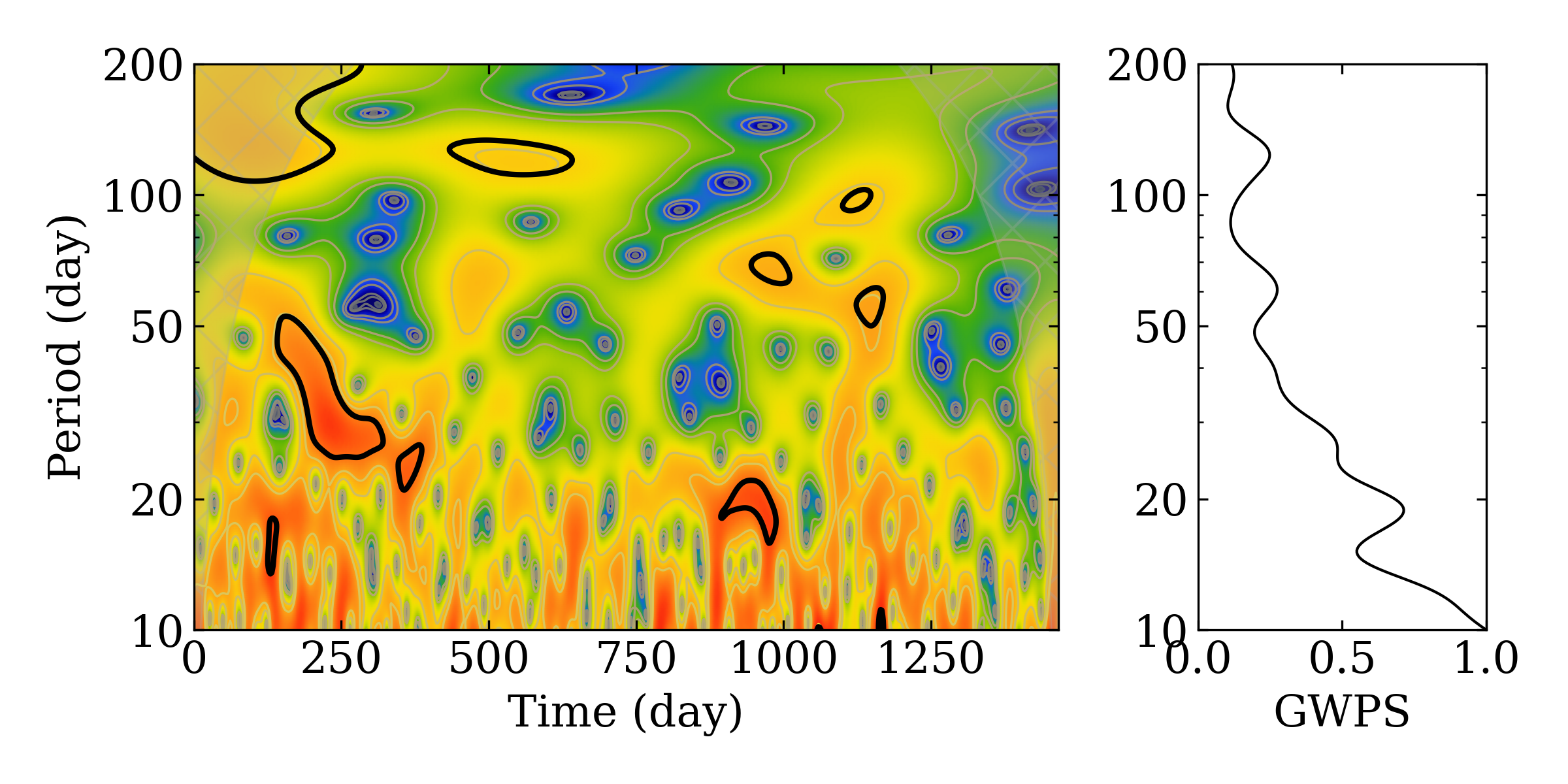}
    \caption{Same as Fig.~\ref{fig:kic3733735_wavelet} for KIC~9226926.}
    \label{fig:kic9226926_wavelet}
\end{figure}

\begin{figure}[ht!]
    \centering
    \includegraphics[width=0.49\textwidth]{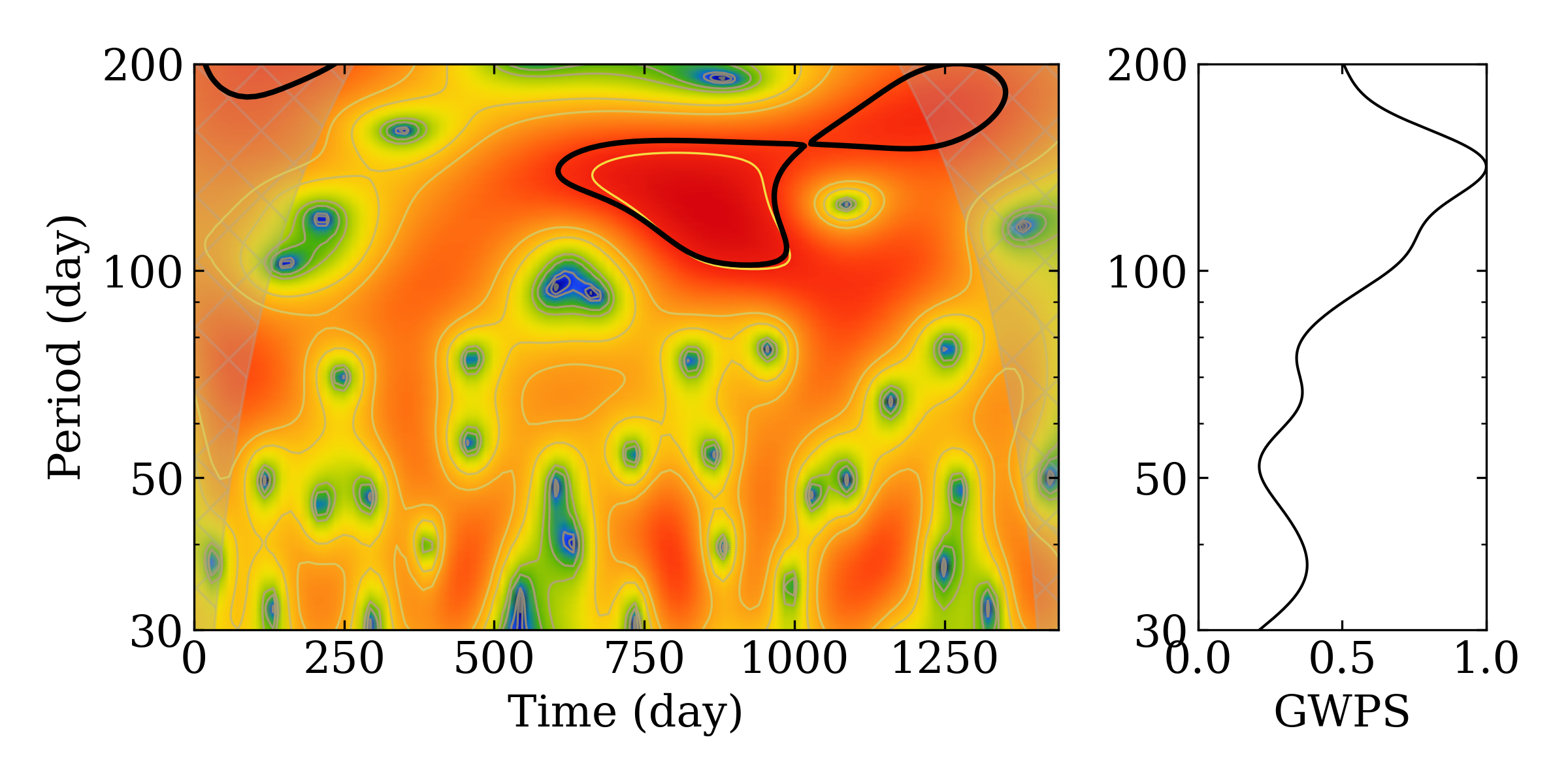}
    \includegraphics[width=0.49\textwidth]{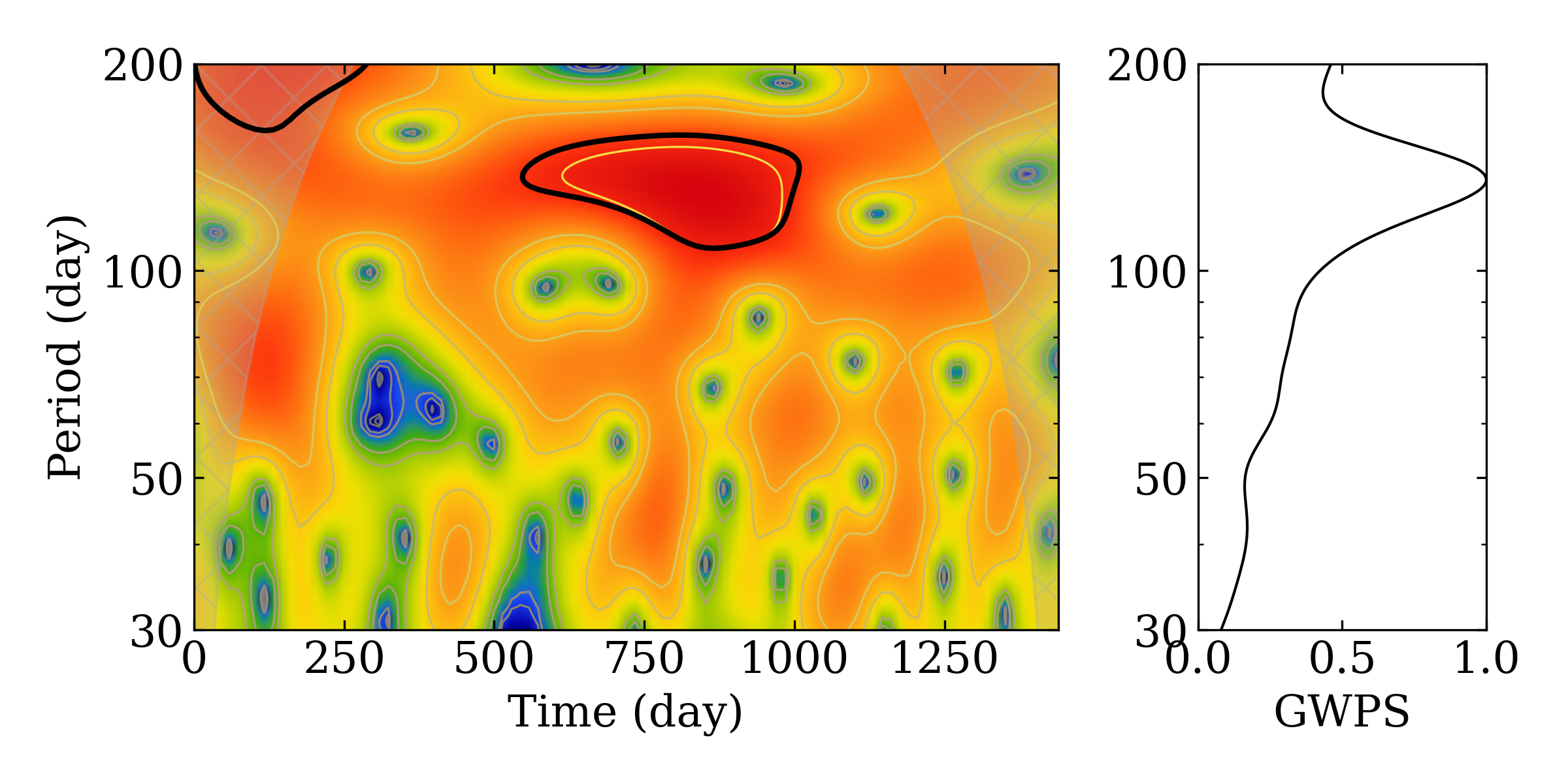}
    \caption{Same as Fig.~\ref{fig:kic3733735_wavelet} for KIC~10454113.}
    \label{fig:kic10454113_wavelet}
\end{figure}

In this section, we compute the wavelet decomposition of the reconstructed spot coverage of the \textit{Kepler} targets following the procedure described in Sect.~\ref{sec:analysing_complete_lc}. Because the \textit{Kepler} time series are shorter than the VIRGO/SPM solar time series, we restricted our analysis to periods shorter than 300 days. Periods that are longer than this are too strongly affected by the edge effects of the wavelet transform and the corresponding cone of influence. 
We only show the results of the wavelet decomposition for the stars for which we obtained a significant signal in the spectrum. We therefore underline that we do not observe a significant modulation in the wavelet decomposition of the spot coverage for KIC~8379927, although it was one of the stars with the strongest active nest pattern. The wavelet decompositions are shown in Fig.~\ref{fig:kic3733735_wavelet} to \ref{fig:kic10454113_wavelet}.

For KIC~3733735 (see Fig.~\ref{fig:kic3733735_wavelet}), we note a strong modulation around 100 days at an epoch that is contemporaneous with the appearance of the stable active nests in our longitudinal maps. 
This 100-day feature dominates the GWPS both for the spots-and-faculae and the spots-only model. 
We recall that for this star, \citet{Mathur2014} detected a low-frequency signal with a similar periodicity when they analysed the $S_\mathrm{ph}$ time series, but they attributed this signal to a beating phenomenon with an expected period of 116 days that was induced by the simultaneous presence of spots at two different latitudes with similar but distinct rotation rates. The phenomenon was thought to be caused by two close peaks related to rotational modulation in the periodogram. 
This evidence of a differential rotation signature is consistent with the group of migrating active nests that we observed in both our models (see Fig.~\ref{fig:kic3733735_longitude_map}). As discussed in Sect.~\ref{sec:analysing_complete_lc}, the most likely explanation for the 100-day feature we see here is the beating phenomenon that affects the reconstructed spot coverage.  

The wavelet decomposition for KIC~6508366, shown in Fig.~\ref{fig:kic6508366_wavelet}, exhibits short-lived modulations at different periodicities of 15 to 150 days. Features with a periodicity between 50 and 80 days are common to the spots-and-faculae and the spot-only models. The GWPS is dominated by short-term modulations with a periodicity shorter than 30 days. 

The main feature in the wavelet decomposition of KIC~7103006 (see Fig.~\ref{fig:kic7103006_wavelet}) is visible as a $\sim$100-day modulation that clearly appears after 750 days of observation. We also note a strong modulation of about 40 days in the time-frequency decomposition and in the GWPS.   

In the diagram obtained for KIC~9025370 (see Fig.~\ref{fig:kic9025370_wavelet}), we observe a strong modulation with a periodicity between 150 and 250~days that lasts for a few hundred days in the interval of time in which the spot coverage is the most important, as shown in Fig.~\ref{fig:kic9025370_longitude_map}. This is also the main feature of the GWPS. 

Just as for KIC~6508366, the GWPS for KIC~9226926 is dominated by short-period modulations (see Fig.~\ref{fig:kic9226926_wavelet}). Nevertheless, we also note a strong modulation between 20 and 50 days at about 250 days of observations, and another modulation between 30 and 120 days in the second half of the \textit{Kepler} mission, between 1000 and 1250 days of observations. As for KIC~3733735, \citet{Mathur2014} found evidence for a beating modulation in the $S_\mathrm{ph}$ time series of KIC~9226926, with an expected period of 513 days. This period lies in the cone of influence of our wavelet decomposition. 

Finally, the feature observed for KIC~10454113 (see Fig.~\ref{fig:kic10454113_wavelet}) is similar to the feature witnessed for KIC~9025370, with a modulation of about 150 days that coincides with the epoch of observation of the strongest active nests modelled in the time series, between 500 and 1000 days of observations (see Fig.~\ref{fig:kic10454113_wavelet}).

\section{Discussion \label{sec:discussion}}

The variety of features observed in the wavelet decomposition of starspot coverage suggests that just as in the Sun, internal processes modulate the surface activity and the magnetic flux emergence on a timescale that is shorter than Schwabe-like activity cycles.
In addition of the specific case of KIC~3733735, for which we favour the interpretation of the 100-day modulation in the wavelet transform as a signature of the beating phenomenon already discussed by \citet{Mathur2014}, the nature of the modulations we detect in five other targets (KIC~6508366, KIC~7103006, KIC~9025370, KIC~9226926, and KIC~1045113) has to be investigated. 
The periodogram inspection for these stars does not reveal evidence of a possible beating in the period range of interest, thus we may assume that these signatures are actual cyclic modulations of the spot coverage, although we emphasise that some power excesses around 90 days might also be related to modulations introduced by the jump between the \textit{Kepler} CCDs at the beginning of each new observation quarter. 
We recall here that numerical magnetohydrodynamics (MHD) simulations from \citet{Brun2022} for instance showed evidence of short cycles with a timescale of one year for stars at low $Ro$, which were presented by the authors as quasi-biennal-oscillation-like magnetic modulations. 
The timescale of these modulations are also compatible with Rieger-like periodicities of the spot coverage. 

\begin{figure*}[ht!]
    \centering
    \includegraphics[width=0.98\textwidth]{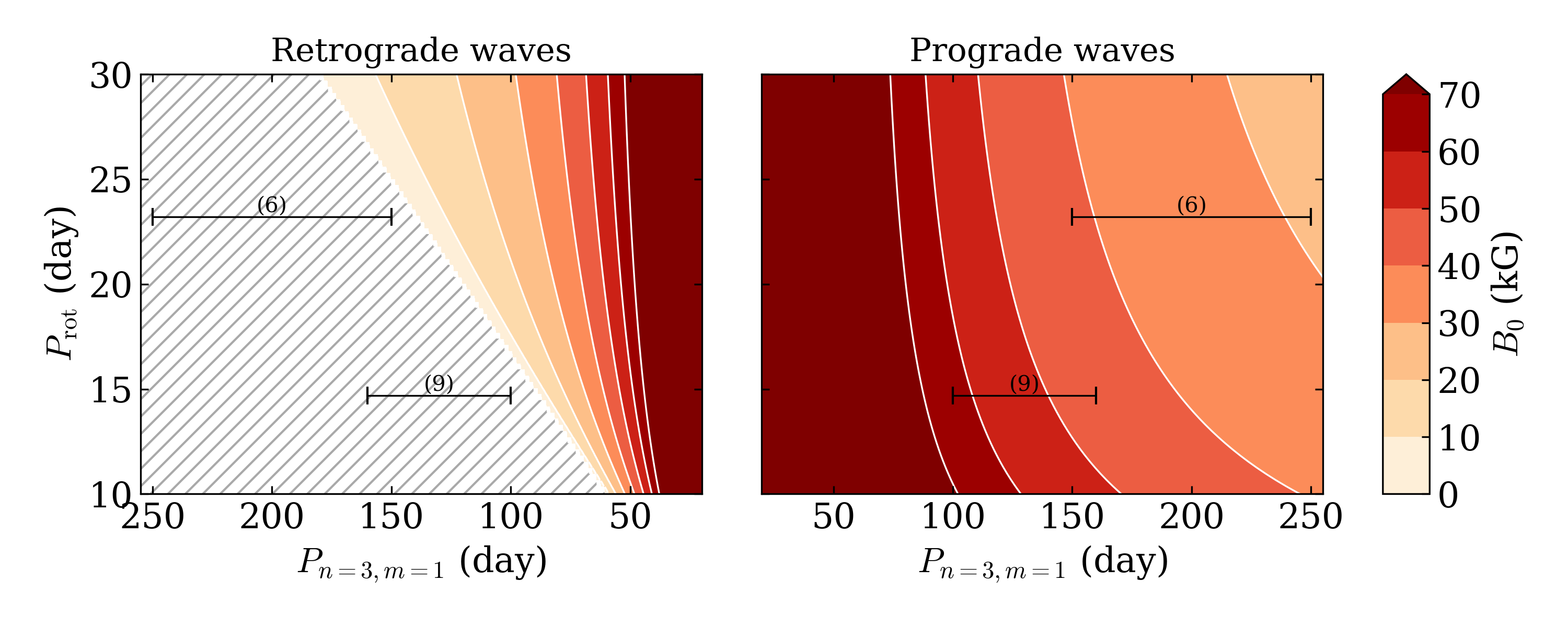}
    \includegraphics[width=0.98\textwidth]{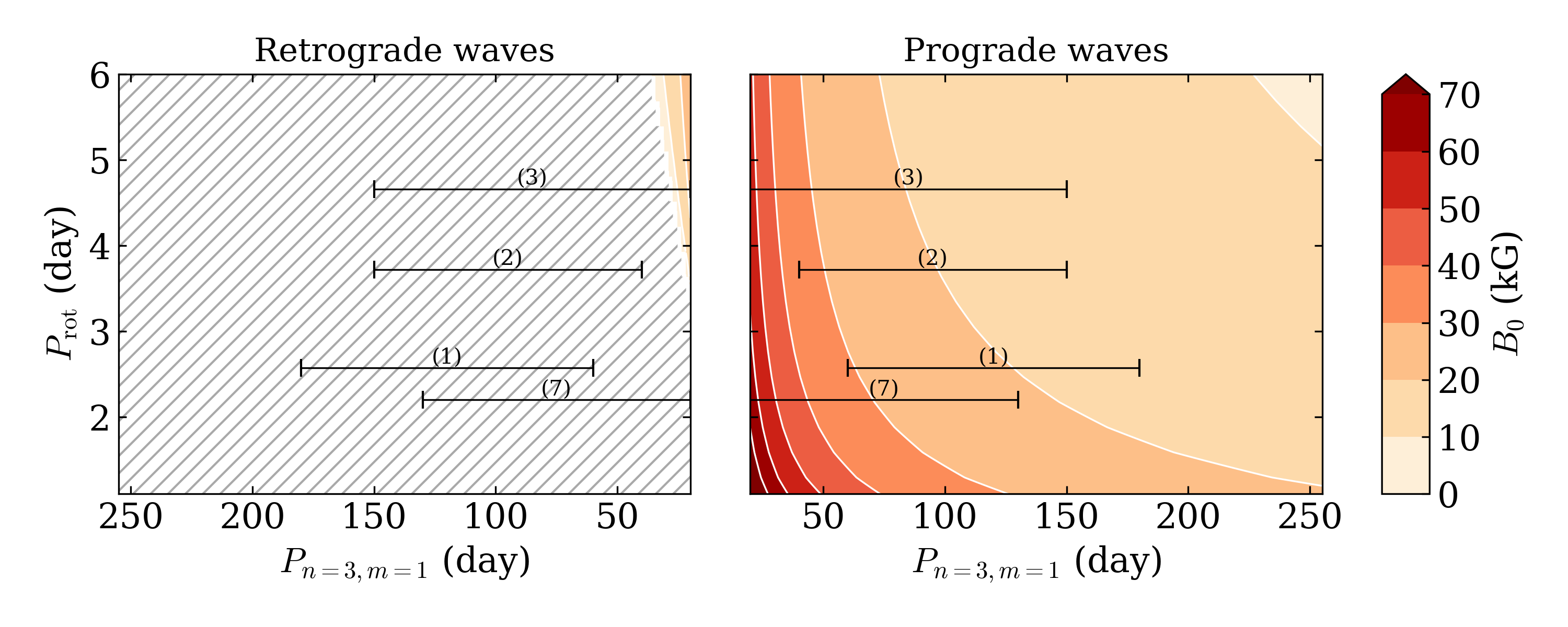}
    \caption{$B_0$ field intensity computed for a $n=3$, $m=1$ magneto-Rossby wave in a given range of rotation period $P_\mathrm{rot}$ and wave periods $P_{nm} = 2\pi / \omega_{nm}$, for $\rho= \num{1e-1}$~g.cm$^{-1}$, $R_{\rm tach} = 0.7 \, \mathrm{R}_\odot$ (\textit{top}) and $\rho= \num{1e-3}$~g.cm$^{-1}$, $R_{\rm tach} = 0.9 \, R_\star$, $R_\star = 1.5 \, \mathrm{R}_\odot$ (\textit{bottom}).
    Retrograde waves are shown in the left panel, and prograde waves are plotted in the right panel.
    The location of the six stars with significant spot coverage modulations is shown in black in the diagram.
    Each star is identified with the identifier provided in Table~\ref{tab:considered_targets}.
    See the main text for the explanation of the range of $P_\mathrm{rot}$ shown in each panel. 
    }
    \label{fig:summary_rossby_n3}
\end{figure*}

\begin{figure*}[ht!]
    \centering
    \includegraphics[width=0.98\textwidth]{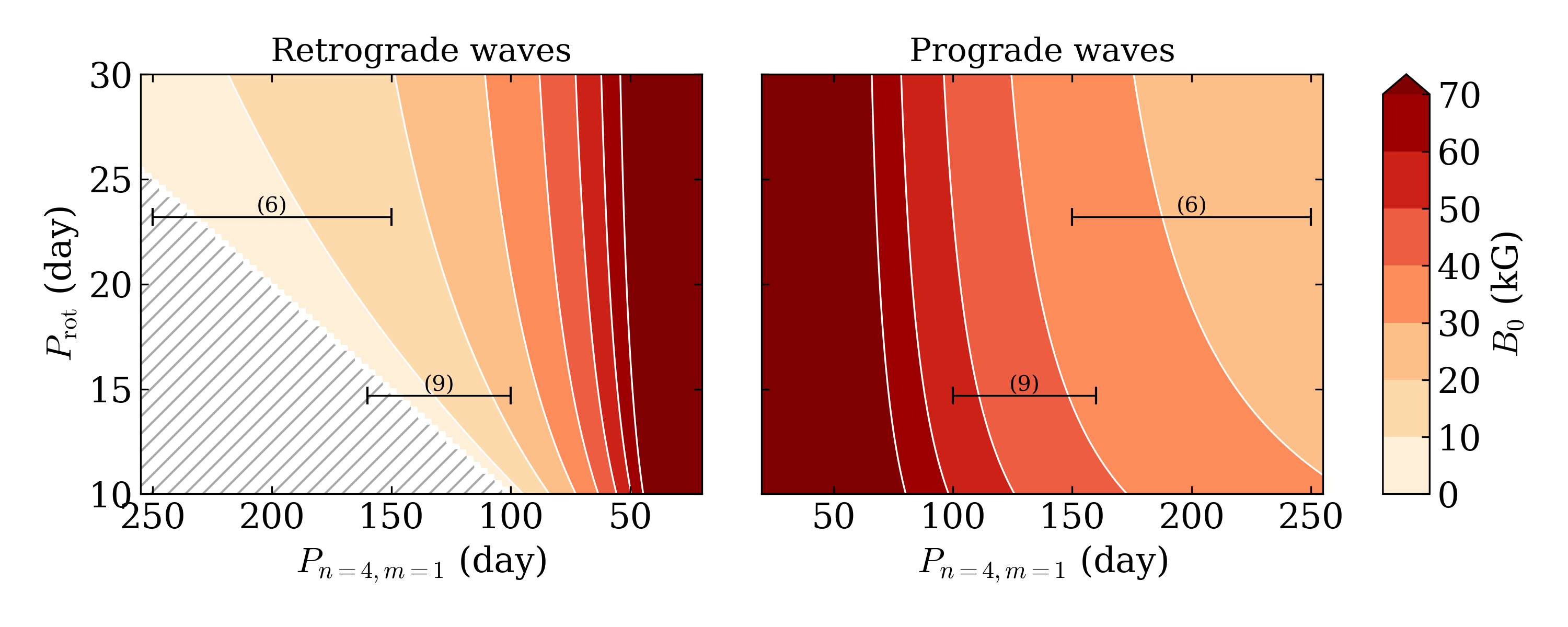}
    \includegraphics[width=0.98\textwidth]{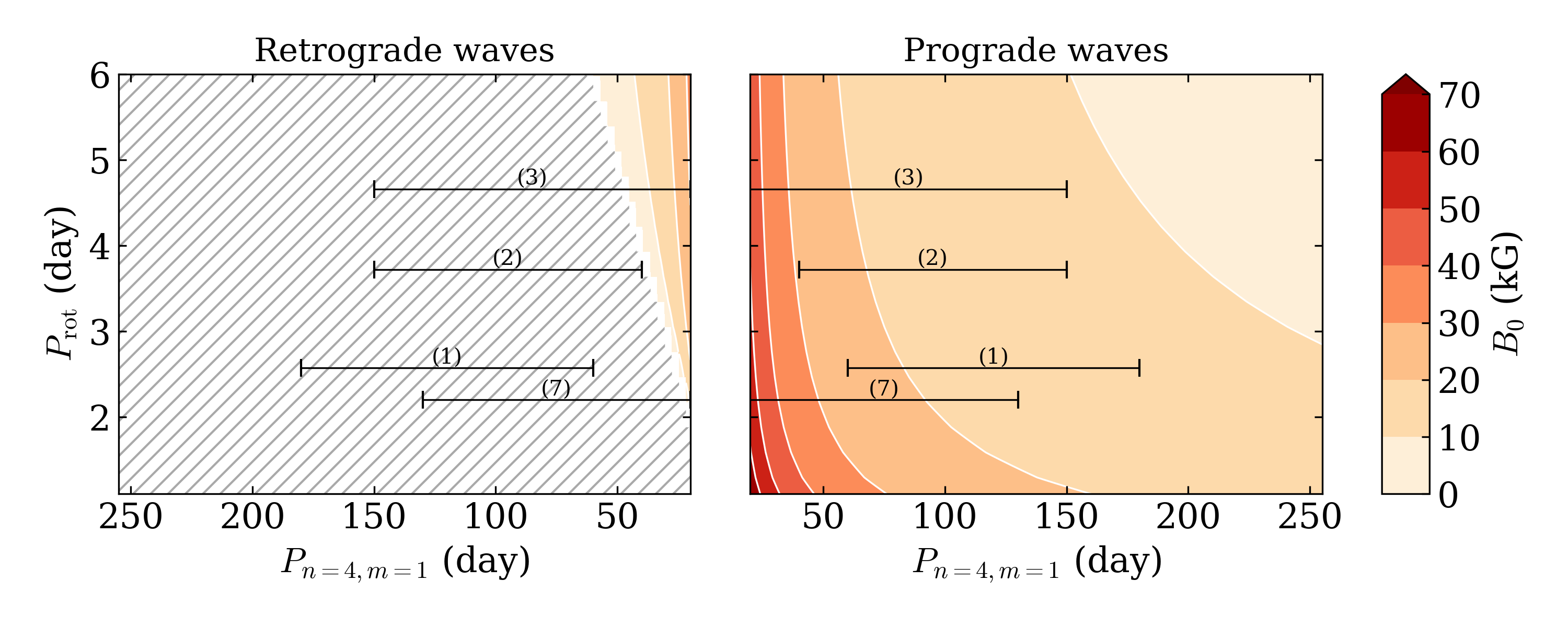}
    \caption{Same as Fig.~\ref{fig:summary_rossby_n3} for a $n=4$, $m=1$ magneto-Rossby wave.}
    \label{fig:summary_rossby_n4}
\end{figure*}

Under this last hypothesis, following \citet{Gurgenashvili2021}, we can use the magneto-Rossby wave dispersion relation derived by \citet{Zaqarashvili2007,Zaqarashvili2018} in the case of an homogeneous toroidal magnetic field with amplitude depending on the co-latitude $\theta$, that is, $B = B_0 \sin \theta$
\begin{equation}
\label{eq:dispersion_magneto_rossby}
    \omega_{n m} = - \frac{m \Omega}{n (n + 1)} \left ( 1 \pm \sqrt{1 - \frac{v_A^2 n (n + 1)}{\Omega^2 R_{\rm tach}^2} (2 - n (n + 1)) } \right ) \; , 
\end{equation}
where $\Omega$ is the stellar rotational angular frequency, $R_{\rm tach}$ is the radius at the tachocline,  $n$ is the poloidal wave number, and $m$ is the azimuthal wave number.
The Alfvèn speed, $v_A$, is given by
\begin{equation}
    v_A = \frac{B_0}{\sqrt{4 \pi \rho}} \; ,
\end{equation}
where $\rho$ is the density of the medium in which the waves propagate. 
Two regimes of waves exist: slow and fast magneto-Rossby waves. They depend on the sign of the operator in front of the square root in Eq.~(\ref{eq:dispersion_magneto_rossby}). 
The fast waves are retrograde waves relative to stellar rotation, while slow waves have a prograde propagation and do not exist in the absence of magnetic field \citep{Dikpati2020}. 
Compared to the traditional case of hydrodynamical Rossby waves with the well-known dispersion relation $\omega_{nm} = - 2m \Omega / [n (n+1)]$ \citep[e.g.][]{Papaloizou1978,Saio1982}, it should be noted that strong magnetic fields allow the existence of magneto-inertial waves with a frequency faster than the Coriolis frequency, $2 \Omega$.
Given Eq.~(\ref{eq:dispersion_magneto_rossby}), $B_0$ can therefore be trivially obtained from this relation: 
\begin{equation}
    B_0^2 = \Omega^2 R_{\rm tach}^2 \frac{4 \pi \rho}{n (n+1) (2 - n (n+1))} 
    \left [ 1 - \Bigg(1 + \frac{n (n+1)}{m \Omega} \omega_{n m} \Bigg)^2  \right]
    \; .
\end{equation}

We computed $B_0$ for a given range of $\Omega$ and $\omega_{nm}$, considering $\rho = \num{1e-1}$~g.cm$^{-3}$ and $\rho = \num{1e-3}$~g.cm$^{-3}$, which typically are the order of magnitude of the tachocline density in a 1~$\rm M_\odot$~star and in a 1.3~$\rm M_\odot$~star, respectively \citep[e.g.][]{Breton2022simuFstars}. 
In the first case, we considered $R_{\rm tach} = 0.7 \, R_\odot$, and in the second case, we considered $R_{\rm tach} = 0.9 \, R_\star$ with $R_\star = 1.5 \, R_\odot$.
Given the uncertainties on the quantities we measure, this choice allowed us to limit what follows to an order-of-magnitude discussion.   
We show the corresponding diagrams in Fig.~\ref{fig:summary_rossby_n3} for $n=3$, $m=1$ waves and in Fig.~\ref{fig:summary_rossby_n4} for $n=4$, $m=1$ waves.
In order to compare the predicted $B_0$ intensities with the observations, we show in the diagrams the $P_\mathrm{rot}$ of each star for which we find a significant modulation in the wavelet transform, with the corresponding period interval.  
KIC~9025370 and KIC~1045113 are shown in the diagram that we computed considering $\rho = \num{1e-1}$~g.cm$^{-3}$, while the location of KIC~3733735, KIC~6508366, KIC~7103006, and KIC~9226926 is shown in the diagrams computed with $\rho = \num{1e-3}$~g.cm$^{-3}$. The $P_\mathrm{rot}$ intervals chosen for each diagram are therefore chosen accordingly. For sake of clarity, we refer to KIC~9025370 and KIC~1045113 as group~1 and to KIC~3733735, KIC~6508366, KIC~7103006, and KIC~9226926 as group~2.  
For $n=3$, $m=1$ waves, we note that the observed modulation lies in a period range in which no retrograde waves are expected for either group~1 or group~2. The corresponding $B_0$ fields that cause these waves in the case of group~1 would have to be between 20 and 70~kG, and they would have to be between 10 and 60~kG for group~2. 
While $n=4$, $m=1$ still excludes retrograde waves for group~2 (except possibly KIC~7103006), it corresponds to a range for group~1 in which $B_0$ is below 25~kG. For these two stars, prograde waves with analogous periodicites would have to be generated by the interaction with stronger magnetic fields, between 20 and 60~kG. 
Similarly to what was found in the $n=3$, $m=1$ case, a $B_0$ field with an intensity between 10 and 50~kG is sufficient for the envelope to host prograde magneto-Rossby waves in the range of periods observed for group~2.

From the comparison between the diagrams and the observed periodicities, it might seem surprising that we obtain similar magnetic field amplitude estimates both the slow and fast rotators. Nevertheless, the stars considered for this analysis have $S_\mathrm{ph}$ activity proxies that are quite similar, corresponding to moderate levels of activity, ranging from 100 to 400~ppm (see Table~\ref{tab:considered_targets}). This can be explained by the fact that the fast rotators in our working sample consist of F-type stars whose $T_\mathrm{eff}$ locates them above the Kraft break at 6250~K \citep{Kraft1967} or close to it, where stars are expected to have reduced magnetic activity because their convective envelope are more shallow.

We finally underline that searching for the signatures of waves like this in 3D~MHD simulations of solar-type stars such as those performed by \citet{Brun2022} would bring more insight into the expected amplitudes of $n$, $m$ modes and therefore would allow us to refine the above discussion by considering the expected dominating modes. A prescription for distinguishing between prograde and retrograde waves would be particularly useful.   

\section{Conclusion \label{section:conclusion}}

We reported the results of an ensemble starspot-modelling analysis applied on the Sun and on a sample of solar-type pulsators observed by the \textit{Kepler} mission. 
By considering both a spots-and-faculae model and a spots-only model, we assessed how the unknown ratio of the faculae-to-spot coverage might affect the properties inferred from the spot models. We showed that the two approaches shared many common features, which allowed us to draw a picture of activity-induced brightness modulations of solar-type stars that depends on models in a limited way. 
In the particular case of the Sun, we were indeed able to show that both the spots-and-faculae and the spots-only model were able to reconstruct the longitudinal distribution of observed sunspots in cycles~23 and 24. Consequently, we demonstrated that stable longitudinal active nests can successfully searched for in moderately active solar-type pulsators. This is an important step to characterise the low-frequency mechanisms in these key targets, especially magneto-inertial waves that propagate in the convective envelope and interact with convective flows. Stable active nests were observed in our sample for different levels of photospheric activity and different $Ro$ regimes.

Furthermore, we used a Morlet wavelet decomposition to search for significant signatures of cyclic modulations in the stellar spot coverage. In the case of the Sun, we showed that the main modulation that is visible after applying the wavelet transform had a timescale that allowed us to identify it with a manifestation of the quasi-biennal oscillation. The analysis of the \textit{Kepler} sample showed a variety of modulations ranging from a few dozen days to several hundreds of days. We recalled that some of these modulations might arise from beating due to stable active regions with distinct rotation rates, as identified by \citet{Mathur2014}, and we discussed the possible nature of the remaining signatures, suggesting that they could be related to quasi-biennal-oscillation- or Rieger-like cycles. 
Adopting the hypothesis that these cycles are connected to the action of magneto-Rossby waves close to the stellar tachocline, we showed that these modulations could arise from the action of internal toroidal fields with an intensity of up to several dozen kiloGauss. 

Starspot modelling opens important applications for the light curves that will be collected by PLATO in the near future. This prospect motivated our choice to consider only stars with detected solar-type acoustic oscillations in this work because moderately active targets like this will represent the core sample of the upcoming space mission. 
In particular, the spectro-polarimetric follow-up of the most interesting PLATO targets will be a strong asset based on which the questions discussed in this work can be explored in more detail. In particular, given the structural constraints available for the stars in our sample \citep[e.g.][]{SilvaAguirre2017}, the longitudinal maps we present here constitute an interesting benchmark for a comparison with numerical MHD simulations of flux emergence in solar-type stars \citep[e.g.][]{Toriumi2012,Stein2012,Isik2018}. Reproducing some of the nest features we observed in self-consistent simulations run in a spherical shell would indeed allow us to proceed in our understanding of the parameter dependence of these structures.

\begin{acknowledgements}
The authors want to thank the anonymous referee for providing constructive comments and suggestions.
S.N.B, A.F.L, and S.M. acknowledge support from PLATO ASI-INAF agreement n.~2015-019-R.1-2018.
The authors also acknowledge R.A~García for providing the calibrated \textit{Kepler} light curves used in this study, and A.~Jímenez for providing the VIRGO/SPM time series. 
The spot models computations were performed with the IRFU/CEA Saclay server facilities, funded by ERC Synergy grant WholeSun No.810218, the P2IO Labex emergence project FlarePredict, and CNES PLATO funds. 
This paper includes data collected by the \textit{Kepler} mission, and obtained from the MAST data archive at the Space Telescope Science Institute (STScI). Funding for the \textit{Kepler} mission is provided by the NASA Science Mission Directorate. STScI is operated by the Association of Universities for Research in Astronomy, Inc., under NASA contract NAS 5–26555.
\\
\textit{Software:} 
\texttt{loupiotes} (this work),
\texttt{numpy} \citep{harris2020array}, \texttt{matplotlib} \citep{Hunter:2007}, \texttt{scipy} \citep{2020SciPy-NMeth}, 
\texttt{astropy } \citep{astropy:2022}, 
\texttt{PyMC} \citep{pymc}.

\\
\end{acknowledgements}

\bibliographystyle{aa} 
\bibliography{biblio.bib} 

\begin{thebibliography}{94}
\expandafter\ifx\csname natexlab\endcsname\relax\def\natexlab#1{#1}\fi

\bibitem[{{Aerts} {et~al.}(2010){Aerts}, {Christensen-Dalsgaard}, \&
  {Kurtz}}]{Aerts2010}
{Aerts}, C., {Christensen-Dalsgaard}, J., \& {Kurtz}, D.~W. 2010,
  {Asteroseismology}

\bibitem[{{Appourchaux} {et~al.}(2012){Appourchaux}, {Chaplin}, {Garc{\'\i}a},
  {Gruberbauer}, {Verner}, {Antia}, {Benomar}, {Campante}, {Davies},
  {Deheuvels}, {Handberg}, {Hekker}, {Howe}, {R{\'e}gulo}, {Salabert},
  {Bedding}, {White}, {Ballot}, {Mathur}, {Silva Aguirre}, {Elsworth}, {Basu},
  {Gilliland}, {Christensen-Dalsgaard}, {Kjeldsen}, {Uddin}, {Stumpe}, \&
  {Barclay}}]{Appourchaux2012}
{Appourchaux}, T., {Chaplin}, W.~J., {Garc{\'\i}a}, R.~A., {et~al.} 2012, \aap,
  543, A54

\bibitem[{{Appourchaux} \& {Pall{\'e}}(2013)}]{Appourchaux2013}
{Appourchaux}, T. \& {Pall{\'e}}, P.~L. 2013, in Astronomical Society of the
  Pacific Conference Series, Vol. 478, Fifty Years of Seismology of the Sun and
  Stars, ed. K.~{Jain}, S.~C. {Tripathy}, F.~{Hill}, J.~W. {Leibacher}, \&
  A.~A. {Pevtsov}, 125

\bibitem[{{Astropy Collaboration} {et~al.}(2022){Astropy Collaboration},
  {Price-Whelan}, {Lim}, {Earl}, {Starkman}, {Bradley}, {Shupe}, {Patil},
  {Corrales}, {Brasseur}, {N{"o}the}, {Donath}, {Tollerud}, {Morris},
  {Ginsburg}, {Vaher}, {Weaver}, {Tocknell}, {Jamieson}, {van Kerkwijk},
  {Robitaille}, {Merry}, {Bachetti}, {G{"u}nther}, {Aldcroft},
  {Alvarado-Montes}, {Archibald}, {B{'o}di}, {Bapat}, {Barentsen}, {Baz{'a}n},
  {Biswas}, {Boquien}, {Burke}, {Cara}, {Cara}, {Conroy}, {Conseil}, {Craig},
  {Cross}, {Cruz}, {D'Eugenio}, {Dencheva}, {Devillepoix}, {Dietrich},
  {Eigenbrot}, {Erben}, {Ferreira}, {Foreman-Mackey}, {Fox}, {Freij}, {Garg},
  {Geda}, {Glattly}, {Gondhalekar}, {Gordon}, {Grant}, {Greenfield}, {Groener},
  {Guest}, {Gurovich}, {Handberg}, {Hart}, {Hatfield-Dodds}, {Homeier},
  {Hosseinzadeh}, {Jenness}, {Jones}, {Joseph}, {Kalmbach}, {Karamehmetoglu},
  {Ka{l}uszy{'n}ski}, {Kelley}, {Kern}, {Kerzendorf}, {Koch}, {Kulumani},
  {Lee}, {Ly}, {Ma}, {MacBride}, {Maljaars}, {Muna}, {Murphy}, {Norman},
  {O'Steen}, {Oman}, {Pacifici}, {Pascual}, {Pascual-Granado}, {Patil},
  {Perren}, {Pickering}, {Rastogi}, {Roulston}, {Ryan}, {Rykoff}, {Sabater},
  {Sakurikar}, {Salgado}, {Sanghi}, {Saunders}, {Savchenko}, {Schwardt},
  {Seifert-Eckert}, {Shih}, {Jain}, {Shukla}, {Sick}, {Simpson},
  {Singanamalla}, {Singer}, {Singhal}, {Sinha}, {Sip{H{o}}cz}, {Spitler},
  {Stansby}, {Streicher}, {{{S}}umak}, {Swinbank}, {Taranu}, {Tewary},
  {Tremblay}, {Val-Borro}, {Van Kooten}, {Vasovi{'c}}, {Verma}, {de Miranda
  Cardoso}, {Williams}, {Wilson}, {Winkel}, {Wood-Vasey}, {Xue}, {Yoachim},
  {Zhang}, {Zonca}, \& {Astropy Project Contributors}}]{astropy:2022}
{Astropy Collaboration}, {Price-Whelan}, A.~M., {Lim}, P.~L., {et~al.} 2022,
  apj, 935, 167

\bibitem[{{Basri} \& {Shah}(2020)}]{Basri2020}
{Basri}, G. \& {Shah}, R. 2020, \apj, 901, 14

\bibitem[{{Basri} {et~al.}(2010){Basri}, {Walkowicz}, {Batalha}, {Gilliland},
  {Jenkins}, {Borucki}, {Koch}, {Caldwell}, {Dupree}, {Latham}, {Meibom},
  {Howell}, \& {Brown}}]{Basri2010}
{Basri}, G., {Walkowicz}, L.~M., {Batalha}, N., {et~al.} 2010, \apjl, 713, L155

\bibitem[{{Bazot} {et~al.}(2018){Bazot}, {Nielsen}, {Mary},
  {Christensen-Dalsgaard}, {Benomar}, {Petit}, {Gizon}, {Sreenivasan}, \&
  {White}}]{Bazot2018}
{Bazot}, M., {Nielsen}, M.~B., {Mary}, D., {et~al.} 2018, \aap, 619, L9

\bibitem[{{Bekki} {et~al.}(2022){Bekki}, {Cameron}, \& {Gizon}}]{Bekki2022b}
{Bekki}, Y., {Cameron}, R.~H., \& {Gizon}, L. 2022, \aap, 666, A135

\bibitem[{{Belkacem} {et~al.}(2022){Belkacem}, {Pin{\c{c}}on}, \&
  {Buldgen}}]{Belkacem2022}
{Belkacem}, K., {Pin{\c{c}}on}, C., \& {Buldgen}, G. 2022, \solphys, 297, 147

\bibitem[{{Benomar} {et~al.}(2018){Benomar}, {Bazot}, {Nielsen}, {Gizon},
  {Sekii}, {Takata}, {Hotta}, {Hanasoge}, {Sreenivasan}, \&
  {Christensen-Dalsgaard}}]{Benomar2018}
{Benomar}, O., {Bazot}, M., {Nielsen}, M.~B., {et~al.} 2018, Science, 361, 1231

\bibitem[{{Berdyugina} \& {Usoskin}(2003)}]{Berdyugina2003}
{Berdyugina}, S.~V. \& {Usoskin}, I.~G. 2003, \aap, 405, 1121

\bibitem[{{B{\'e}trisey} {et~al.}(2023){B{\'e}trisey}, {Eggenberger},
  {Buldgen}, {Benomar}, \& {Bazot}}]{Betrisey2023}
{B{\'e}trisey}, J., {Eggenberger}, P., {Buldgen}, G., {Benomar}, O., \&
  {Bazot}, M. 2023, \aap, 673, L11

\bibitem[{{Borucki} {et~al.}(2010){Borucki}, {Koch}, {Basri}, {Batalha},
  {Brown}, {Caldwell}, {Caldwell}, {Christensen-Dalsgaard}, {Cochran},
  {DeVore}, {Dunham}, {Dupree}, {Gautier}, {Geary}, {Gilliland}, {Gould},
  {Howell}, {Jenkins}, {Kondo}, {Latham}, {Marcy}, {Meibom}, {Kjeldsen},
  {Lissauer}, {Monet}, {Morrison}, {Sasselov}, {Tarter}, {Boss}, {Brownlee},
  {Owen}, {Buzasi}, {Charbonneau}, {Doyle}, {Fortney}, {Ford}, {Holman},
  {Seager}, {Steffen}, {Welsh}, {Rowe}, {Anderson}, {Buchhave}, {Ciardi},
  {Walkowicz}, {Sherry}, {Horch}, {Isaacson}, {Everett}, {Fischer}, {Torres},
  {Johnson}, {Endl}, {MacQueen}, {Bryson}, {Dotson}, {Haas}, {Kolodziejczak},
  {Van Cleve}, {Chandrasekaran}, {Twicken}, {Quintana}, {Clarke}, {Allen},
  {Li}, {Wu}, {Tenenbaum}, {Verner}, {Bruhweiler}, {Barnes}, \&
  {Prsa}}]{Borucki2010}
{Borucki}, W.~J., {Koch}, D., {Basri}, G., {et~al.} 2010, Science, 327, 977

\bibitem[{{Breton} {et~al.}(2022){Breton}, {Brun}, \&
  {Garc{\'\i}a}}]{Breton2022simuFstars}
{Breton}, S.~N., {Brun}, A.~S., \& {Garc{\'\i}a}, R.~A. 2022, \aap, 667, A43

\bibitem[{{Breton} {et~al.}(2023){Breton}, {Dhouib}, {Garc{\'\i}a}, {Brun},
  {Mathis}, {P{\'e}rez Hern{\'a}ndez}, {Mathur}, {Dyrek}, {Santos}, \&
  {Pall{\'e}}}]{Breton2023}
{Breton}, S.~N., {Dhouib}, H., {Garc{\'\i}a}, R.~A., {et~al.} 2023, \aap, 679,
  A104

\bibitem[{{Breton} {et~al.}(2021){Breton}, {Santos}, {Bugnet}, {Mathur},
  {Garc{\'\i}a}, \& {Pall{\'e}}}]{Breton2021}
{Breton}, S.~N., {Santos}, A.~R.~G., {Bugnet}, L., {et~al.} 2021, \aap, 647,
  A125

\bibitem[{Brun \& Browning(2017)}]{Brun2017b}
Brun, A.~S. \& Browning, M.~K. 2017, Living Rev. Solar Phys., 14, 4

\bibitem[{{Brun} {et~al.}(2022){Brun}, {Strugarek}, {Noraz}, {Perri}, {Varela},
  {Augustson}, {Charbonneau}, \& {Toomre}}]{Brun2022}
{Brun}, A.~S., {Strugarek}, A., {Noraz}, Q., {et~al.} 2022, \apj, 926, 21

\bibitem[{{Brun} {et~al.}(2017){Brun}, {Strugarek}, {Varela}, {Matt},
  {Augustson}, {Emeriau}, {DoCao}, {Brown}, \& {Toomre}}]{Brun2017a}
{Brun}, A.~S., {Strugarek}, A., {Varela}, J., {et~al.} 2017, \apj, 836, 192

\bibitem[{{Carbonell} \& {Ballester}(1990)}]{Carbonell1990}
{Carbonell}, M. \& {Ballester}, J.~L. 1990, \aap, 238, 377

\bibitem[{{Chaplin} {et~al.}(2011{\natexlab{a}}){Chaplin}, {Bedding},
  {Bonanno}, {Broomhall}, {Garc{\'\i}a}, {Hekker}, {Huber}, {Verner}, {Basu},
  {Elsworth}, {Houdek}, {Mathur}, {Mosser}, {New}, {Stevens}, {Appourchaux},
  {Karoff}, {Metcalfe}, {Molenda-{\.Z}akowicz}, {Monteiro}, {Thompson},
  {Christensen-Dalsgaard}, {Gilliland}, {Kawaler}, {Kjeldsen}, {Ballot},
  {Benomar}, {Corsaro}, {Campante}, {Gaulme}, {Hale}, {Handberg}, {Jarvis},
  {R{\'e}gulo}, {Roxburgh}, {Salabert}, {Stello}, {Mullally}, {Li}, \&
  {Wohler}}]{Chaplin2011b}
{Chaplin}, W.~J., {Bedding}, T.~R., {Bonanno}, A., {et~al.} 2011{\natexlab{a}},
  \apjl, 732, L5

\bibitem[{{Chaplin} {et~al.}(2014){Chaplin}, {Elsworth}, {Davies}, {Campante},
  {Handberg}, {Miglio}, \& {Basu}}]{Chaplin2014}
{Chaplin}, W.~J., {Elsworth}, Y., {Davies}, G.~R., {et~al.} 2014, \mnras, 445,
  946

\bibitem[{{Chaplin} {et~al.}(2011{\natexlab{b}}){Chaplin}, {Kjeldsen},
  {Christensen-Dalsgaard}, {Basu}, {Miglio}, {Appourchaux}, {Bedding},
  {Elsworth}, {Garc{\'\i}a}, {Gilliland}, {Girardi}, {Houdek}, {Karoff},
  {Kawaler}, {Metcalfe}, {Molenda-{\.Z}akowicz}, {Monteiro}, {Thompson},
  {Verner}, {Ballot}, {Bonanno}, {Brand{\~a}o}, {Broomhall}, {Bruntt},
  {Campante}, {Corsaro}, {Creevey}, {Do{\u{g}}an}, {Esch}, {Gai}, {Gaulme},
  {Hale}, {Handberg}, {Hekker}, {Huber}, {Jim{\'e}nez}, {Mathur}, {Mazumdar},
  {Mosser}, {New}, {Pinsonneault}, {Pricopi}, {Quirion}, {R{\'e}gulo},
  {Salabert}, {Serenelli}, {Silva Aguirre}, {Sousa}, {Stello}, {Stevens},
  {Suran}, {Uytterhoeven}, {White}, {Borucki}, {Brown}, {Jenkins}, {Kinemuchi},
  {Van Cleve}, \& {Klaus}}]{Chaplin2011}
{Chaplin}, W.~J., {Kjeldsen}, H., {Christensen-Dalsgaard}, J., {et~al.}
  2011{\natexlab{b}}, Science, 332, 213

\bibitem[{Christensen-Dalsgaard(2014)}]{ChristensenDaalsgardLectureNotes}
Christensen-Dalsgaard, J. 2014, Lecture Notes on Stellar Oscillations, fifth
  edition

\bibitem[{{Corsaro} {et~al.}(2021){Corsaro}, {Bonanno}, {Mathur},
  {Garc{\'\i}a}, {Santos}, {Breton}, \& {Khalatyan}}]{Corsaro2021}
{Corsaro}, E., {Bonanno}, A., {Mathur}, S., {et~al.} 2021, \aap, 652, L2

\bibitem[{{Creevey} {et~al.}(2017){Creevey}, {Metcalfe}, {Schultheis},
  {Salabert}, {Bazot}, {Th{\'e}venin}, {Mathur}, {Xu}, \&
  {Garc{\'\i}a}}]{Creevey2017}
{Creevey}, O.~L., {Metcalfe}, T.~S., {Schultheis}, M., {et~al.} 2017, \aap,
  601, A67

\bibitem[{{de Toma} {et~al.}(2000){de Toma}, {White}, \& {Harvey}}]{deToma2000}
{de Toma}, G., {White}, O.~R., \& {Harvey}, K.~L. 2000, \apj, 529, 1101

\bibitem[{{Deheuvels} {et~al.}(2016){Deheuvels}, {Brand{\~a}o}, {Silva
  Aguirre}, {Ballot}, {Michel}, {Cunha}, {Lebreton}, \&
  {Appourchaux}}]{Deheuvels2016}
{Deheuvels}, S., {Brand{\~a}o}, I., {Silva Aguirre}, V., {et~al.} 2016, \aap,
  589, A93

\bibitem[{{Dikpati} {et~al.}(2020){Dikpati}, {Gilman}, {Chatterjee},
  {McIntosh}, \& {Zaqarashvili}}]{Dikpati2020}
{Dikpati}, M., {Gilman}, P.~A., {Chatterjee}, S., {McIntosh}, S.~W., \&
  {Zaqarashvili}, T.~V. 2020, \apj, 896, 141

\bibitem[{{Distefano} {et~al.}(2017){Distefano}, {Lanzafame}, {Lanza},
  {Messina}, \& {Spada}}]{Distefano2017}
{Distefano}, E., {Lanzafame}, A.~C., {Lanza}, A.~F., {Messina}, S., \& {Spada},
  F. 2017, \aap, 606, A58

\bibitem[{{Domingo} {et~al.}(1995){Domingo}, {Fleck}, \&
  {Poland}}]{Domingo1995}
{Domingo}, V., {Fleck}, B., \& {Poland}, A.~I. 1995, \solphys, 162, 1

\bibitem[{{Duane} {et~al.}(1987){Duane}, {Kennedy}, {Pendleton}, \&
  {Roweth}}]{Duane1987}
{Duane}, S., {Kennedy}, A.~D., {Pendleton}, B.~J., \& {Roweth}, D. 1987,
  Physics Letters B, 195, 216

\bibitem[{{Foukal} {et~al.}(1991){Foukal}, {Harvey}, \& {Hill}}]{Foukal1991}
{Foukal}, P., {Harvey}, K., \& {Hill}, F. 1991, \apjl, 383, L89

\bibitem[{{Fr{\"o}hlich} {et~al.}(1995){Fr{\"o}hlich}, {Romero}, {Roth},
  {Wehrli}, {Andersen}, {Appourchaux}, {Domingo}, {Telljohann}, {Berthomieu},
  {Delache}, {Provost}, {Toutain}, {Crommelynck}, {Chevalier}, {Fichot},
  {D{\"a}ppen}, {Gough}, {Hoeksema}, {Jim{\'e}nez}, {G{\'o}mez}, {Herreros},
  {Cort{\'e}s}, {Jones}, {Pap}, \& {Willson}}]{Frohlich1995}
{Fr{\"o}hlich}, C., {Romero}, J., {Roth}, H., {et~al.} 1995, \solphys, 162, 101

\bibitem[{{Garc{\'\i}a} {et~al.}(2011){Garc{\'\i}a}, {Hekker}, {Stello},
  {Guti{\'e}rrez-Soto}, {Handberg}, {Huber}, {Karoff}, {Uytterhoeven},
  {Appourchaux}, {Chaplin}, {Elsworth}, {Mathur}, {Ballot},
  {Christensen-Dalsgaard}, {Gilliland}, {Houdek}, {Jenkins}, {Kjeldsen},
  {McCauliff}, {Metcalfe}, {Middour}, {Molenda-Zakowicz}, {Monteiro}, {Smith},
  \& {Thompson}}]{Garcia2011}
{Garc{\'\i}a}, R.~A., {Hekker}, S., {Stello}, D., {et~al.} 2011, \mnras, 414,
  L6

\bibitem[{{Garc{\'\i}a} {et~al.}(2014){Garc{\'\i}a}, {Mathur}, {Pires},
  {R{\'e}gulo}, {Bellamy}, {Pall{\'e}}, {Ballot}, {Barcel{\'o} Forteza},
  {Beck}, \& {Bedding}}]{Garcia2014}
{Garc{\'\i}a}, R.~A., {Mathur}, S., {Pires}, S., {et~al.} 2014, \aap, 568, A10

\bibitem[{{Garc{\'\i}a} {et~al.}(2005){Garc{\'\i}a}, {Turck-Chi{\`e}ze},
  {Boumier}, {Robillot}, {Bertello}, {Charra}, {Dzitko}, {Gabriel},
  {Jim{\'e}nez-Reyes}, {Pall{\'e}}, {Renaud}, {Roca Cort{\'e}s}, \&
  {Ulrich}}]{Garcia2005}
{Garc{\'\i}a}, R.~A., {Turck-Chi{\`e}ze}, S., {Boumier}, P., {et~al.} 2005,
  \aap, 442, 385

\bibitem[{{Gizon} {et~al.}(2021){Gizon}, {Cameron}, {Bekki}, {Birch}, {Bogart},
  {Brun}, {Damiani}, {Fournier}, {Hyest}, {Jain}, {Lekshmi}, {Liang}, \&
  {Proxauf}}]{Gizon2021}
{Gizon}, L., {Cameron}, R.~H., {Bekki}, Y., {et~al.} 2021, \aap, 652, L6

\bibitem[{{Gurgenashvili} {et~al.}(2017){Gurgenashvili}, {Zaqarashvili},
  {Kukhianidze}, {Oliver}, {Ballester}, {Dikpati}, \&
  {McIntosh}}]{Gurgenashvili2017}
{Gurgenashvili}, E., {Zaqarashvili}, T.~V., {Kukhianidze}, V., {et~al.} 2017,
  \apj, 845, 137

\bibitem[{{Gurgenashvili} {et~al.}(2016){Gurgenashvili}, {Zaqarashvili},
  {Kukhianidze}, {Oliver}, {Ballester}, {Ramishvili}, {Shergelashvili},
  {Hanslmeier}, \& {Poedts}}]{Gurgenashvili2016}
{Gurgenashvili}, E., {Zaqarashvili}, T.~V., {Kukhianidze}, V., {et~al.} 2016,
  \apj, 826, 55

\bibitem[{{Gurgenashvili} {et~al.}(2021){Gurgenashvili}, {Zaqarashvili},
  {Kukhianidze}, {Reiners}, {Oliver}, {Lanza}, \&
  {Reinhold}}]{Gurgenashvili2021}
{Gurgenashvili}, E., {Zaqarashvili}, T.~V., {Kukhianidze}, V., {et~al.} 2021,
  \aap, 653, A146

\bibitem[{{Gurgenashvili} {et~al.}(2022){Gurgenashvili}, {Zaqarashvili},
  {Kukhianidze}, {Reiners}, {Reinhold}, \& {Lanza}}]{Gurgenashvili2022}
{Gurgenashvili}, E., {Zaqarashvili}, T.~V., {Kukhianidze}, V., {et~al.} 2022,
  \aap, 660, A33

\bibitem[{{Hall} {et~al.}(2021){Hall}, {Davies}, {van Saders}, {Nielsen},
  {Lund}, {Chaplin}, {Garc{\'\i}a}, {Amard}, {Breimann}, {Khan}, {See}, \&
  {Tayar}}]{Hall2021}
{Hall}, O.~J., {Davies}, G.~R., {van Saders}, J., {et~al.} 2021, Nature
  Astronomy, 5, 707

\bibitem[{Harris {et~al.}(2020)Harris, Millman, van~der Walt, Gommers,
  Virtanen, Cournapeau, Wieser, Taylor, Berg, Smith, Kern, Picus, Hoyer, van
  Kerkwijk, Brett, Haldane, del R{\'{i}}o, Wiebe, Peterson,
  G{\'{e}}rard-Marchant, Sheppard, Reddy, Weckesser, Abbasi, Gohlke, \&
  Oliphant}]{harris2020array}
Harris, C.~R., Millman, K.~J., van~der Walt, S.~J., {et~al.} 2020, Nature, 585,
  357

\bibitem[{Hunter(2007)}]{Hunter:2007}
Hunter, J.~D. 2007, Computing in Science \& Engineering, 9, 90

\bibitem[{{I{\c{s}}{\i}k} {et~al.}(2018){I{\c{s}}{\i}k}, {Solanki}, {Krivova},
  \& {Shapiro}}]{Isik2018}
{I{\c{s}}{\i}k}, E., {Solanki}, S.~K., {Krivova}, N.~A., \& {Shapiro}, A.~I.
  2018, \aap, 620, A177

\bibitem[{{Karoff} {et~al.}(2018){Karoff}, {Metcalfe}, {Santos}, {Montet},
  {Isaacson}, {Witzke}, {Shapiro}, {Mathur}, {Davies}, {Lund}, {Garcia},
  {Brun}, {Salabert}, {Avelino}, {van Saders}, {Egeland}, {Cunha}, {Campante},
  {Chaplin}, {Krivova}, {Solanki}, {Stritzinger}, \& {Knudsen}}]{Karoff2018}
{Karoff}, C., {Metcalfe}, T.~S., {Santos}, {\^A}. R.~G., {et~al.} 2018, \apj,
  852, 46

\bibitem[{{Kostyuchenko} \& {Bruevich}(2021)}]{Kostyuchenko2021}
{Kostyuchenko}, I. \& {Bruevich}, E. 2021, \solphys, 296, 8

\bibitem[{{Kraft}(1967)}]{Kraft1967}
{Kraft}, R.~P. 1967, \apj, 150, 551

\bibitem[{{Lanza}(2016)}]{Lanza2016}
{Lanza}, A.~F. 2016, in Lecture Notes in Physics, Berlin Springer Verlag, ed.
  J.-P. {Rozelot} \& C.~{Neiner}, Vol. 914, 43

\bibitem[{{Lanza} {et~al.}(2007){Lanza}, {Bonomo}, \& {Rodon{\`o}}}]{Lanza2007}
{Lanza}, A.~F., {Bonomo}, A.~S., \& {Rodon{\`o}}, M. 2007, \aap, 464, 741

\bibitem[{{Lanza} {et~al.}(2019){Lanza}, {Netto}, {Bonomo}, {Parviainen},
  {Valio}, \& {Aigrain}}]{Lanza2019}
{Lanza}, A.~F., {Netto}, Y., {Bonomo}, A.~S., {et~al.} 2019, \aap, 626, A38

\bibitem[{{Lanza} {et~al.}(2009){Lanza}, {Pagano}, {Leto}, {Messina},
  {Aigrain}, {Alonso}, {Auvergne}, {Baglin}, {Barge}, {Bonomo}, {Boumier},
  {Collier Cameron}, {Comparato}, {Cutispoto}, {de Medeiros}, {Foing},
  {Kaiser}, {Moutou}, {Parihar}, {Silva-Valio}, \& {Weiss}}]{Lanza2009}
{Lanza}, A.~F., {Pagano}, I., {Leto}, G., {et~al.} 2009, \aap, 493, 193

\bibitem[{{Lean} \& {Brueckner}(1989)}]{Lean1989}
{Lean}, J.~L. \& {Brueckner}, G.~E. 1989, \apj, 337, 568

\bibitem[{Liu {et~al.}(2007)Liu, San~Liang, \& Weisberg}]{Liu2007}
Liu, Y., San~Liang, X., \& Weisberg, R.~H. 2007, Journal of Atmospheric and
  Oceanic Technology, 24, 2093

\bibitem[{{L{\"o}ptien} {et~al.}(2018){L{\"o}ptien}, {Gizon}, {Birch}, {Schou},
  {Proxauf}, {Duvall}, {Bogart}, \& {Christensen}}]{Loptien2018}
{L{\"o}ptien}, B., {Gizon}, L., {Birch}, A.~C., {et~al.} 2018, Nature
  Astronomy, 2, 568

\bibitem[{{Luger} {et~al.}(2021){Luger}, {Foreman-Mackey}, {Hedges}, \&
  {Hogg}}]{Luger2021}
{Luger}, R., {Foreman-Mackey}, D., {Hedges}, C., \& {Hogg}, D.~W. 2021, \aj,
  162, 123

\bibitem[{{Lund} {et~al.}(2017){Lund}, {Silva Aguirre}, {Davies}, {Chaplin},
  {Christensen-Dalsgaard}, {Houdek}, {White}, {Bedding}, {Ball}, {Huber},
  {Antia}, {Lebreton}, {Latham}, {Handberg}, {Verma}, {Basu}, {Casagrande},
  {Justesen}, {Kjeldsen}, \& {Mosumgaard}}]{Lund2017}
{Lund}, M.~N., {Silva Aguirre}, V., {Davies}, G.~R., {et~al.} 2017, \apj, 835,
  172

\bibitem[{{Mathur} {et~al.}(2014{\natexlab{a}}){Mathur}, {Garc{\'\i}a},
  {Ballot}, {Ceillier}, {Salabert}, {Metcalfe}, {R{\'e}gulo}, {Jim{\'e}nez}, \&
  {Bloemen}}]{Mathur2014}
{Mathur}, S., {Garc{\'\i}a}, R.~A., {Ballot}, J., {et~al.} 2014{\natexlab{a}},
  \aap, 562, A124

\bibitem[{{Mathur} {et~al.}(2019){Mathur}, {Garc{\'\i}a}, {Bugnet}, {Santos},
  {Santiago}, \& {Beck}}]{Mathur2019}
{Mathur}, S., {Garc{\'\i}a}, R.~A., {Bugnet}, L., {et~al.} 2019, Frontiers in
  Astronomy and Space Sciences, 6, 46

\bibitem[{{Mathur} {et~al.}(2017){Mathur}, {Huber}, {Batalha}, {Ciardi},
  {Bastien}, {Bieryla}, {Buchhave}, {Cochran}, {Endl}, {Esquerdo}, {Furlan},
  {Howard}, {Howell}, {Isaacson}, {Latham}, {MacQueen}, \&
  {Silva}}]{Mathur2017}
{Mathur}, S., {Huber}, D., {Batalha}, N.~M., {et~al.} 2017, \apjs, 229, 30

\bibitem[{{Mathur} {et~al.}(2014{\natexlab{b}}){Mathur}, {Salabert},
  {Garc{\'\i}a}, \& {Ceillier}}]{Mathur2014b}
{Mathur}, S., {Salabert}, D., {Garc{\'\i}a}, R.~A., \& {Ceillier}, T.
  2014{\natexlab{b}}, Journal of Space Weather and Space Climate, 4, A15

\bibitem[{{Mittag} {et~al.}(2019){Mittag}, {Schmitt}, {Hempelmann}, \&
  {Schr{\"o}der}}]{Mittag2019}
{Mittag}, M., {Schmitt}, J.~H.~M.~M., {Hempelmann}, A., \& {Schr{\"o}der},
  K.~P. 2019, \aap, 621, A136

\bibitem[{{Montalto} {et~al.}(2021){Montalto}, {Piotto}, {Marrese},
  {Nascimbeni}, {Prisinzano}, {Granata}, {Marinoni}, {Desidera}, {Ortolani},
  {Aerts}, {Alei}, {Altavilla}, {Benatti}, {B{\"o}rner}, {Cabrera}, {Claudi},
  {Deleuil}, {Fabrizio}, {Gizon}, {Goupil}, {Heras}, {Magrin}, {Malavolta},
  {Mas-Hesse}, {Pagano}, {Paproth}, {Pertenais}, {Pollacco}, {Ragazzoni},
  {Ramsay}, {Rauer}, \& {Udry}}]{Montalto2021}
{Montalto}, M., {Piotto}, G., {Marrese}, P.~M., {et~al.} 2021, \aap, 653, A98

\bibitem[{{Neal}(2011)}]{Neal2011}
{Neal}, R. 2011, in Handbook of Markov Chain Monte Carlo, 113--162

\bibitem[{{Papaloizou} \& {Pringle}(1978)}]{Papaloizou1978}
{Papaloizou}, J. \& {Pringle}, J.~E. 1978, \mnras, 182, 423

\bibitem[{{Philidet} \& {Gizon}(2023)}]{Philidet2023}
{Philidet}, J. \& {Gizon}, L. 2023, \aap, 673, A124

\bibitem[{{Pires} {et~al.}(2015){Pires}, {Mathur}, {Garc{\'\i}a}, {Ballot},
  {Stello}, \& {Sato}}]{Pires2015}
{Pires}, S., {Mathur}, S., {Garc{\'\i}a}, R.~A., {et~al.} 2015, \aap, 574, A18

\bibitem[{{Rauer} {et~al.}(2014){Rauer}, {Catala}, {Aerts}, {Appourchaux},
  {Benz}, {Brandeker}, {Christensen-Dalsgaard}, {Deleuil}, {Gizon}, {Goupil},
  {G{\"u}del}, {Janot-Pacheco}, {Mas-Hesse}, {Pagano}, {Piotto}, {Pollacco},
  {Santos}, {Smith}, {Su{\'a}rez}, {Szab{\'o}}, {Udry}, {Adibekyan}, {Alibert},
  {Almenara}, {Amaro-Seoane}, {Eiff}, {Asplund}, {Antonello}, {Barnes},
  {Baudin}, {Belkacem}, {Bergemann}, {Bihain}, {Birch}, {Bonfils}, {Boisse},
  {Bonomo}, {Borsa}, {Brand {\~a}o}, {Brocato}, {Brun}, {Burleigh}, {Burston},
  {Cabrera}, {Cassisi}, {Chaplin}, {Charpinet}, {Chiappini}, {Church},
  {Csizmadia}, {Cunha}, {Damasso}, {Davies}, {Deeg}, {D{\'\i}az}, {Dreizler},
  {Dreyer}, {Eggenberger}, {Ehrenreich}, {Eigm{\"u}ller}, {Erikson}, {Farmer},
  {Feltzing}, {de Oliveira Fialho}, {Figueira}, {Forveille}, {Fridlund},
  {Garc{\'\i}a}, {Giommi}, {Giuffrida}, {Godolt}, {Gomes da Silva}, {Granzer},
  {Grenfell}, {Grotsch-Noels}, {G{\"u}nther}, {Haswell}, {Hatzes},
  {H{\'e}brard}, {Hekker}, {Helled}, {Heng}, {Jenkins}, {Johansen},
  {Khodachenko}, {Kislyakova}, {Kley}, {Kolb}, {Krivova}, {Kupka}, {Lammer},
  {Lanza}, {Lebreton}, {Magrin}, {Marcos-Arenal}, {Marrese}, {Marques},
  {Martins}, {Mathis}, {Mathur}, {Messina}, {Miglio}, {Montalban}, {Montalto},
  {Monteiro}, {Moradi}, {Moravveji}, {Mordasini}, {Morel}, {Mortier},
  {Nascimbeni}, {Nelson}, {Nielsen}, {Noack}, {Norton}, {Ofir}, {Oshagh},
  {Ouazzani}, {P{\'a}pics}, {Parro}, {Petit}, {Plez}, {Poretti}, {Quirrenbach},
  {Ragazzoni}, {Raimondo}, {Rainer}, {Reese}, {Redmer}, {Reffert},
  {Rojas-Ayala}, {Roxburgh}, {Salmon}, {Santerne}, {Schneider}, {Schou},
  {Schuh}, {Schunker}, {Silva-Valio}, {Silvotti}, {Skillen}, {Snellen}, {Sohl},
  {Sousa}, {Sozzetti}, {Stello}, {Strassmeier}, {{\v{S}}vanda}, {Szab{\'o}},
  {Tkachenko}, {Valencia}, {Van Grootel}, {Vauclair}, {Ventura}, {Wagner},
  {Walton}, {Weingrill}, {Werner}, {Wheatley}, \& {Zwintz}}]{Rauer2014}
{Rauer}, H., {Catala}, C., {Aerts}, C., {et~al.} 2014, Experimental Astronomy,
  38, 249

\bibitem[{{Reinhold} {et~al.}(2019){Reinhold}, {Bell}, {Kuszlewicz}, {Hekker},
  \& {Shapiro}}]{Reinhold2019}
{Reinhold}, T., {Bell}, K.~J., {Kuszlewicz}, J., {Hekker}, S., \& {Shapiro},
  A.~I. 2019, \aap, 621, A21

\bibitem[{{Rieger} {et~al.}(1984){Rieger}, {Share}, {Forrest}, {Kanbach},
  {Reppin}, \& {Chupp}}]{Rieger1984}
{Rieger}, E., {Share}, G.~H., {Forrest}, D.~J., {et~al.} 1984, \nat, 312, 623

\bibitem[{{Rodon{\`o}} {et~al.}(2000){Rodon{\`o}}, {Messina}, {Lanza},
  {Cutispoto}, \& {Teriaca}}]{Rodono2000}
{Rodon{\`o}}, M., {Messina}, S., {Lanza}, A.~F., {Cutispoto}, G., \& {Teriaca},
  L. 2000, \aap, 358, 624

\bibitem[{{Roettenbacher} {et~al.}(2013){Roettenbacher}, {Monnier}, {Harmon},
  {Barclay}, \& {Still}}]{Roettenbacher2013}
{Roettenbacher}, R.~M., {Monnier}, J.~D., {Harmon}, R.~O., {Barclay}, T., \&
  {Still}, M. 2013, \apj, 767, 60

\bibitem[{{Saio}(1982)}]{Saio1982}
{Saio}, H. 1982, \apj, 256, 717

\bibitem[{{Salabert} {et~al.}(2016{\natexlab{a}}){Salabert}, {Garc{\'\i}a},
  {Beck}, {Egeland}, {Pall{\'e}}, {Mathur}, {Metcalfe}, {do Nascimento},
  {Ceillier}, {Andersen}, \& {Trivi{\~n}o Hage}}]{Salabert2016b}
{Salabert}, D., {Garc{\'\i}a}, R.~A., {Beck}, P.~G., {et~al.}
  2016{\natexlab{a}}, \aap, 596, A31

\bibitem[{{Salabert} {et~al.}(2017){Salabert}, {Garc{\'\i}a}, {Jim{\'e}nez},
  {Bertello}, {Corsaro}, \& {Pall{\'e}}}]{Salabert2017}
{Salabert}, D., {Garc{\'\i}a}, R.~A., {Jim{\'e}nez}, A., {et~al.} 2017, \aap,
  608, A87

\bibitem[{{Salabert} {et~al.}(2016{\natexlab{b}}){Salabert}, {R{\'e}gulo},
  {Garc{\'\i}a}, {Beck}, {Ballot}, {Creevey}, {P{\'e}rez Hern{\'a}ndez}, {do
  Nascimento}, {Corsaro}, {Egeland}, {Mathur}, {Metcalfe}, {Bigot}, {Ceillier},
  \& {Pall{\'e}}}]{Salabert2016}
{Salabert}, D., {R{\'e}gulo}, C., {Garc{\'\i}a}, R.~A., {et~al.}
  2016{\natexlab{b}}, \aap, 589, A118

\bibitem[{{Santos} {et~al.}(2021){Santos}, {Breton}, {Mathur}, \&
  {Garc{\'\i}a}}]{Santos2021}
{Santos}, A.~R.~G., {Breton}, S.~N., {Mathur}, S., \& {Garc{\'\i}a}, R.~A.
  2021, \apjs, 255, 17

\bibitem[{{Santos} {et~al.}(2018){Santos}, {Campante}, {Chaplin}, {Cunha},
  {Lund}, {Kiefer}, {Salabert}, {Garc{\'\i}a}, {Davies}, {Elsworth}, \&
  {Howe}}]{Santos2018}
{Santos}, A.~R.~G., {Campante}, T.~L., {Chaplin}, W.~J., {et~al.} 2018, \apjs,
  237, 17

\bibitem[{{Santos} {et~al.}(2019){Santos}, {Garc{\'\i}a}, {Mathur}, {Bugnet},
  {van Saders}, {Metcalfe}, {Simonian}, \& {Pinsonneault}}]{Santos2019}
{Santos}, A.~R.~G., {Garc{\'\i}a}, R.~A., {Mathur}, S., {et~al.} 2019, \apjs,
  244, 21

\bibitem[{{Silva Aguirre} {et~al.}(2017){Silva Aguirre}, {Lund}, {Antia},
  {Ball}, {Basu}, {Christensen-Dalsgaard}, {Lebreton}, {Reese}, {Verma},
  {Casagrande}, {Justesen}, {Mosumgaard}, {Chaplin}, {Bedding}, {Davies},
  {Handberg}, {Houdek}, {Huber}, {Kjeldsen}, {Latham}, {White}, {Coelho},
  {Miglio}, \& {Rendle}}]{SilvaAguirre2017}
{Silva Aguirre}, V., {Lund}, M.~N., {Antia}, H.~M., {et~al.} 2017, \apj, 835,
  173

\bibitem[{{Sing}(2010)}]{Sing2010}
{Sing}, D.~K. 2010, \aap, 510, A21

\bibitem[{{Spada} \& {Lanzafame}(2020)}]{Spada2020}
{Spada}, F. \& {Lanzafame}, A.~C. 2020, \aap, 636, A76

\bibitem[{{Stein} \& {Nordlund}(2012)}]{Stein2012}
{Stein}, R.~F. \& {Nordlund}, {\r{A}}. 2012, \apjl, 753, L13

\bibitem[{{Thomas} {et~al.}(2019){Thomas}, {Chaplin}, {Davies}, {Howe},
  {Santos}, {Elsworth}, {Miglio}, {Campante}, \& {Cunha}}]{Thomas2019}
{Thomas}, A. E.~L., {Chaplin}, W.~J., {Davies}, G.~R., {et~al.} 2019, \mnras,
  485, 3857

\bibitem[{{Toriumi} \& {Yokoyama}(2012)}]{Toriumi2012}
{Toriumi}, S. \& {Yokoyama}, T. 2012, \aap, 539, A22

\bibitem[{{Torrence} \& {Compo}(1998)}]{Torrence1998}
{Torrence}, C. \& {Compo}, G.~P. 1998, Bulletin of the American Meteorological
  Society, 79, 61

\bibitem[{{Usoskin} {et~al.}(2007){Usoskin}, {Berdyugina}, {Moss}, \&
  {Sokoloff}}]{Usoskin2007}
{Usoskin}, I.~G., {Berdyugina}, S.~V., {Moss}, D., \& {Sokoloff}, D.~D. 2007,
  Advances in Space Research, 40, 951

\bibitem[{Virtanen {et~al.}(2020)Virtanen, Gommers, Oliphant, Haberland, Reddy,
  Cournapeau, Burovski, Peterson, Weckesser, Bright, {van der Walt}, Brett,
  Wilson, Millman, Mayorov, Nelson, Jones, Kern, Larson, Carey, Polat, Feng,
  Moore, {VanderPlas}, Laxalde, Perktold, Cimrman, Henriksen, Quintero, Harris,
  Archibald, Ribeiro, Pedregosa, {van Mulbregt}, \& {SciPy 1.0
  Contributors}}]{2020SciPy-NMeth}
Virtanen, P., Gommers, R., Oliphant, T.~E., {et~al.} 2020, Nature Methods, 17,
  261

\bibitem[{{Weber} {et~al.}(2013){Weber}, {Fan}, \& {Miesch}}]{Weber2013}
{Weber}, M.~A., {Fan}, Y., \& {Miesch}, M.~S. 2013, \apj, 770, 149

\bibitem[{Wiecki {et~al.}(2023)Wiecki, Salvatier, Vieira, Kochurov, Patil,
  Osthege, Willard, \& Engels}]{pymc}
Wiecki, T., Salvatier, J., Vieira, R., {et~al.} 2023, pymc-devs/pymc: v5.4.0

\bibitem[{{Zaqarashvili} {et~al.}(2021){Zaqarashvili}, {Albekioni},
  {Ballester}, {Bekki}, {Biancofiore}, {Birch}, {Dikpati}, {Gizon},
  {Gurgenashvili}, {Heifetz}, {Lanza}, {McIntosh}, {Ofman}, {Oliver},
  {Proxauf}, {Umurhan}, \& {Yellin-Bergovoy}}]{Zaqarashvili2021}
{Zaqarashvili}, T.~V., {Albekioni}, M., {Ballester}, J.~L., {et~al.} 2021,
  \ssr, 217, 15

\bibitem[{{Zaqarashvili} \& {Gurgenashvili}(2018)}]{Zaqarashvili2018}
{Zaqarashvili}, T.~V. \& {Gurgenashvili}, E. 2018, Frontiers in Astronomy and
  Space Sciences, 5, 7

\bibitem[{{Zaqarashvili} {et~al.}(2007){Zaqarashvili}, {Oliver}, {Ballester},
  \& {Shergelashvili}}]{Zaqarashvili2007}
{Zaqarashvili}, T.~V., {Oliver}, R., {Ballester}, J.~L., \& {Shergelashvili},
  B.~M. 2007, \aap, 470, 815

\end{thebibliography}

\appendix


\end{document}